**ROHDE&SCHWARZ**

ROHDE & SCHWARZ UK Ltd • Ancells Business Park • Fleet • GU51 2UZ

**ROHDE & SCHWARZ UK Ltd**

| | |
|---|---|
| General | + 44 (0) 1252 811377 |
| Sales | + 44 (0) 1252 818888 |
| Service | + 44 (0) 1252 818818 |
| Customer Support | + 44 (0) 1252 818900 |
| Fax | + 44 (0) 1252 811447 |
| Email | sales@rsuk.rohde-schwarz.com<br>service@rsuk.rohde-schwarz.com |

Reg No. 539607

Date: 18th August 2011

To: Mr. Ardavan Rahimian

| Your ref./dated | Our ref. | Responsible | Direct Line<br>01252 818 - |
|---|---|---|---|

**<u>Reference: Confirmation Letter for Rohde & Schwarz Technology Prize 2009</u>**

Dear Mr Rahimian,

I can confirm that you were the winner of the Rohde & Schwarz Technology Prize in 2009 at the University of Birmingham for your poster presentation titled “Steerable Antennas for Automotive Communication Systems”.

A wide variety of Engineering poster presentations were on display from MEng & BEng students in the Dept of Electronic, Electrical and Computer Engineering. The criteria for awarding you the prize was based around best oral overview of the project and defence of results, presentation layout and relevance to industry.

Good luck with your future Engineering career.

Kind regards

Phil McCluskey
University Business Development

# Steerable Antennas for Automotive Communication Systems

by

Ardavan Rahimian

(915288)

Dissertation

Submitted to the School of Electronic, Electrical and Computer Engineering

in Partial Fulfilment of the Requirements for the Degree of

Master of Engineering in Electronic and Communications Engineering

at the

UNIVERSITY OF BIRMINGHAM

April 2009

Supervisor: Dr Peter Gardner

# ACKNOWLEDGEMENTS

First and foremost I would like to thank my parents. If it was not their love, support, sacrifices, and the great interest that they took in my education, I would not be able to accomplish what I did. It is for all of this that I dedicate this dissertation to them.

My greatest appreciation surely belongs to Dr Peter Gardner. I appreciate the opportunity that he gave me to continue my studies under his supervision in the research field that I truly love. I certainly could not have asked for a better supervisor or a greater opportunity. He has been an outstanding teacher and his academic guidance, continuous encouragement, motivation, advice, and support throughout this research project have taught a lot to me.

I would like to thank Mr Alan Yates for offering his kind help, support, and assistance during the fabrication process.

I also would like to thank my assessors, Prof Mike Cherniakov and Dr Sandra I. Woolley for useful feedbacks and time and consideration they have put to read throughout the dissertation.

Ardavan Rahimian

*Birmingham, United Kingdom*

*06 April, 2009*

# ABSTRACT

This research project undertakes a comprehensive analysis of RF beamforming techniques for design, simulation, fabrication, and measurement of Butler Matrix and Rotman Lens beamforming networks. It is aimed to develop novel and well–established designs for steerable antenna systems that can be used in vehicular telematics and automotive communication systems based on microwave and millimeter–wave techniques.

# TABLE OF CONTENTS

# CHAPTER 1
# INTRODUCTION AND BACKGROUND

## 1.1 INTRODUCTION

Generating multiple beams using an array along with having wide bandwidth and beam steering capability are of crucial importance for modern radar and communication systems. For this purpose, various multiple beamforming networks are introduced to have control over the amplitude and phase at each element of the antenna array. Microwave passive networks form an important class of these multiple beamforming networks. Two well–known examples of such networks are Butler Matrix (circuit beamformer) and Rotman Lens (lens–based beamformer). In this contribution, the mentioned beamforming networks will be introduced, simulated, fabricated, and measured to develop well–established designs for systems that can be used in vehicular telematics and ITS applications based on microwave and millimeter–wave techniques.

## 1.2 BACKGROUND

An Intelligent Transportation System (ITS) is a system based on wireless communication which has been investigated for many years in Europe, North America, and Japan in order to provide new technologies able to improve safety and efficiency of road transportation with integrated vehicle and road systems. It combines all aspects of technology and systems designs concepts in order to develop and improve transportation system of all kinds [1].

In the last 20 years Europe has had several large–scale programs towards the Road Transport Informantics (RTI). From this scheme, the two main projects that had been undertaken are Dedicated Road Infrastructures for Vehicle Safety in Europe (DRIVE) and Program for European Traffic with Highest Efficiency and Unprecedented Safety (PROMETHEUS). DRIVE which is under the control of the Commission of European Communities (CEC) tends to concentrate on human behaviour issues and implementation of the systems into the entire European community. PROMETHEUS which was started in 1986 as part of the European Research Coordination Agency (EUREKA) tries to develop the actual systems and apply the following four categories: Improved driver information, Active driver support, Cooperative driving, and Traffic and fleet management. But all of these projects aimed to improve road transportation and reduce traffic problems in Europe [2].

In addition, FATCAT and MILTRANS projects have also been accomplished in the United Kingdom towards the ITS mission. The projects were being carried out by BAE Systems Advanced Technology Centre, QinetiQ, Panorama Antennas and University of Birmingham as the foresight vehicle projects funded by UK Highways Agency, DTI and company investment based on recent advances in MMIC technology and antenna options in the millimetric frequency bands [3].

The inter–vehicle (IVC) system is used to allow a direct contact between vehicles and exchange of information related to their position, speed and etc. But the roadside–to–vehicle (RVC) system is a system that communicates between the drivers and the control system. The system is configured from wireless facilities such as base–transmit stations installed along the road to on–board units installed on vehicles [4].

The MILTRANS project, aimed to design and demonstrate the operation of a high capacity IVC and RVC data link operating in the 63–64 GHz frequency band that will be suitable for the next generation of advanced data services and traffic management systems [5]. The concept of IVC and RVC can be seen in Figure 1.1.

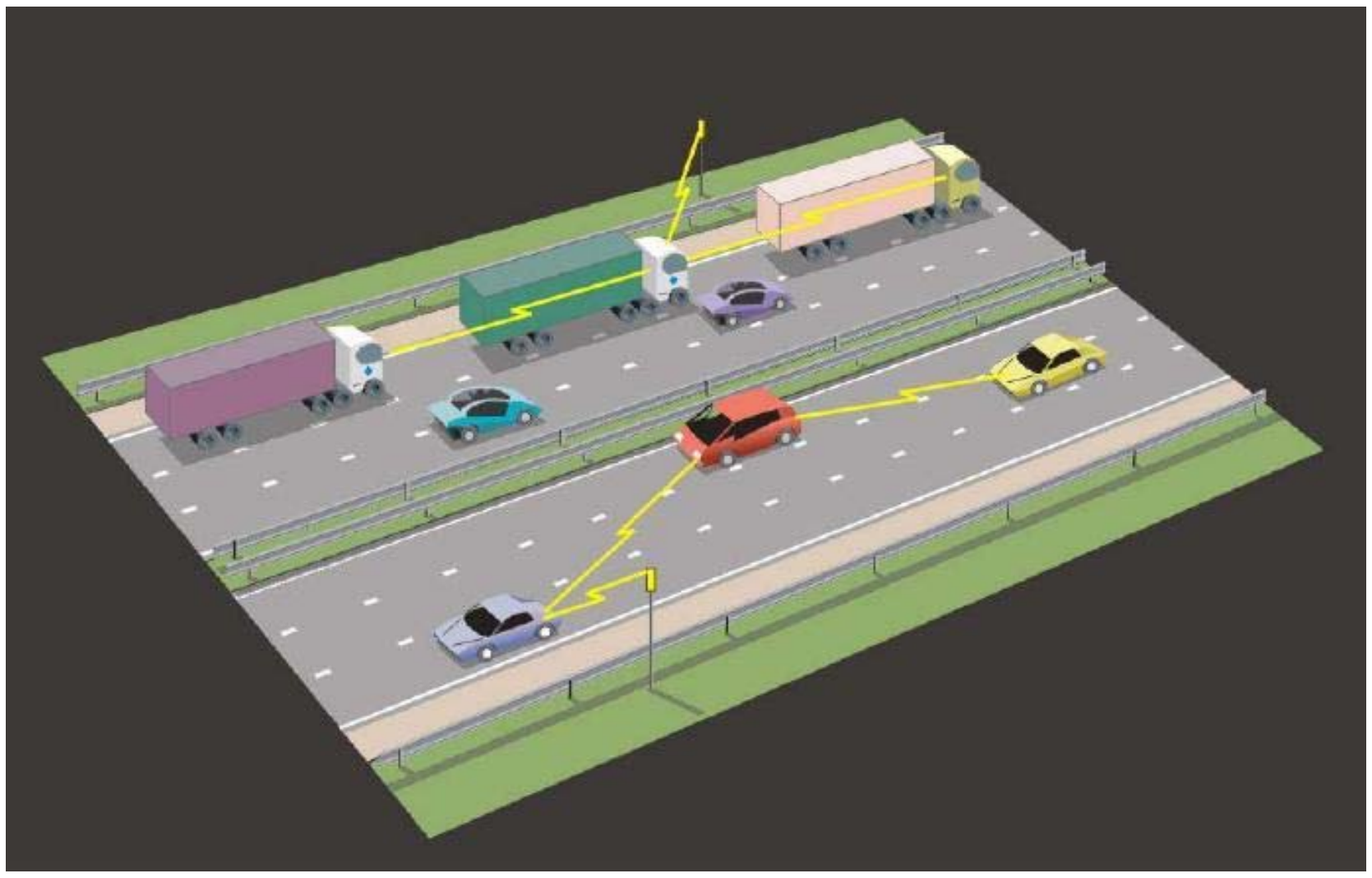

Figure 1.1 IVC and RVC Concepts, taken from [3].

## 1.3 AIMS AND OBJECTIVES

This project aims to envisage the radio communications links between vehicles which leads to the formation of ad–hoc networks between clusters of vehicles and roadside beacons. Several approaches are available for the improvement of the system performance including mechanical steering of a fixed array, switching between fixed patterns using Butler or Blass Matrices or Rotman or Ruze lenses, or continuous steering using a splitter and multiple phase shifters driving the elements of an array.

The primary tasks are formulating the specifications of the system to choose the appropriate justified design approaches to simulate, fabricate, and measure three competing beamforming networks based on the Butler Matrix (as the circuit beamformer) in the working frequency of 3.15 GHz and the Rotman Lenses (as the lens–based beamformer) in the working frequencies of 3.15 GHz and 6.3 GHz. There will be several detailed computer simulations using AWR Microwave Office 2007 (MWO), Remcom Rotman Lens Designer (RLD), and MATLAB software packages for the selected designs. After the fabrication of the frequency scaled models, the networks will be tested using the network analyser for the beam steering concept verification.

## 1.4 DISSERTATION ORGANISATION

This dissertation is organised into seven chapters to cover all the important and required materials. The summary of each chapter is as follow:

*Chapter 1 – Introduction and Background*: This chapter covers the project introduction and background, aims and objectives, and organsiation of the dissertation.

*Chapter 2 – Literature Survey*: This chapter introduces the critical RF front–end elements including antenna arrays, multibeam antennas, and RF beamforming terminology and concepts along with its different techniques like circuit and lens–based beamforming.

*Chapter 3 – The Butler Matrix Simulations*: This chapter covers the design and simulation procedures of the *4x4* Butler Matrix with its elements including the branch line coupler, 0dB crossover, and phase shifters and their output characteristics.

*Chapter 4 – The Butler Matrix Measurements*: This chapter presents measurements of the *4x4* Butler Matrix including insertion loss, output phase, and computed array factor radiation patterns of the output ports.

*Chapter 5 – The Rotman Lens Simulations*: This chapter introduces the Rotman Lens theory and design equations and simulation of the *4x4* and *8x8* Rotman Lenses along with their output characteristics including beam to array phase errors, beam to array couplings, and array factor radiation patterns.

*Chapter 6 – The Rotman Lens Measurements*: This chapter shows the measured results of the fabricated *8x8* Rotman Lens including the array factor radiation patterns corresponding each port and relative phase differences of the output ports.

*Chapter 7 – Conclusion and Future Work*: This chapter makes a comprehensive review of the whole materials covered in the dissertation that were undertaken during this research project along with some suggested further improvements.

# CHAPTER 2
# LITERATURE SURVEY

## 2.1 INTRODUCTION

This chapter will cover the sections introducing some RF front–end elements like antenna arrays, array factor, and beamforming techniques for multibeam antennas.

## 2.2 ANTENNA ARRAYS

An antenna array is an entity consisting of two or more element antennas. Antenna arrays may have many good properties, which cannot be achieved with a single element, such as high gain, narrow beam, shaped beam, scanning beam, or adaptive beam [6].

Figure 2.1 shows an array that consists of two elements having a separation of *d*. Let us assume that the far–field patterns of the antennas are $E_1(\varphi)$ and $E_2(\varphi)$ and that the phase difference of the feed currents is $\delta$. The total field produced by the array is:

$$E(\varphi) = E_1(\varphi)\, e^{-j(kd\cos\varphi + \delta)} + E_2(\varphi) \tag{2.1}$$

The path length difference of $d\cos\varphi$ in free space produces the phase difference of $kd\cos\varphi$. If the elements of the Figure 2.1 are similar, $E_1(\varphi) = E_2(\varphi)$, and they fed in phase so $\delta = 0$ and the above equation can be written as:

$$E(\varphi) = E_1(\varphi)\,(1 + e^{-jkd\cos\varphi}) \tag{2.2}$$

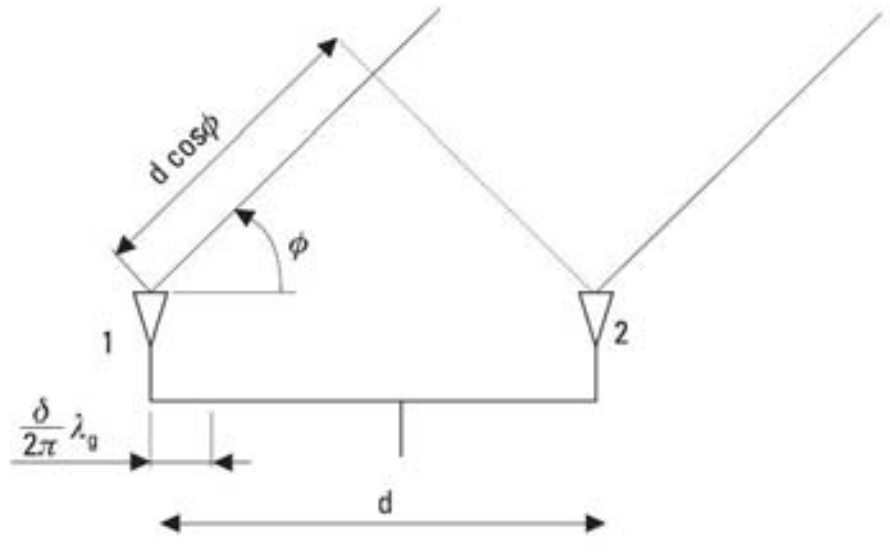

Figure 2.1 An Array of Two Elements, taken from [6].

The array pattern is the product of the element pattern and the array factor. The array factor (*AF*) is a function of the geometry of the antenna array element and the excitation phase and it quantifies the effect of combining radiating elements in an array without the element specific radiation pattern taken into account. The array factor of an *N* element array antenna is given and simplified as the following equations:

$$AF = a_1\, e^{+j(kd\cos\varphi + \beta_1)} + a_2\, e^{+j2(kd\cos\varphi + \beta_2)} + \ldots + a_N\, e^{+j(N-1)(kd\cos\varphi + \beta_N)}$$

$$AF = \sum_{n=1}^{N} a_n\, e^{+j(n-1)(kd\cos\varphi + \beta_n)} \tag{2.3}$$

where $\varphi$ is the beam angle, $d$ is the displacement of the elements, $n$ is the number of radiating element, $\beta$ is the difference in phase excitation between the elements and $a_n$'s are the excitation coefficients of the array elements. The array factor depends on the positions, amplitudes, and phases of the elements. The array may be linear, planar, or conformal (shaped according to the surface) [6].

An array is referred to as a broadside array when it has a maximum radiation in the direction perpendicular to the axis of the array (i.e. when $\varphi = 90^\circ$). An array is referred to as an end–fire array when it has a maximum radiation in the direction along the axis of the array (i.e. when $\varphi = 0^\circ$ or $\varphi = 180^\circ$).

The direction of the radiation for the main beam depends on the phase difference $\beta$ between the elements of the array. Hence it is possible to continuously steer the main beam in any direction by varying the progressive phase between the elements. This type of array in which the main beam is steered to the desired direction is referred to as a phased array [7].

The pattern of a two–element array antenna may look like that shown in Figure 2.2. It equals the element pattern multiplied by the array factor [6].

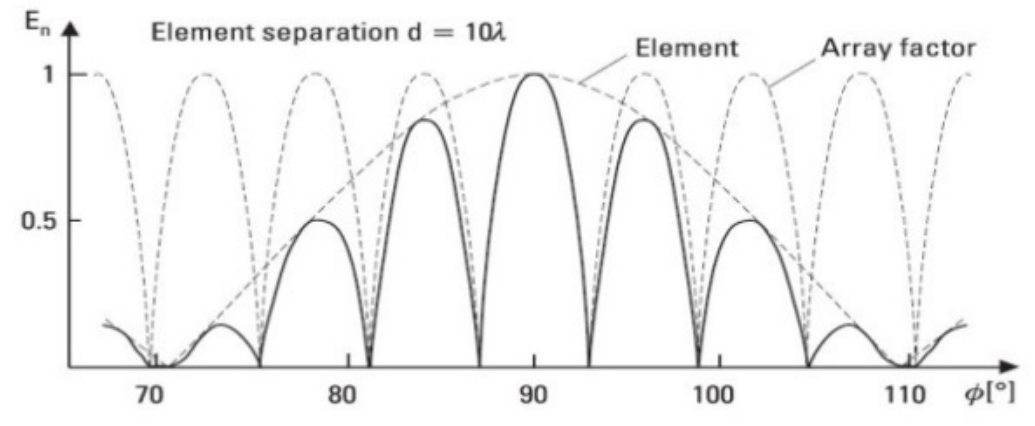

Figure 2.2 Pattern of a Two–element Array, taken from [6].

If there are electronically controlled phase shifters in the feed network of an array, the direction of the beam can be changed rapidly without rotating the antenna. This kind of electronic scanning is much faster than mechanical scanning. A phased array combined with digital signal processing may operate as a 'smart' antenna in which the pattern changes according to the electromagnetic environment [6].

## 2.2 BEAMFORMING TECHNIQUES FOR MULTIBEAM ANTENNAS

A multibeam antenna is one with a capability to form many beams in different directions from the same aperture. It usually consists of separate numbers of ports for different beams and these ports are well isolated. The beam ports can be connected to a transmit/receive system through a multiple way switch giving a sequentially scanning antenna. The field of RF beamforming techniques encompasses two major areas namely quasi–optic types, and circuit types. The quasi–optic types involve a hybrid arrangement of either a reflector or lens objective with a feed array whereas the circuit types use circuit to shape the beams [8].

### 2.2.1 LENS-BASED BEAMFORMERS

With a quasi–optic beamforming networks (BFN) the beam angles are fixed, while beam–widths change accordingly with frequency, causing crossover levels to change with frequency [9]. The examples of the lens based BFNs are the Ruze and Rotman lenses.

#### 2.2.1.1 THE RUZE LENS

The Ruze lens beamforming technique is shown in Figure 2.3. A Ruze lens consists of metal parallel plates that are capable of constraining energy to travel parallel to the axis of the lens and Rays within the lens travel parallel to path P–Q. The electrical path within the lens can be formed in a variety of ways such as waveguide, coaxial line, stripline, or microstrip [8]. Using ray tracing through points P–Q and the lens origin, two focal points $F_1$ and $F_2$ are assumed along the focal arc [4]. Various types of Ruze lenses can be designed by changing the shape of the inner contour and the thickness of the lens.

#### 2.2.1 .2 THE ROTMAN LENS

As the Figure 2.3 shows, the Rotman lens is similar to the Ruze lens except it uses distributed flexible transmission lines instead of parallel plate metals so that the lens interconnections are not parallel to the lens axis (a connection P–Q' is specified). This configuration is called the bootlace concept and allows the inputs and outputs to be displaced in order to optimize performance of the lens [9].

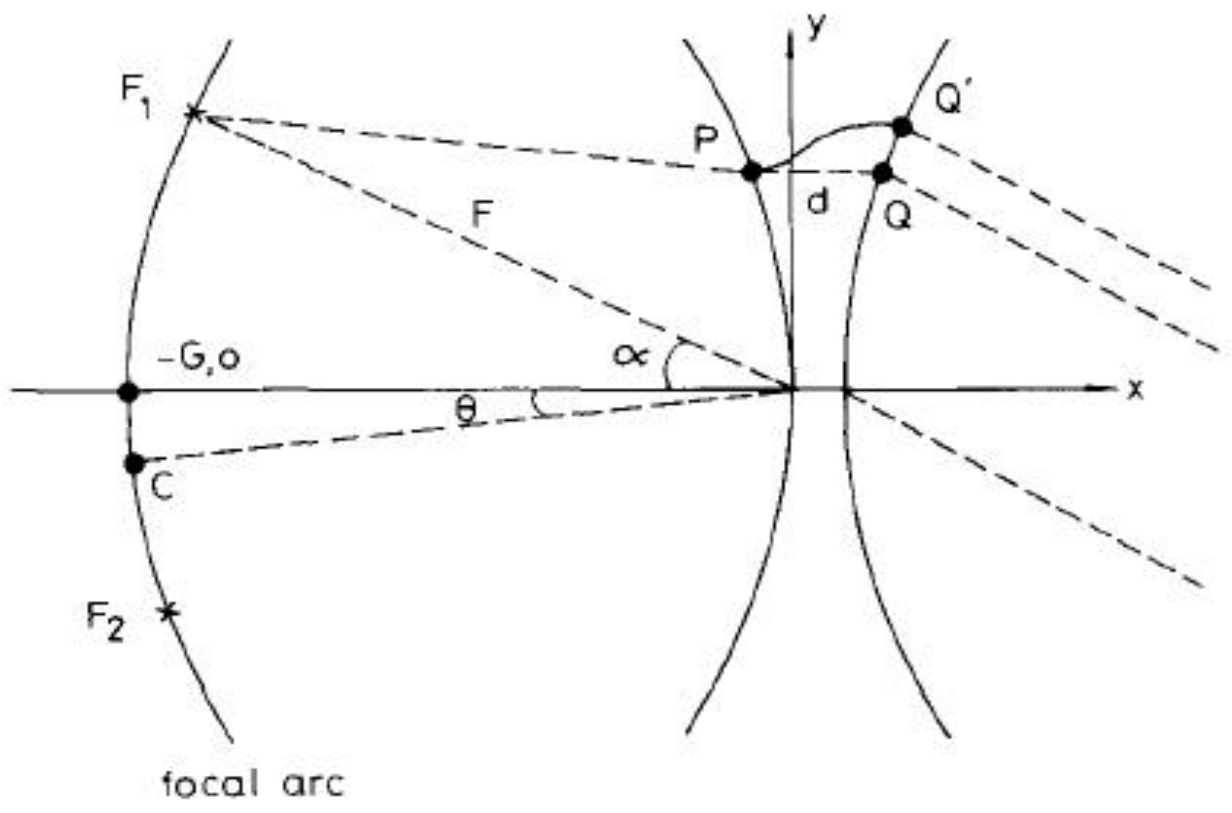

Figure 2.3 Geometry of the Constrained Lens[1], taken from [8].
(P–Q specifies Ruze Lens form whilst P–Q' specifies Rotman Lens)

### 2.2.2 CIRCUIT BEAMFORMERS

The circuit type beamformers use couplers, phase shifters, connecting power splitters, and transmission lines in order to shape the beams and create BFNs. Butler and Blass Matrices are examples of circuit beamformers.

#### 2.2.2.1 THE BLASS MATRIX

The Blass Matrix is another way of providing multiple beams for antenna systems. It consists of a number of travelling wave feed lines connected to a linear antenna array through another set of element lines as shown in Figure 2.4. These lines are interconnected by directional couplers at their crossover points and they are terminated with matched load at each line end [8]. A Blass Matrix can be designed for use with any numbers of elements. However, it is a lossy network because of the resistive terminations [10].

[1] The term *constrained lens* refers to the way the electromagnetic energy passes the lens face.

A signal travelling from the input will progress along the feed line to the termination end. At each crossover point a small signal will be coupled into each element line which excites the corresponding radiating element. The phase difference between each input terminal and the radiating element gives the radiated beam direction. The amplitude distribution of the beam generated can be controlled by having different coupling coefficients form the directional couplers. The structure shown in Figure 2.4 is a phase delay Blass Matrix, a true time–delay type [8]. The time delay Blass Matrix is a configuration that allows the path length from the input to all the radiating elements to be the same. This configuration avoids large delay in exciting the end elements and is suitable where the number of radiating elements is large [9].

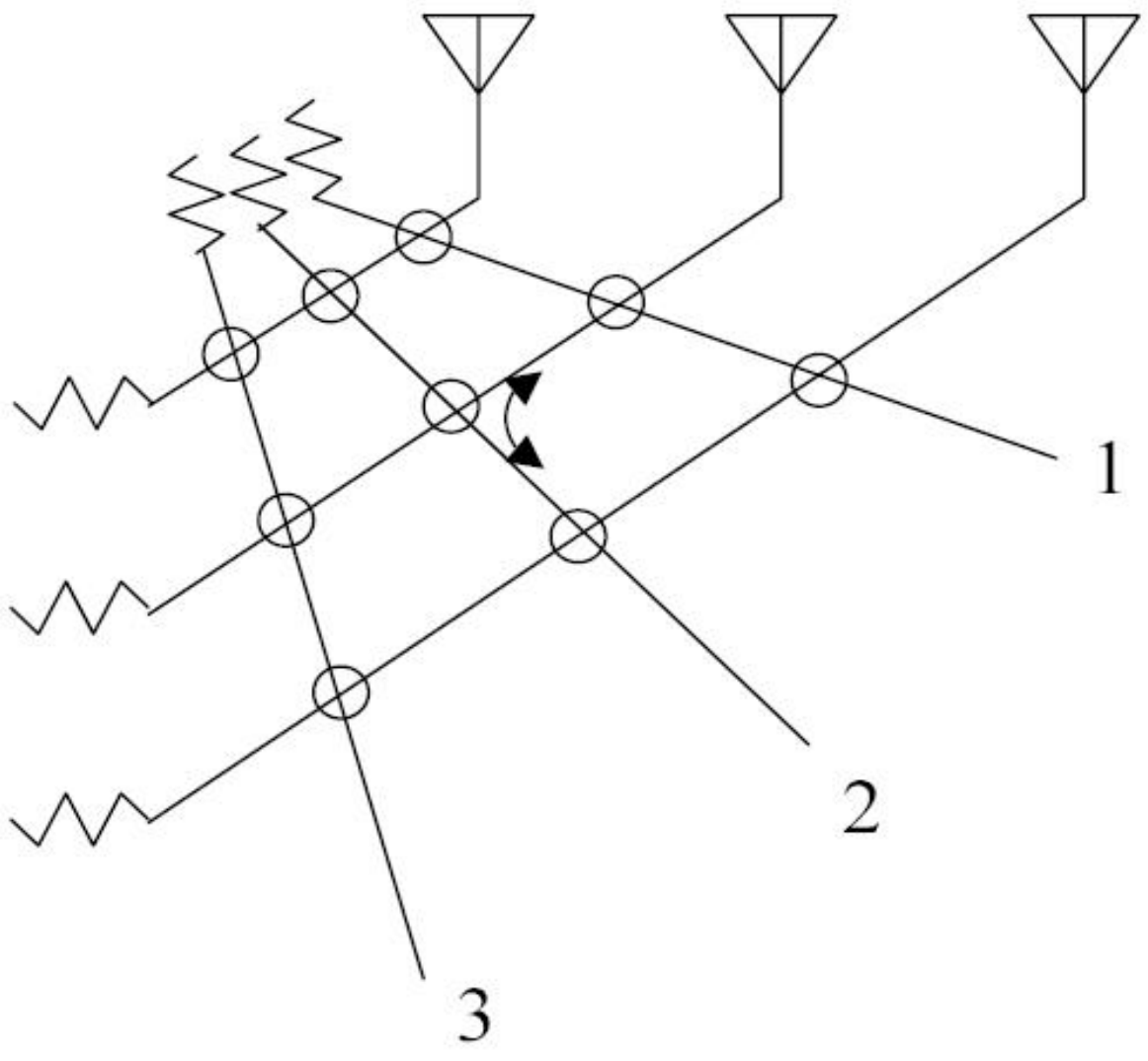

Figure 2.4 Blass Matrix (The Circles are Directional Couplers), taken from [10].

#### 2.2.2.2 THE BUTLER MATRIX

The Butler Matrix (BM) is a circuit beamformer consisting of interconnected 3dB hybrid couplers and fixed phase shifters [8]. It is a method of feeding an array antenna with a uniform distribution and constant phase difference between elements.

Butler matrix is a passive feeding *NxN* network with beam steering capabilities for phased array antennas with *N* outputs connected to antenna elements and *N* inputs or beam ports [11]. *N* must be an integer power of 2 (i.e. $N = 2^n$ where *n* is a positive integer) to form the network and for an *NxN* matrix, *N* possible beam directions can be formed. A total of $(N/2)log_2N$ hybrid couplers and $(N/2)log_2(N-1)$ fixed phase shifters are required to form the network.

An 8–element Butler Matrix is shown in Figure 2.5. A *4x4* matrix is the network generating four independent beams but more complicated *8x8*, *16x16*, and *32x32* networks can be designed [12]. A *64x64* matrix is perhaps the largest possible using microstrip technology on $\varepsilon_r = 2.6$ substrate [8].

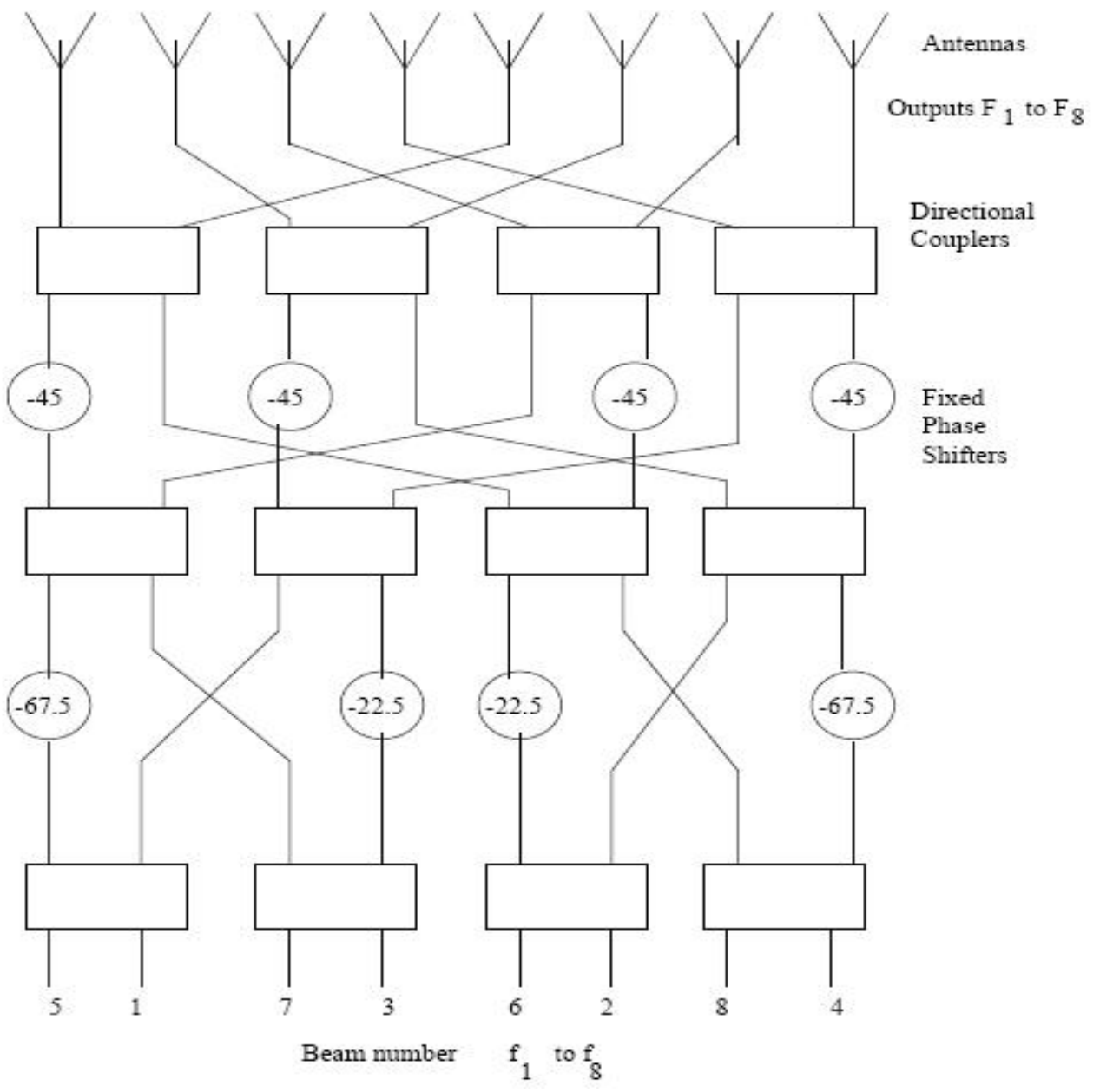

Figure 2.5 Eight–element Butler Matrix, taken from [8].

A Butler Matrix can also be considered as a multiport component which when loaded with radiating elements enables production of a beam of microwave energy in a specified direction in transmission line, dependent upon which input port is activated [13]. If the Butler matrix is connected to an antenna array, then the matrix will act so that the array will have a uniform amplitude distribution and constant phase difference between neighbouring elements. This will then generate the beams depending on which input port is used. The direction of the beam depends on the wavelength of the transmitted signal. If port one is used, then all other ports are terminated [14]. According to Moody, a Butler Matrix provides progressive phase variation at the output ports with a phase difference between radiating elements [15].

$$Phase\ Difference = \pm \frac{2n-1}{N} \times 180^{\circ} \quad (2.4)$$

where $N$ is the order of the Butler Matrix and $n$ is an integer varying from 1 to $N$. Hence, the phase differences for the *4x4* Butler Matrix are ±45˚ for output from ports 1 and 4 and ±135˚ for output from ports 2 and 3. By using the narrow beams available from the Butler Matrix, it is possible for a vehicle to increase gain in the desired signal directions and reduce the gain in interference directions. This is achieved by various phase shift characteristics at the output of the Butler Matrix when signals are fed into different input ports. Hence, high gain–narrow beam signals for long distance application are produced from the output of the first Butler Matrix. On the other hand, by cascading the Butler Matrix, reconstructing the antenna patterns of the individual radiating elements, have high linearity and broad–beam signals that can be used for short distance communication [4].

Figure 2.6 indicates the block structure of a *4x4* Butler Matrix consisting two 90˚ hybrid couplers (branch line couplers), two 0dB crossovers, two 45˚ phase shifters, and two output phase shifters which is going to be simulated and measured in later chapters.

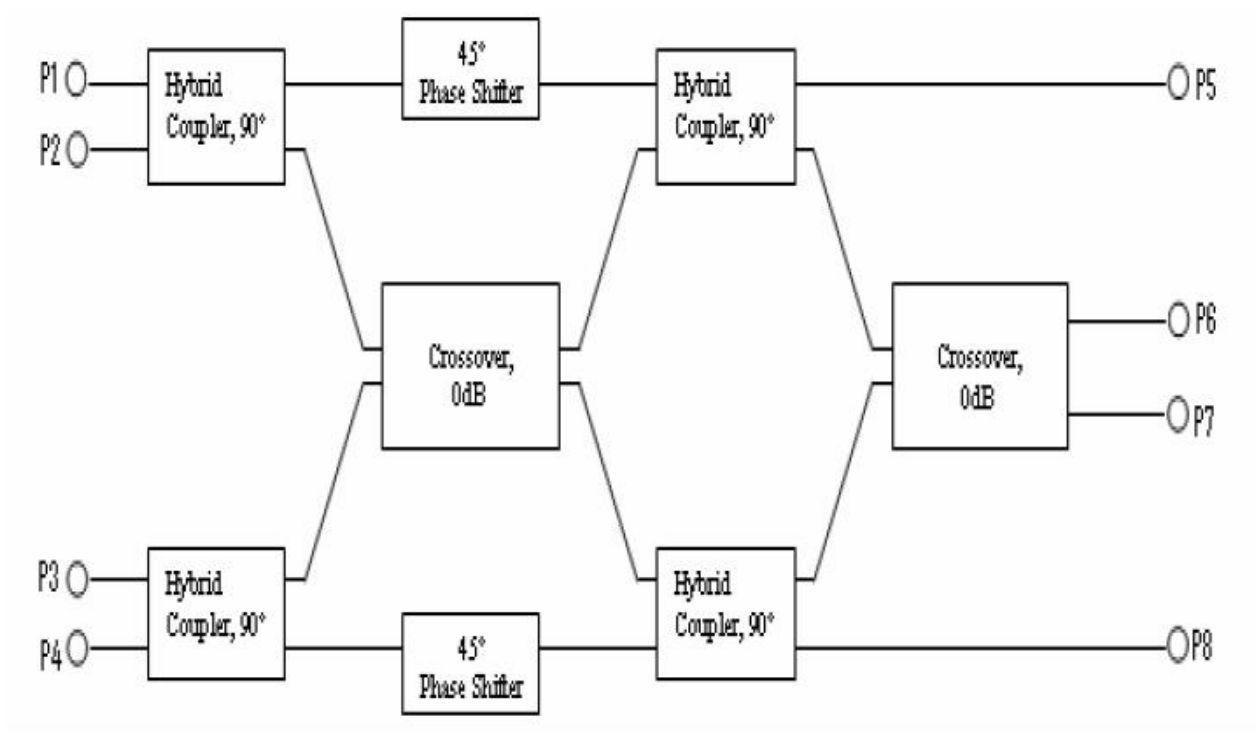

Figure 2.6 Block Structure of a *4x4* Butler Matrix, taken from [13].

## 2.3 SUMMARY

This chapter covered the introductory materials that form the basis of the Bulter Matrix and Rotman Lens beamforming networks. The covered materials include the antenna arrays, array factor, and different beamforming techniques using circuit and lens–based beamformers.

# CHAPTER 3
# THE BUTLER MATRIX SIMULATIONS

## 3.1 INTRODUCTION

In this chapter, design and simulation of the *4x4* Butler Matrix elements along with their definition and properties will be covered. These elements include Branch Line Coupler, 0dB Crossover, Phase Shifters, and their combination to form the complete Butler Matrix structure.

## 3.2 BRANCH LINE COUPLER

The Branch Line Coupler (BLC), also known as 'Quadrature Hybrid', is a four port 3dB power divider/combiner, with one input, two equal split outputs having a 90˚ phase difference between them, and an isolation port.

The structure consists of four quarter–wavelength transmission lines. Its capability of equal power split makes it an important circuit element in many applications. The fact that it can be made planar and it is a relatively simple structure makes it a common module used in microwave mixers and as input and output port in balanced microwave amplifier circuit. As it can be seen in Figure 3.1, the coupler is usually made in microstrip or stripline form.

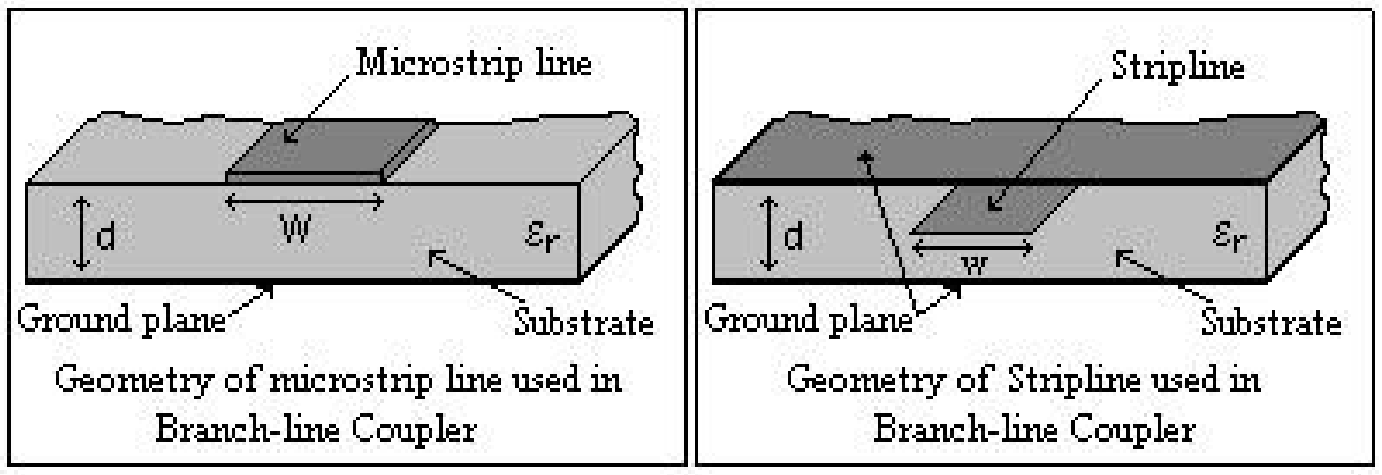

Figure 3.1 Geometry of Microstrip and Stripline used in BLC, taken from [16].

The branch line coupler has the disadvantage of narrow bandwidth, which is limited to 10–20%. This is mainly because the coupler is designed using quarter–wavelength transmission lines and it is perfectly matched at only one frequency [4].

The bandwidth of the coupler can be increased by using multiple sections in cascaded. Paul had reported the technique to overcome the bandwidth limitation [17]. The disadvantage of the method proposed by Paul is that the size of the coupler is increased. This bulky coupler reduces the efficiency in term of space utilisation.

As the Figure 3.2 illustrates, the coupler has a high degree of symmetry. Hence, any port can be used as an input port. Also, it shows that the isolation port (port 4) is in the same side as the input port while the output ports (ports 2 and 3) are on the opposite side of the input and isolation ports and each of the transmission lines forming the square shape is $\frac{\lambda}{4}$.

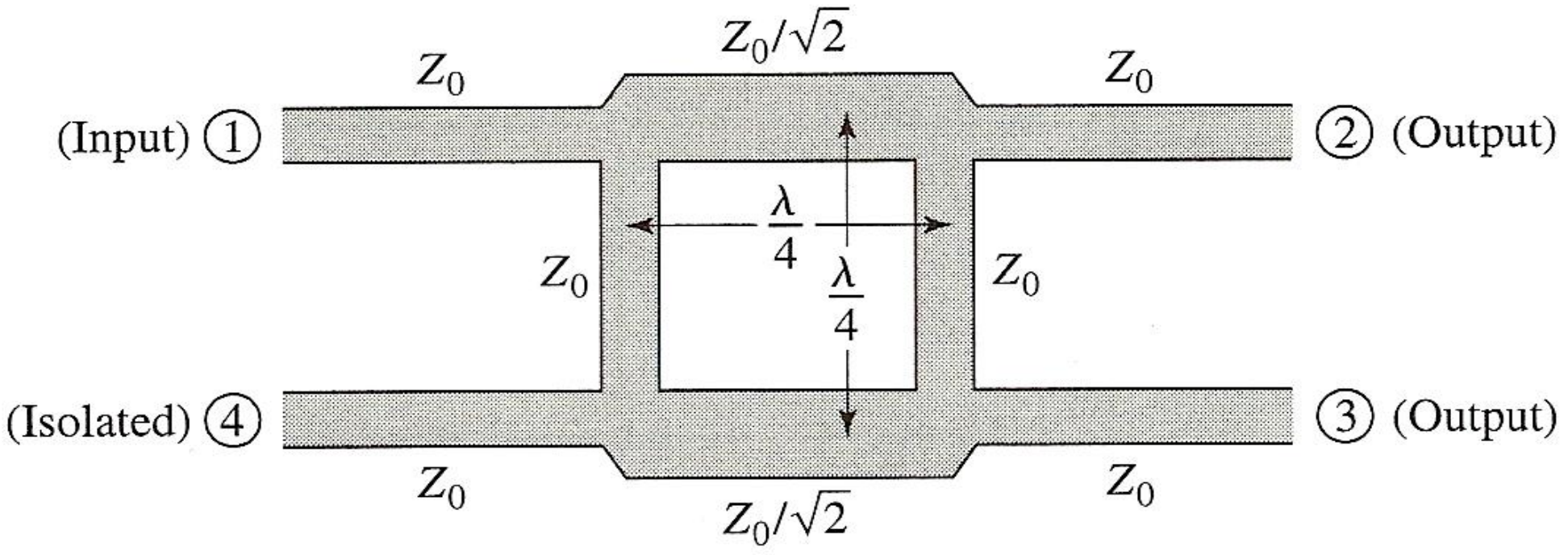

Figure 3.2 Branch Line Coupler Structure, taken from [18].

The characteristics impedances $Z_{01}$, $Z_{02}$, and $Z_{03}$ of the lines are given by the following equations:

$$Z_{01} = \frac{Z_{in}}{k} \tag{3.1a}$$

$$Z_{02} = \sqrt{\frac{Z_{in} Z_{out}}{1+k^2}} \tag{3.1b}$$

$$Z_{03} = \frac{Z_{out}}{k} \tag{3.1c}$$

where

$Z_{in}$ is the input impedance

$Z_{out}$ is the output impedance

$k$ is the coupling factor which is defined as $|\frac{S_{31}}{S_{21}}|$

In the case of an equal power split coupler, $k$ = 1 and $Z_{in} = Z_{out}$, so the above equations can be simplified as:

$$Z_{01} = Z_{03} = Z_0 \tag{3.2a}$$

$$Z_{02} = \frac{Z_0}{\sqrt{2}} \tag{3.2b}$$

where

$Z_0$ is the characteristic impedance

The branch line coupler's symmetry is reflected in the scattering matrix as each row can be obtained as a transmission of the first row. The scattering parameters are power wave descriptors that define the input–output relations of a network in terms of incident and reflected power waves [19]. The incident power wave is the point at which a signal is applied (point of excitation), and reflected power wave is the signal which is reflected back due to the load mismatch. The scattering matrix of the branch line coupler is given by [18]:

$$[S] = \frac{-1}{\sqrt{2}}\begin{bmatrix} 0 & j & 1 & 0 \\ j & 0 & 0 & 1 \\ 1 & 0 & 0 & j \\ 0 & 1 & j & 0 \end{bmatrix} \tag{3.3}$$

### 3.2.1 EVEN-ODD MODE ANALYSIS

The branch line coupler can be analysed by even–odd mode analysis. The initial characteristic impedance of the coupler normalized to $Z_0$ is shown in Figure 3.3.

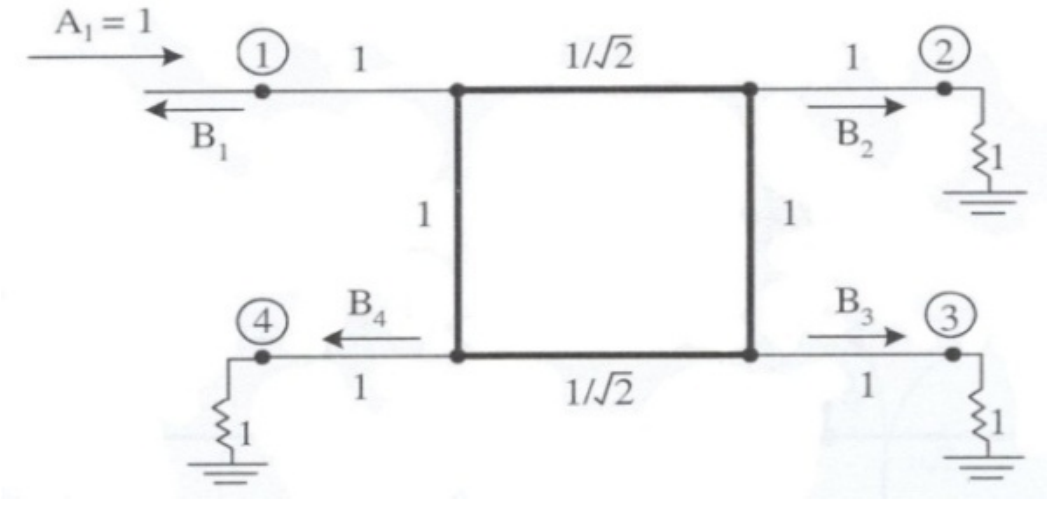

Figure 3.3 Normalised Circuit Diagram of the Branch Line Coupler, taken from [9].

Assume that an incident wave of uniform amplitude arrives at port 1. Due to the symmetrical structure of the coupler, the circuit can be decomposed into the superposition of an even–mode excitation and odd–mode excitation as shown in Figure 3.4.

The original excitation of Figure 3.3 can be produced by adding both even–odd mode excitations and the actual response is the sum of the responses to the even and odd mode excitations. The four–port network is decomposed into a set of two–port networks in Figure 3.4.

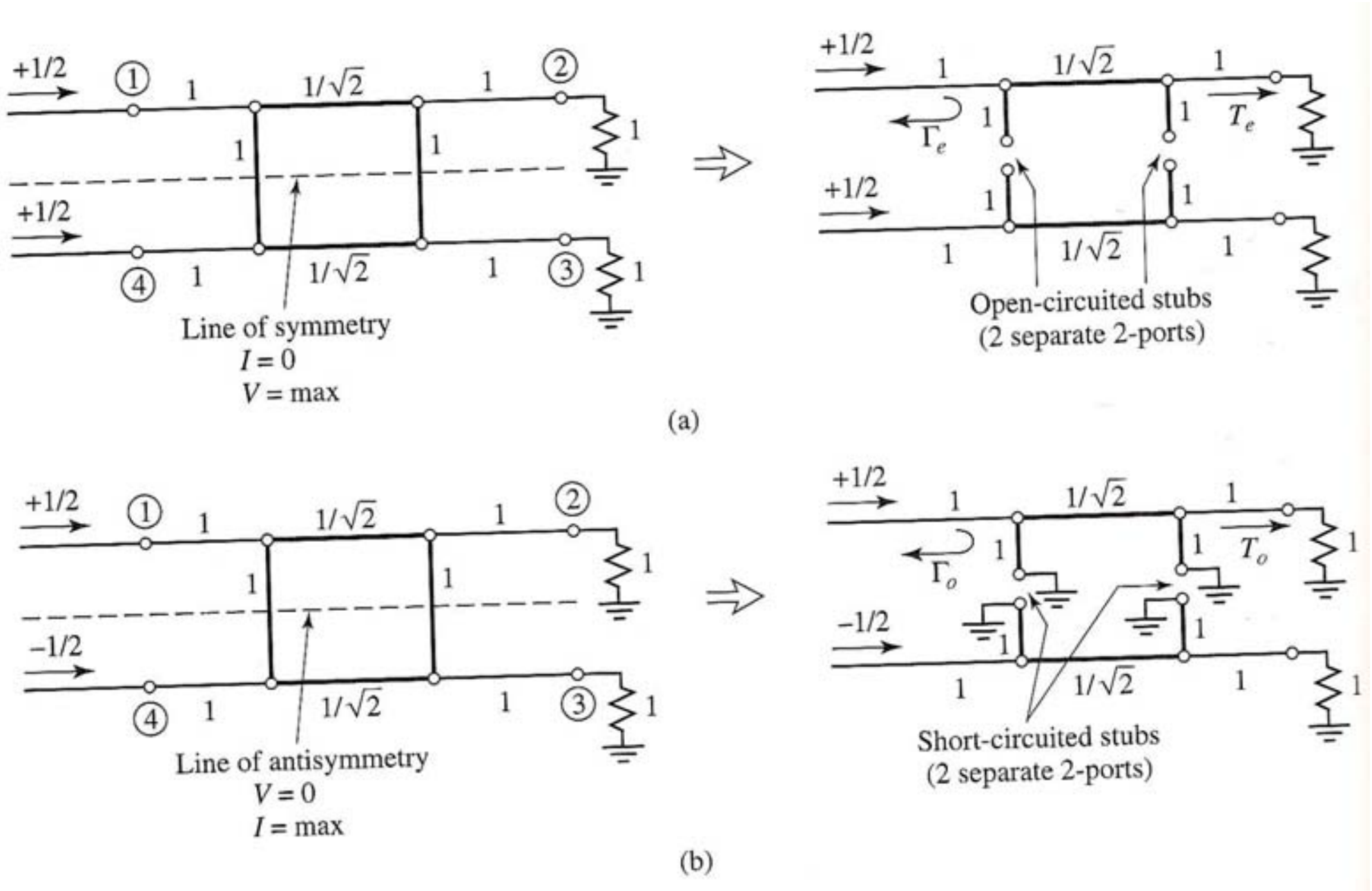

Figure 3.4 Decomposition of BLC into Even–Odd Mode Excitations (a) Even Mode, (b) Odd Mode, taken from [18].

The amplitudes of the incident waves for these two–port networks are $+\frac{1}{2}$ and $-\frac{1}{2}$ respectively, the amplitudes of the emerging wave at each port of the quadrature hybrid are:

$$B_1 = \frac{1}{2}\Gamma_e + \frac{1}{2}\Gamma_o \quad (3.4a)$$

$$B_2 = \frac{1}{2}T_e + \frac{1}{2}T_o \quad (3.4b)$$

$$B_3 = \frac{1}{2}T_e - \frac{1}{2}T_o \quad (3.4c)$$

$$B_4 = \frac{1}{2}\Gamma_e - \frac{1}{2}\Gamma_o \quad (3.4d)$$

where

$T_{e,o}$ is the even and odd mode transmission coefficients

$\Gamma_{e,o}$ is the even and odd mode reflection coefficients

Consider the even mode two–port circuit, $T_e$ and $\Gamma_e$ can be found by obtaining the scattering parameters. According to Pozar [18], this can best be done by first multiplying the *ABCD* matrices of each cascade component in that circuit, such that:

$$\begin{bmatrix} A & B \\ C & D \end{bmatrix}_e = \begin{bmatrix} 1 & 0 \\ j & 1 \end{bmatrix} \begin{bmatrix} 0 & j/\sqrt{2} \\ j/\sqrt{2} & 0 \end{bmatrix} = \frac{1}{\sqrt{2}} \begin{bmatrix} -1 & j \\ j & -1 \end{bmatrix} \tag{3.5}$$

where Y is the admittance of the $\frac{\lambda}{8}$ shunt open circuit stubs, given by:

$$Y = j \tan \beta l = j \tag{3.6}$$

The *ABCD* parameters can be converted to the S–Parameters that are equivalent to the reflection and transmission coefficients of the even mode.

$$\Gamma_e = \frac{A+B-C-D}{A+B+C+D} = \frac{(-1+j-j+1)/\sqrt{2}}{(-1+j+j-1)/\sqrt{2}} = 0 \tag{3.7a}$$

$$T_e = \frac{2}{A+B+C+D} = \frac{2}{(-1+j+j-1)/\sqrt{2}} = \frac{-1}{\sqrt{2}}(1+j) \tag{3.7b}$$

Likewise, for the odd mode two port circuit:

$$\begin{bmatrix} A & B \\ C & D \end{bmatrix}_o = \frac{1}{\sqrt{2}} \begin{bmatrix} 1 & j \\ j & 1 \end{bmatrix} \tag{3.8}$$

which results in:

$$T_o = \frac{1}{\sqrt{2}}(1-j) \tag{3.9a}$$

$$\Gamma_o = 0 \tag{3.9b}$$

Finally, using above equations the following results can be obtained that describe the fundamental operations of the branch line coupler.

$$B_1 = 0 \quad \text{(port 1 is well matched)} \tag{3.10a}$$

$$B_2 = \frac{-j}{2} \quad \text{(half–power, –90° phase shift from port 1 to 2)} \tag{3.10b}$$

$$B_3 = \frac{-1}{\sqrt{2}} \quad \text{(half–power, –180° phase shift from port 1 to 3)} \tag{3.10c}$$

$$B_4 = 0 \quad \text{(no power to port 4)} \tag{3.10d}$$

The analysis shows the fundamental principle of the branch line coupler, with a –3dB power split between the output ports and there is a $\frac{\pi}{2}$ phase difference between them.

### 3.2.2 BRANCH LINE COUPLER SIMULATIONS

In order to illustrate the characteristics of the *4x4* Butler Matrix components, each of them will be separately simulated using MWO in the following sections.

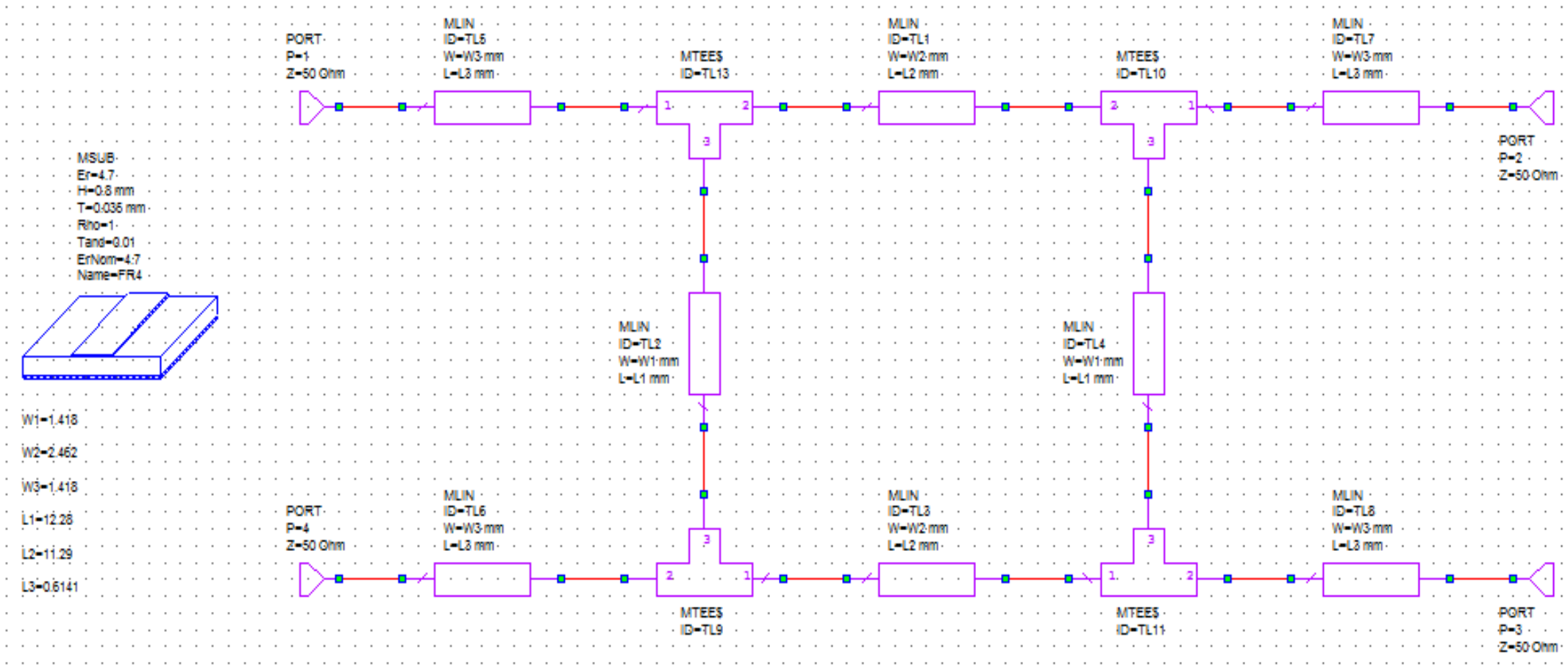

Figure 3.6 BLC Schematic Design.

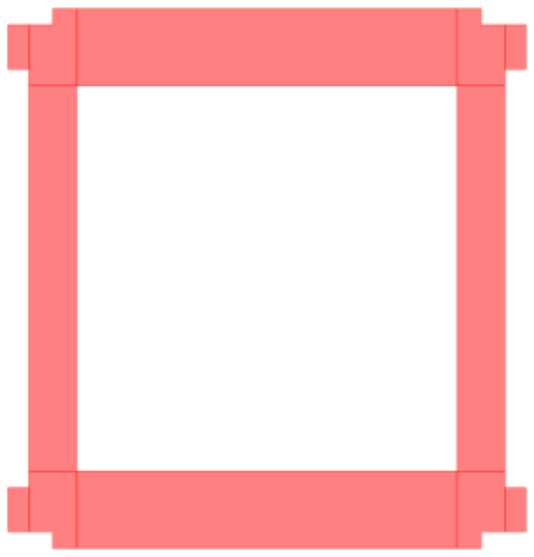

Figure 3.7 BLC 2D Layout.

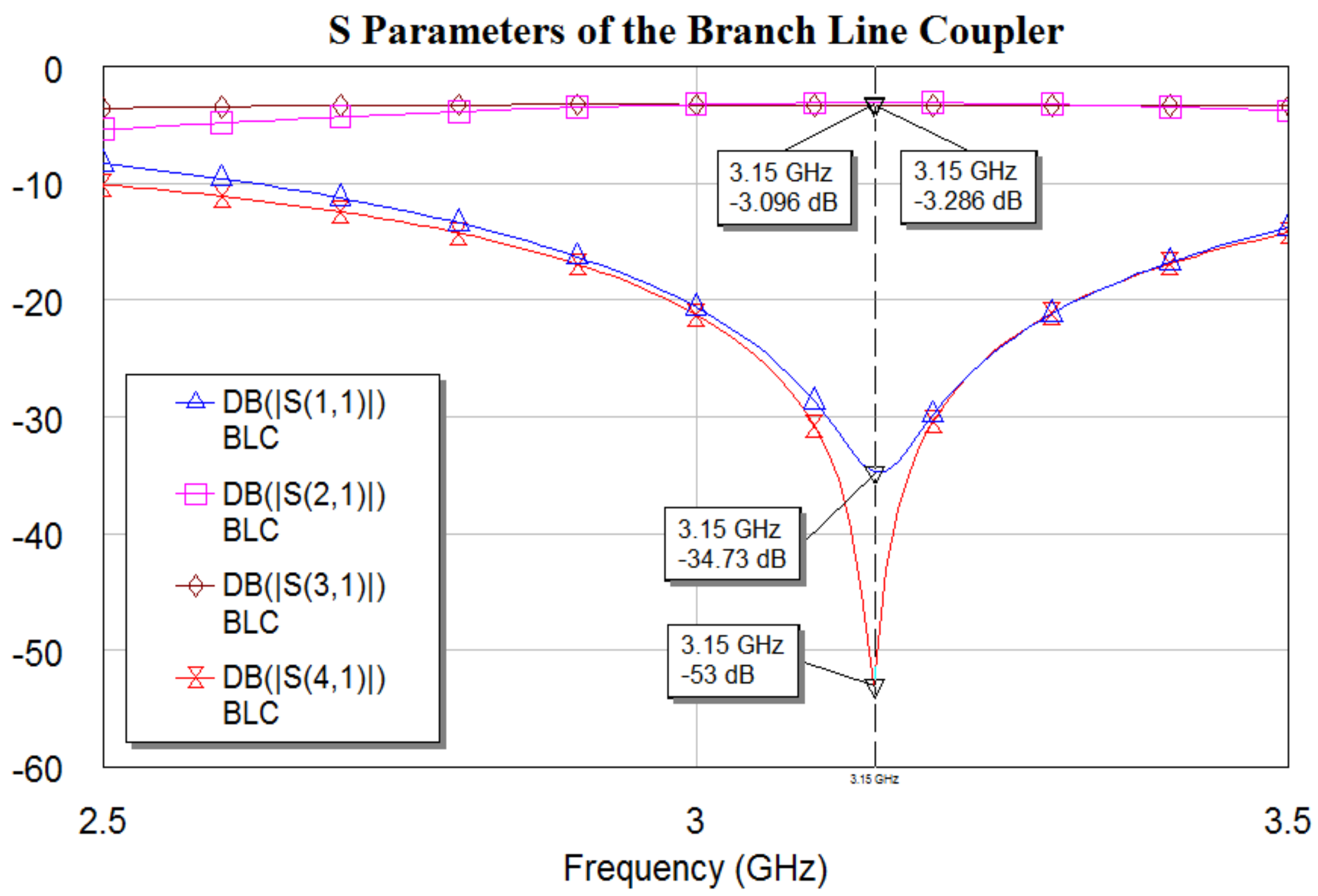

Figure 3.8 Simulated S–Parameters of the BLC.

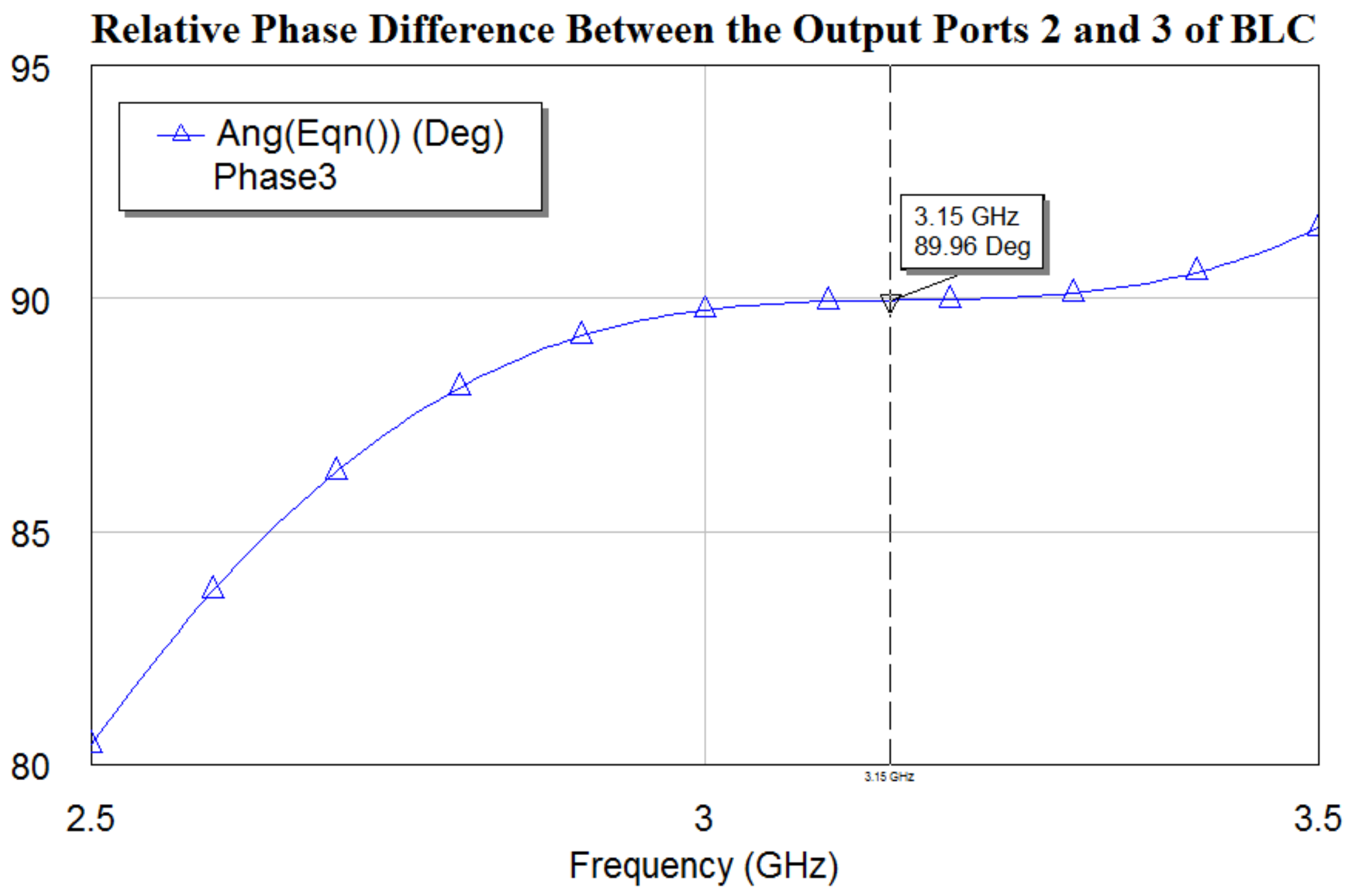

Figure 3.9 Relative Phase Difference Between Output Ports of the BLC.

Figure 3.8 indicates that an average of –3dB of transmission power is obtained from $S_{21}$ and $S_{31}$ at 3.15 GHz. Also, Figure 3.9 indicates that the output ports have a phase difference of 90˚ (89.96˚). The branch line coupler is chosen to be the fundamental module in the overall Butler Matrix design.

## 3.3 BM 0DB CROSSOVER

The 0dB crossover is one of the elements in designing the Butler Matrix, in order to obtain an entirely planar version of the *4x4* matrix.

It consists of two back–to–back connected 3dB quadrature hybrids. The circuit could be analysed using the even–odd mode analysis as well.

It functions in a way that when a signal is entering to input port 1 of the 0dB crossover is split to half by the first 3dB coupler. The second 3dB coupler then recombines the split signal. As a result, signal is produced at the opposite output port. This concept is shown in Figure 3.10.

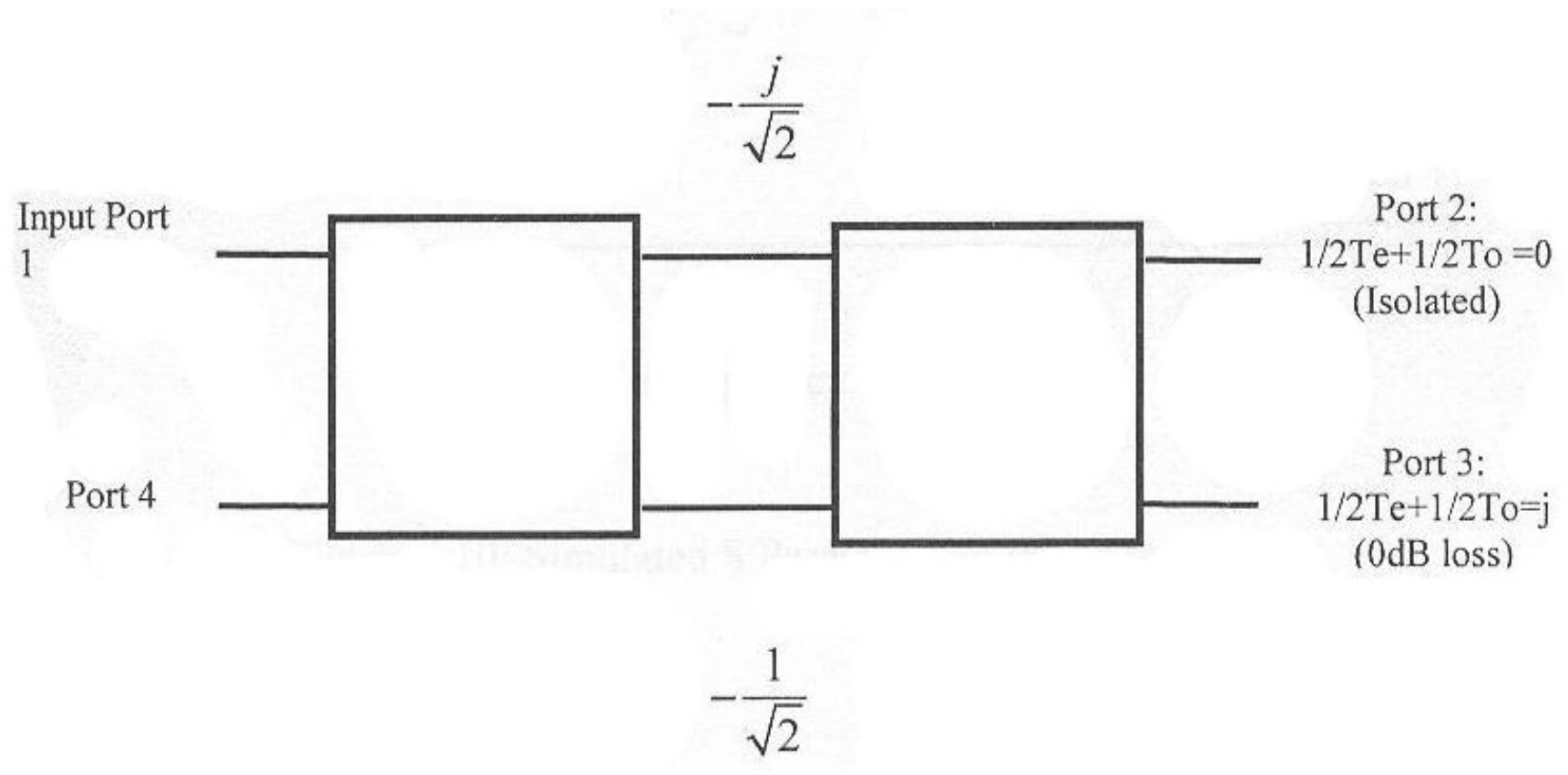

Figure 3.10 BM 0dB Crossover Using Cascaded Branch Line Couplers, taken from [4].

## 3.3.1 BM 0DB CROSSOVER SIMULATIONS

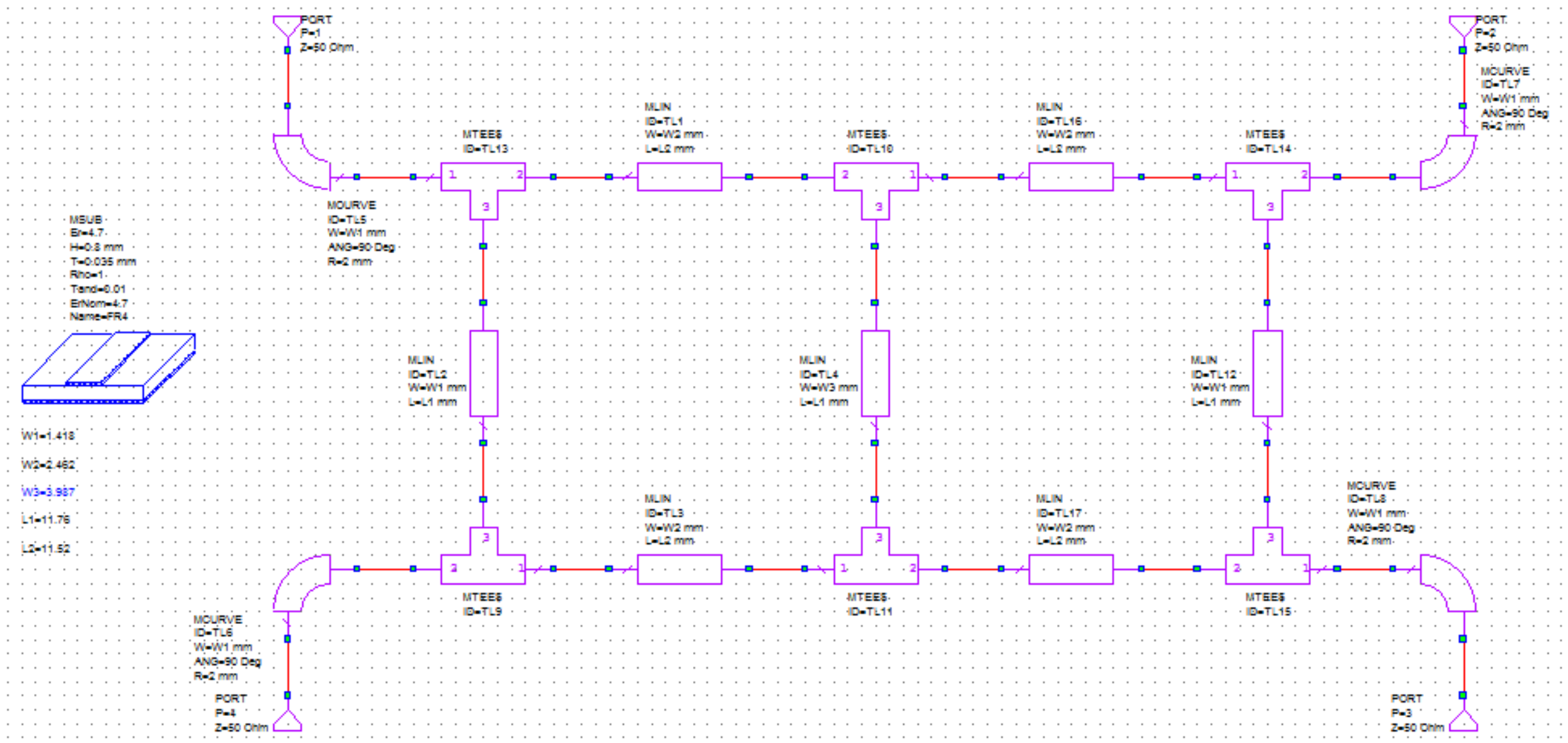

Figure 3.11 BM 0dB Crossover Schematic Design.

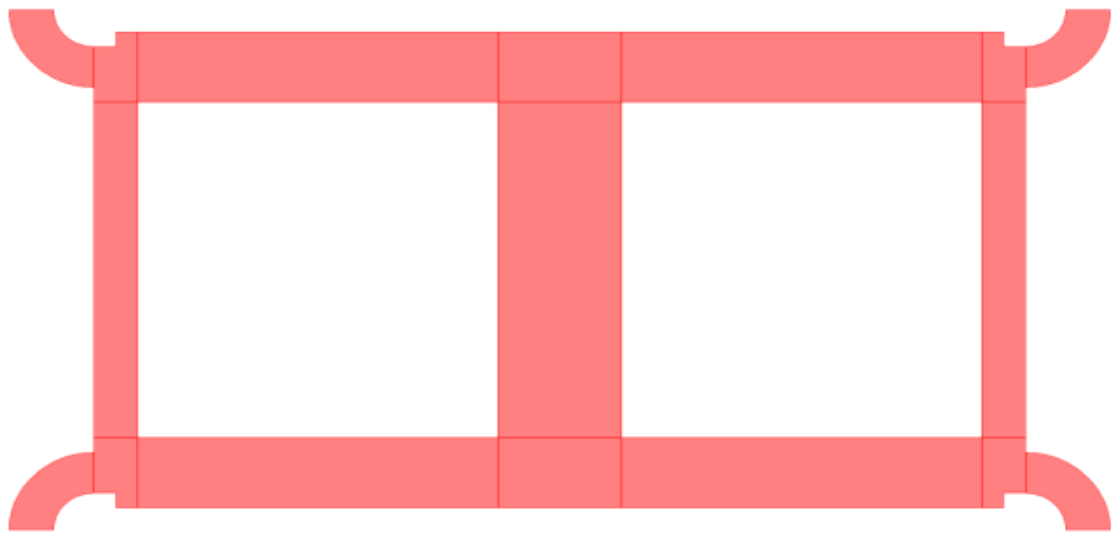

Figure 3.12 BM 0dB Crossover 2D Layout.

Figure 3.13 indicates that the 0dB crossover is well matched at the operating frequency of 3.15 GHz. The $S_{31}$ approximately has a 0dB insertion loss (–0.4262 dB); hence it works properly as a crossover. The 3.15 GHz frequency is the scaled frequency of operation for the automotive communication systems (63–64 GHz), and it is scaled down to $\frac{1}{20}^{th}$ of this frequency.

Also, Figure 3.14 shows that the phase progression from port 1 to port 3 is 23.56˚. This will later be used for extracting the length of the phase shifters.

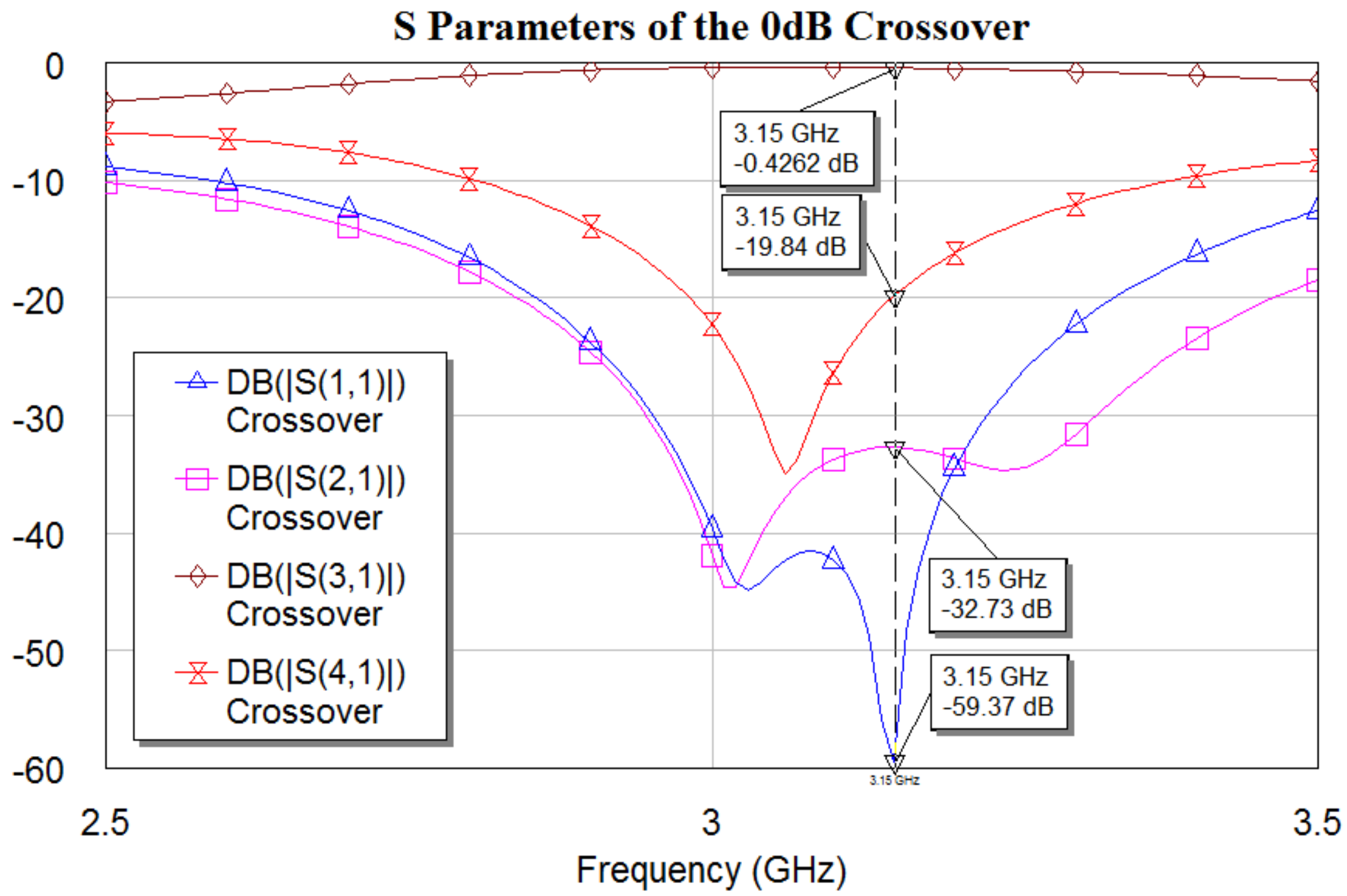

Figure 3.13 Simulated S–Parameters of the 0dB Crossover.

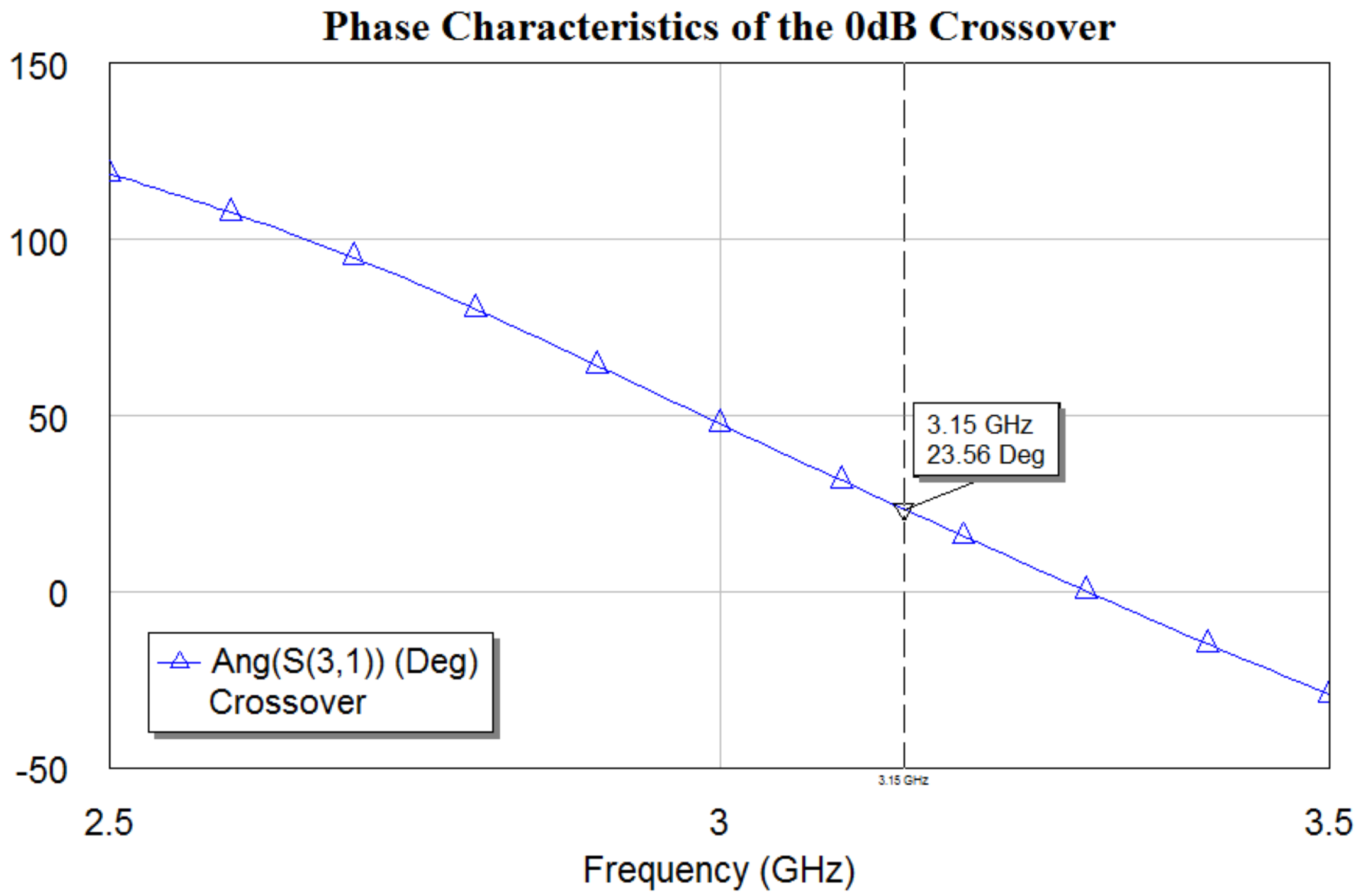

Figure 3.14 Simulated Output Phase of the 0dB Crossover.

## 3.4 BM PHASE SHIFTERS

Phase Shifter is another element in the Butler Matrix design that produces a change in the phase angle of the wave travelling through it.

To obtain the length of the 45˚ phase shifter, the relative phase difference between $S_{31}$ of the first 0dB crossover and $S_{21}$ of the 45˚ phase shifter should be –45˚. The procedure used to obtain the exact length of the phase shifter can be summarised as: 1) By phase output characteristic of the crossover (23.56˚), the approximate length of the phase shifter can be obtained, 2) By the tuner tool in MWO environment, the relative phase difference between the crossover and the phase shifter should be adjusted in a way that in the operation frequency of 3.15 GHz the relative phase of –45˚ is achieved. This concept is shown in Figure 3.17.

The same approach has been conducted for the output phase shifter. But this time, the relative phase difference between $S_{31}$ of the second 0dB crossover and $S_{21}$ of the output phase shifter should be 0˚ ($S_{21}/S_{31}$). This is shown in Figure 3.18.

The formulas needed for calculating the length are as follow:

$$\theta = \frac{360\, l}{\lambda} \tag{3.11}$$

$$\lambda = \frac{c}{f\sqrt{\varepsilon_{eff}}} \tag{3.12}$$

where

$\lambda$ is the wavelength in the transmission line

$\varepsilon_{eff}$ is the effective dielectric constant

In this case, first the value of $\varepsilon_{eff}$ is obtained based on the substrate parameters ($\varepsilon_{eff}$ = 3.5023) using TXLine 2003 tool in MWO environment (Microstrip tab). Then $\lambda$ is found from equation 3.12 ($\lambda$ = 50.89 mm). Finally the procedure for finding the length of phase shifters is carried out using 3.11 and the mentioned procedure of finding $\theta$.

## 3.4.1 BM PHASE SHIFTERS SIMULATIONS

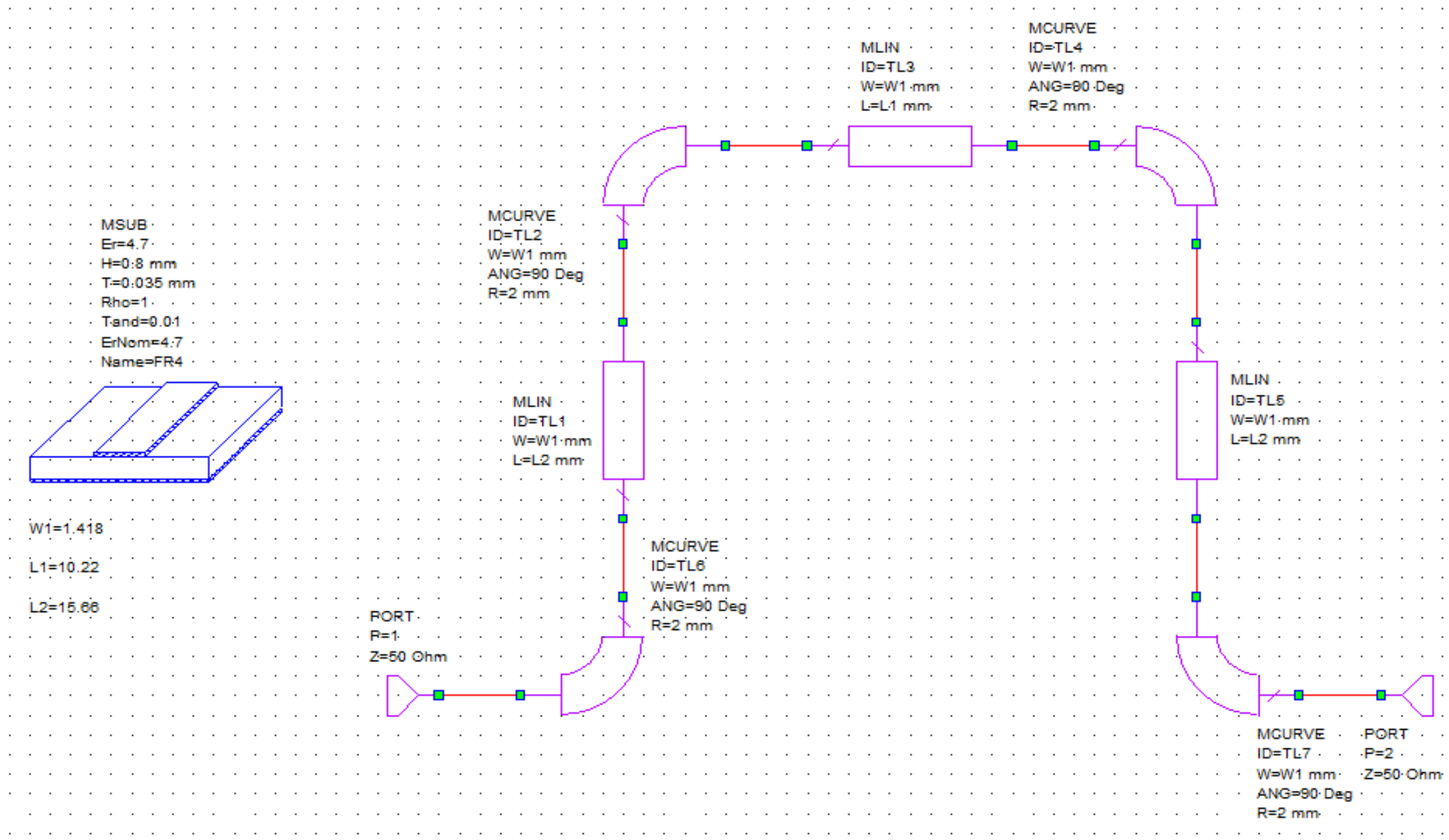

Figure 3.15 BM 45˚ Phase Shifter Schematic Design.

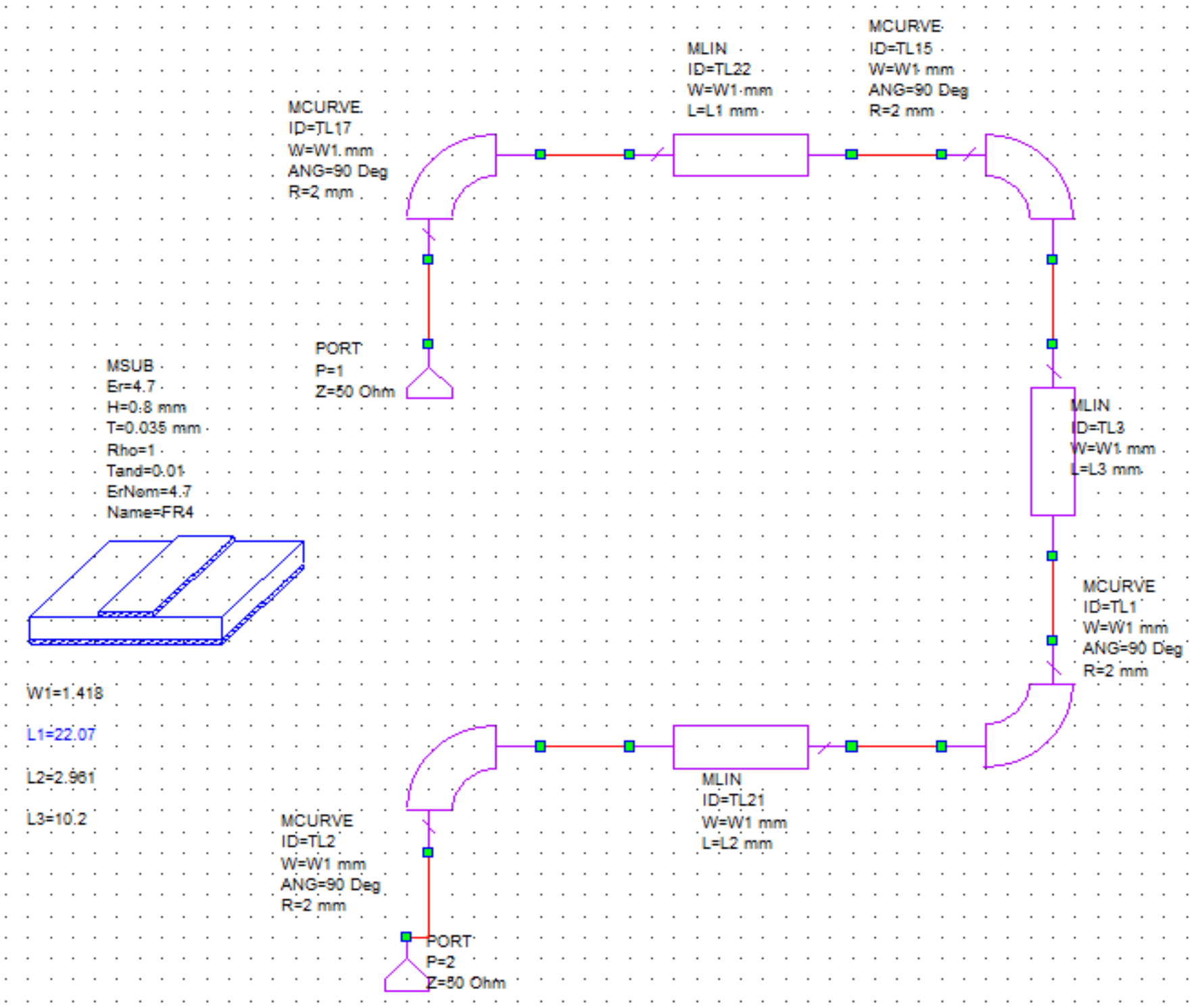

Figure 3.16 BM Output Phase Shifter Schematic Design.

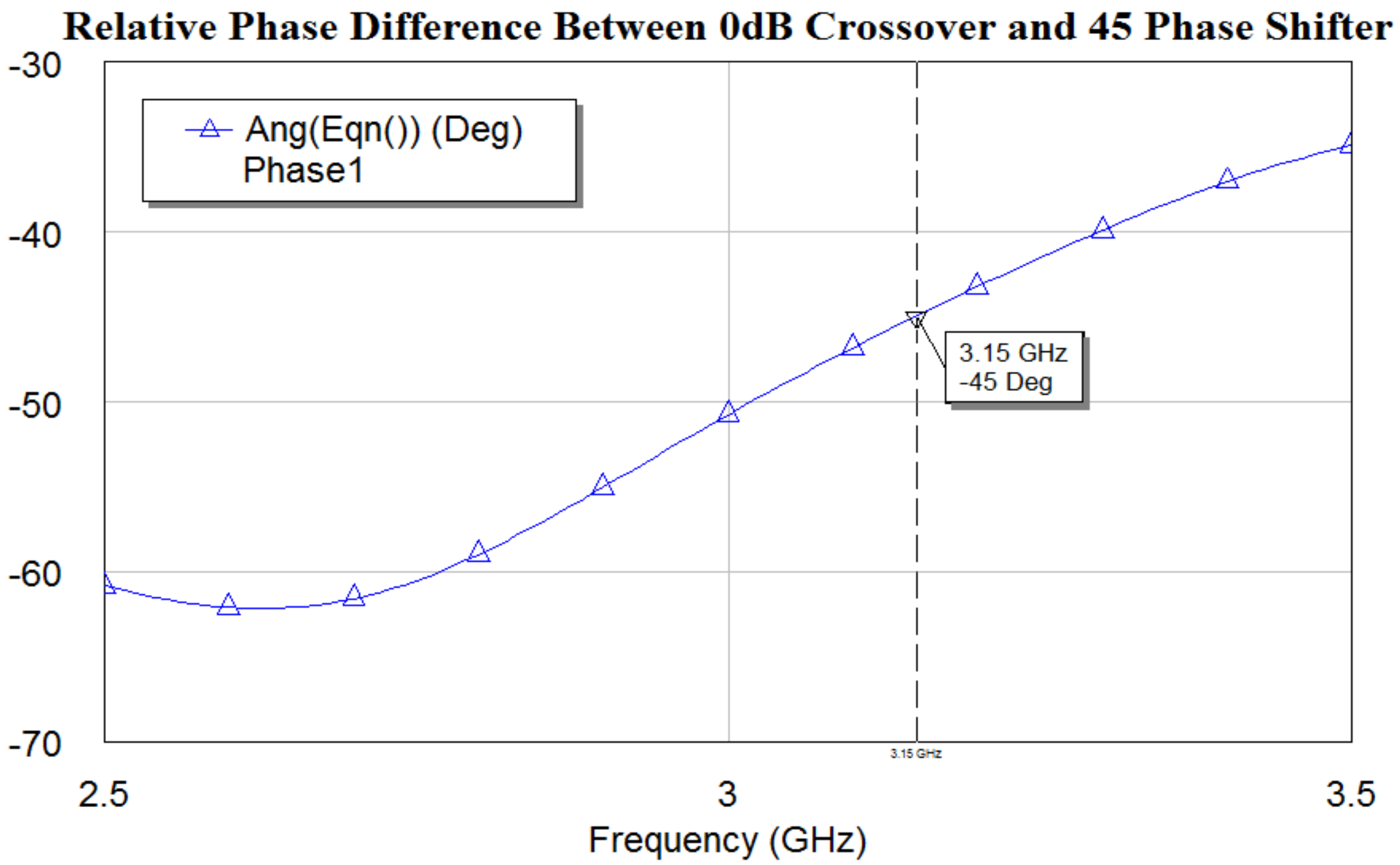

Figure 3.17 Relative Phase Difference Between the 0dB Crossover and 45˚ Phase Shifter.

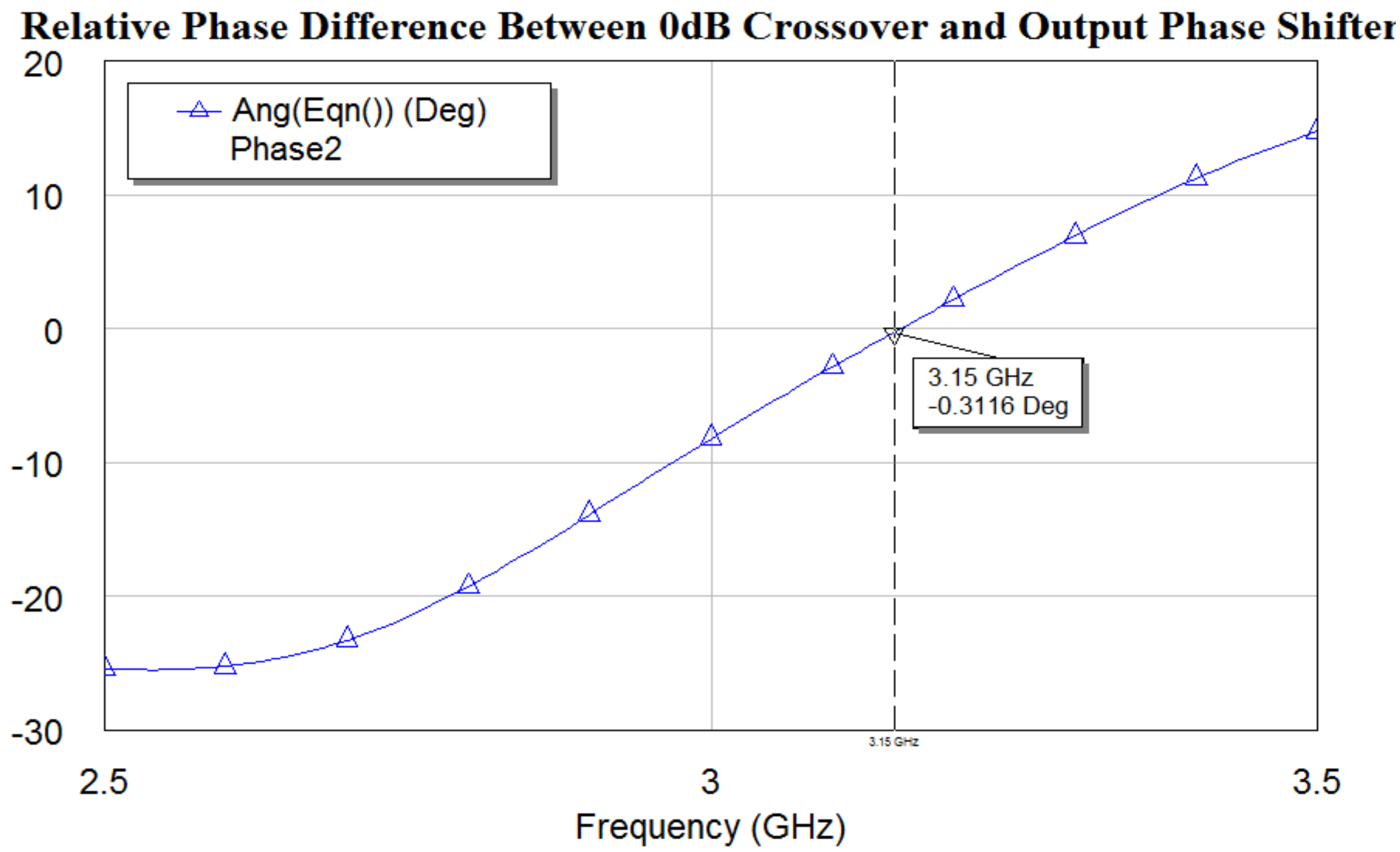

Figure 3.18 Relative Phase Difference Between the 0dB Crossover and Output Phase Shifter.

## 3.5 4X4 BUTLER MATRIX SIMULATIONS

The previous sections elements will form the complete structure of the *4x4* Butler Matrix. As it is mentioned earlier, the Butler Matrix consists of an equal number of input and output ports connected through the phase shifters and branch line couplers such that when a signal is applied to any input port, it produces equal amplitude signals at all the output ports. The following figures show the final Butler Matrix simulated structure, simulated S–Parameters and relative phase differences from port 5 to port 8 respectively.

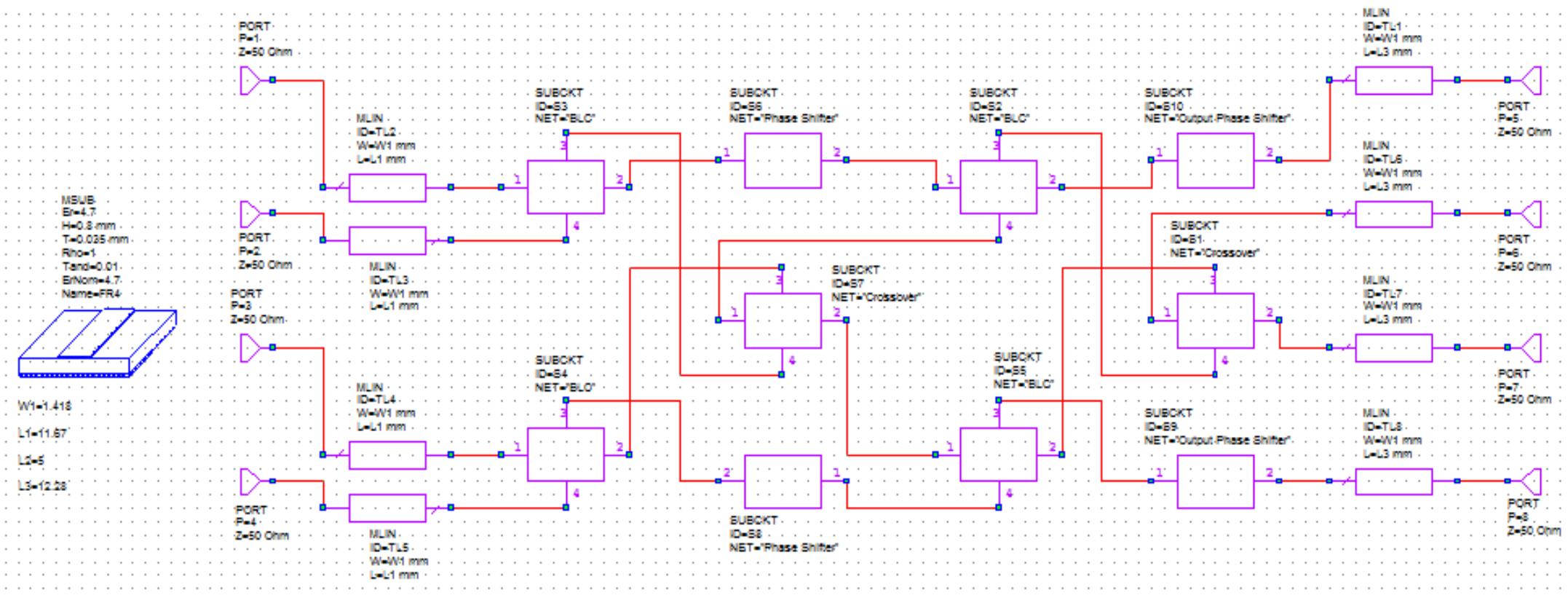

Figure 3.19 *4x4* Butler Matrix Schematic Design.

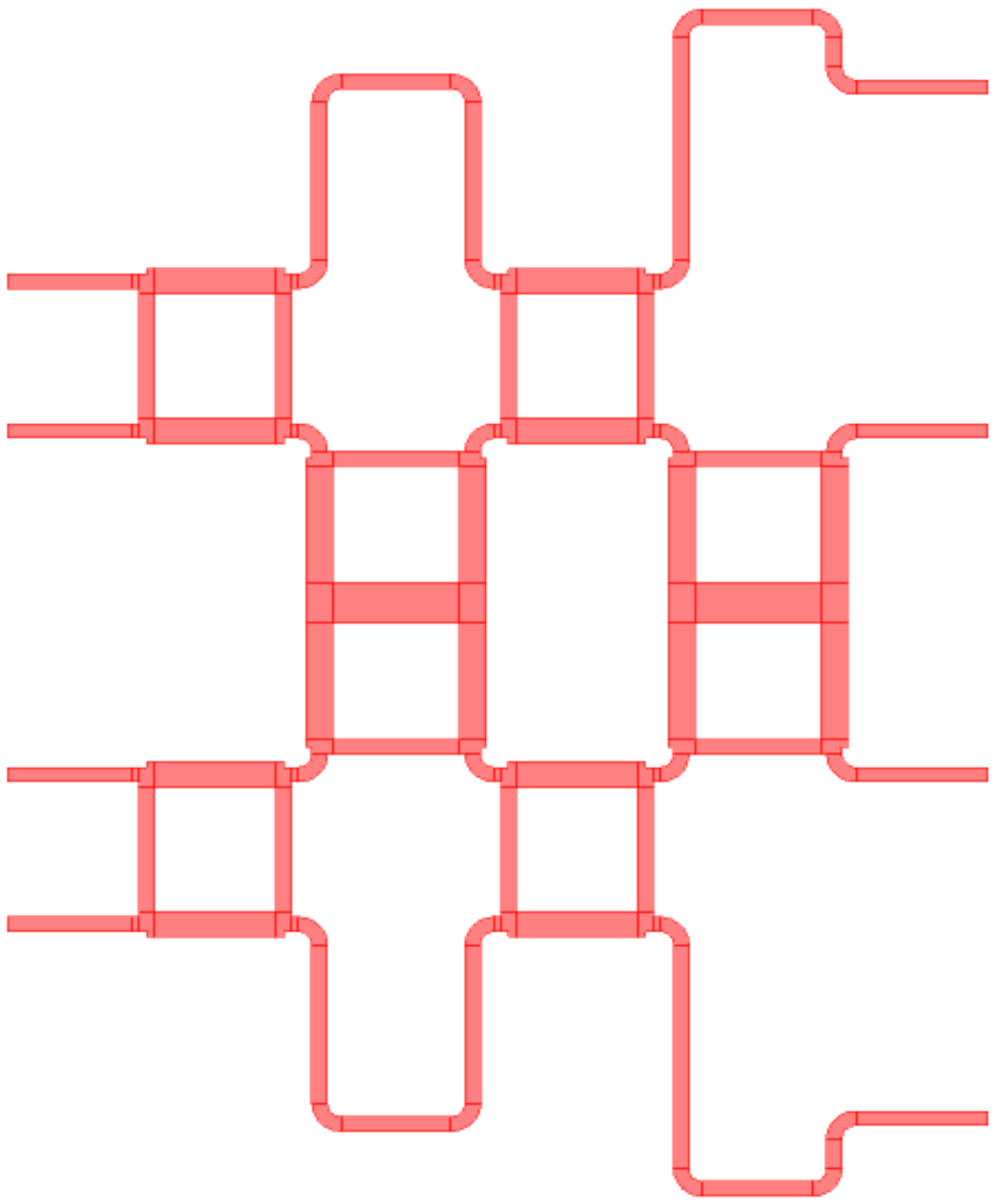

Figure 3.20 *4x4* Butler Matrix 2D Layout.

### 3.5.1 INSERTION LOSS AND RETURN LOSS OF OUTPUT PORTS

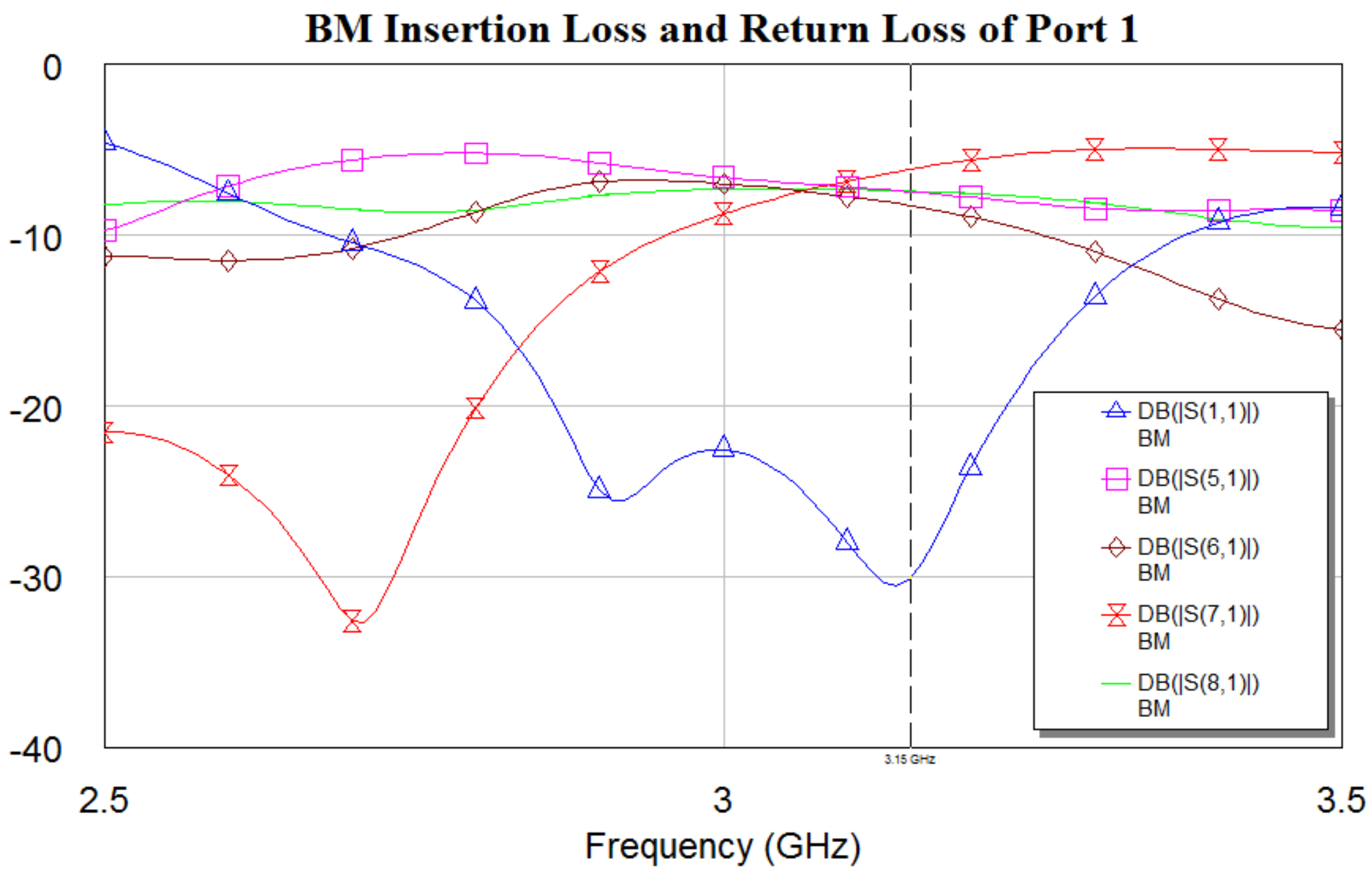

Figure 3.21 Insertion Loss and Return Loss of Port 1.

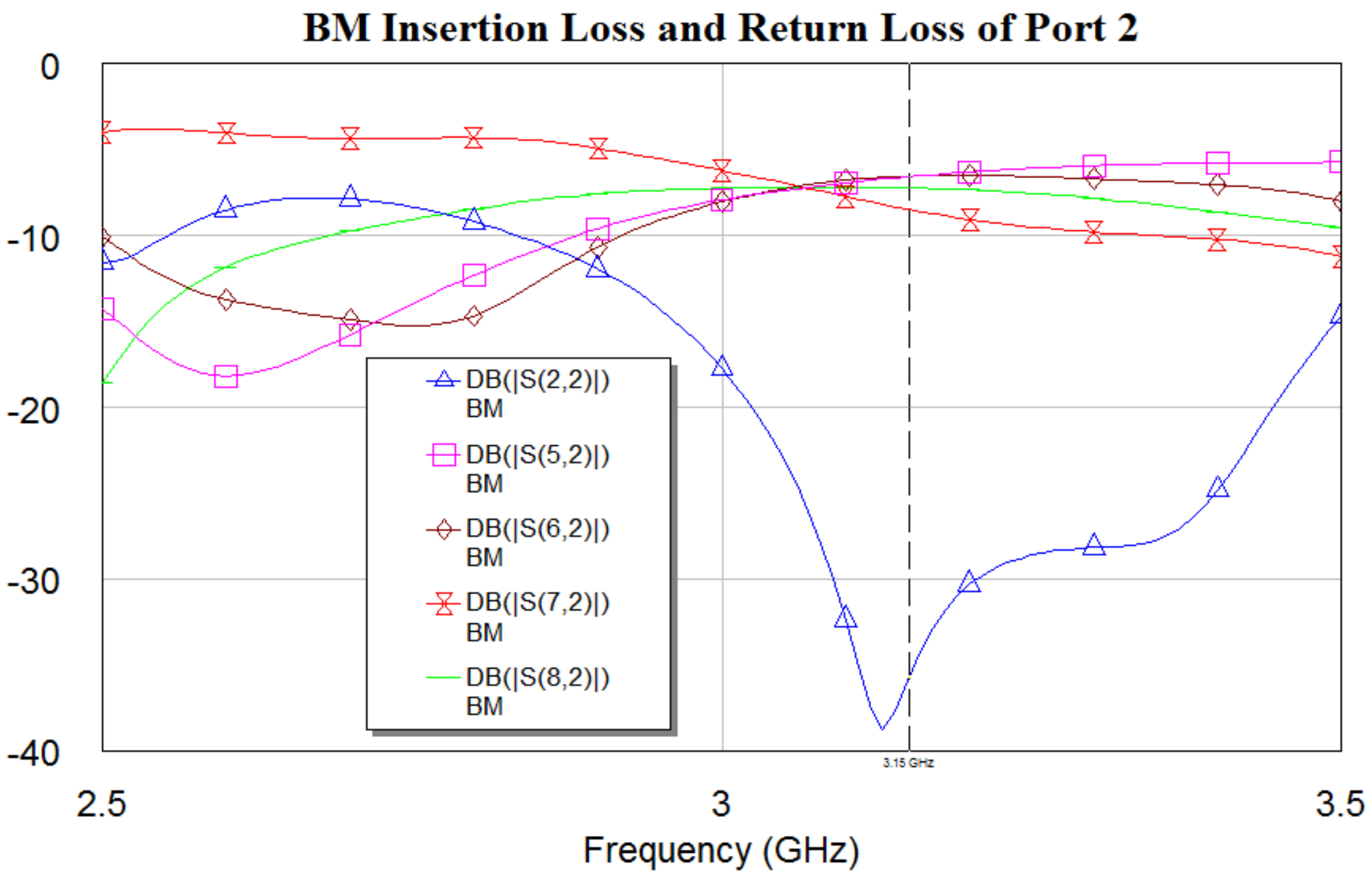

Figure 3.22 Insertion Loss and Return Loss of Port 2.

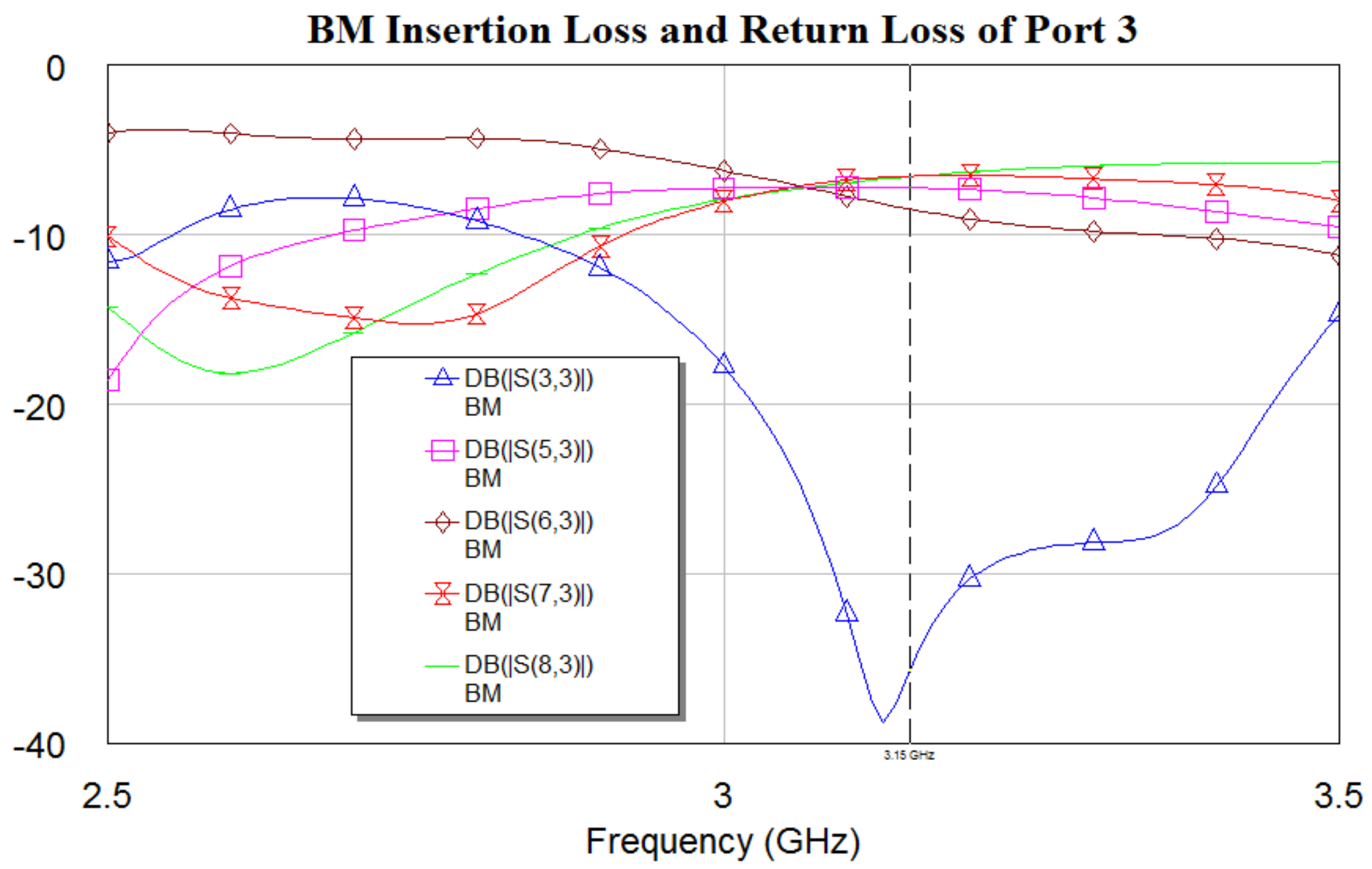

Figure 3.23 Insertion Loss and Return Loss of Port 3.

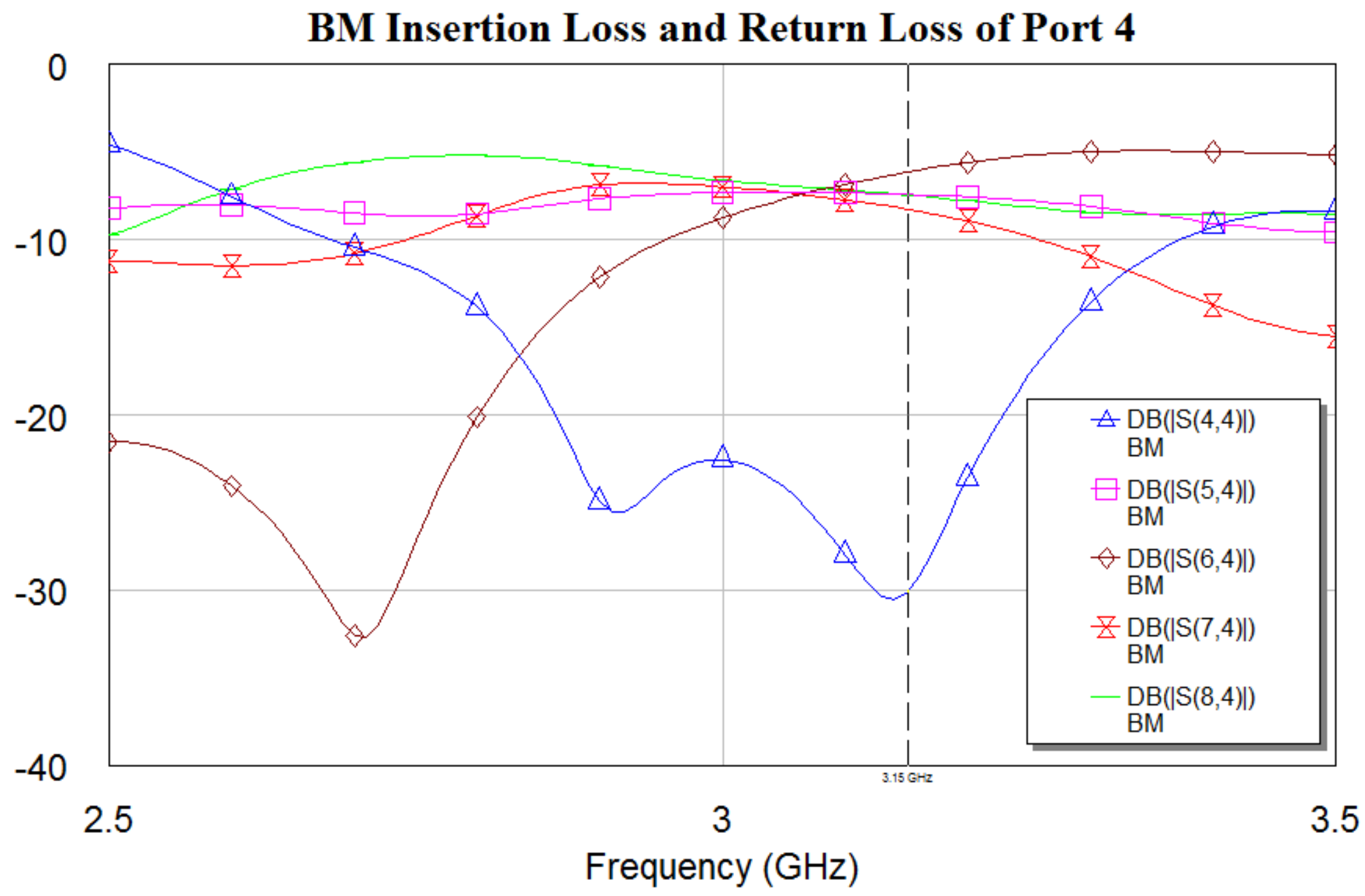

Figure 3.24 Insertion Loss and Return Loss of Port 4.

### 3.5.2 RELATIVE PHASE DIFFERENCES OF OUTPUT PORTS

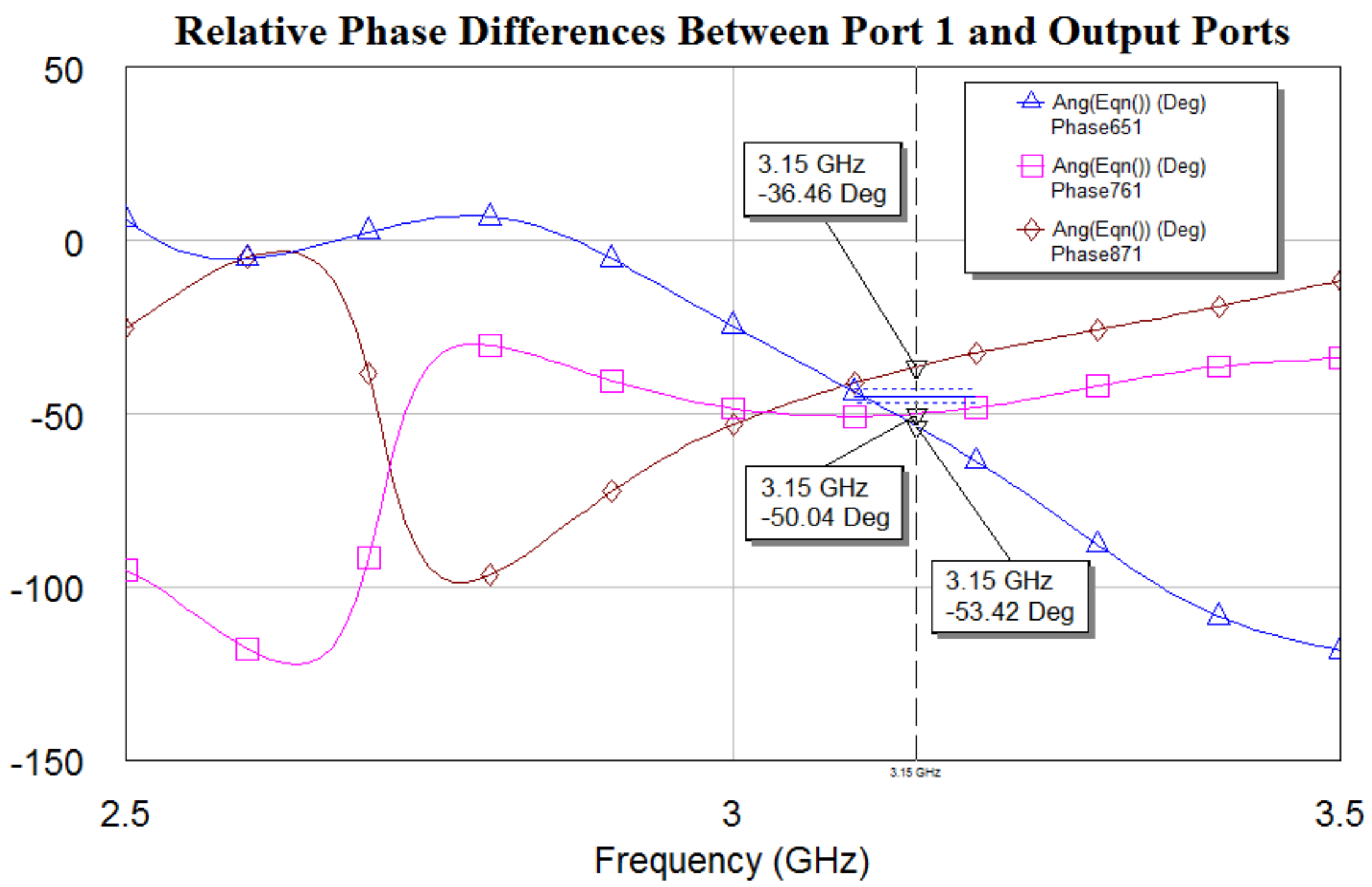

Figure 3.25 Relative Phase Difference Between Port 1 and Output Ports.

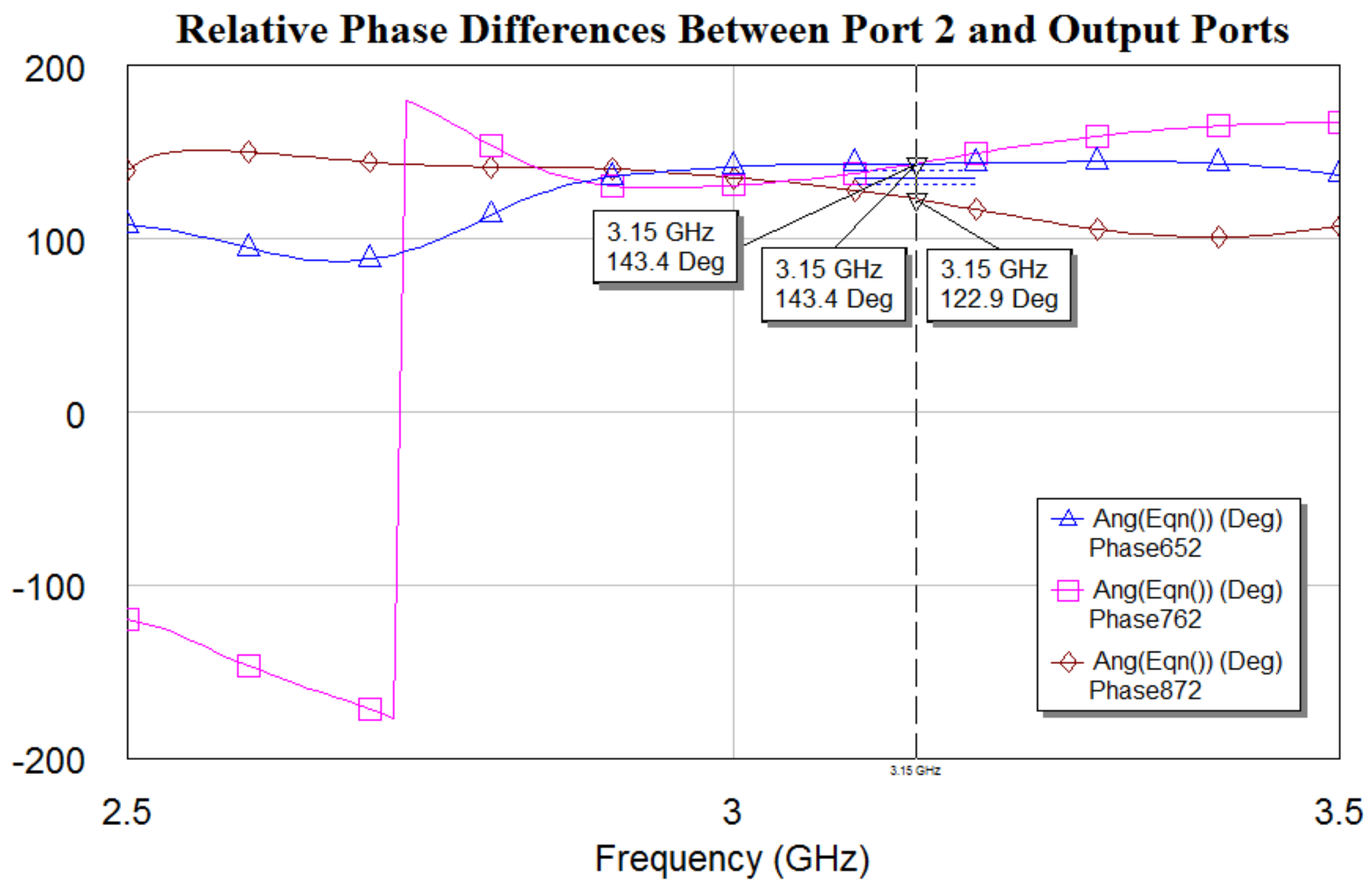

Figure 3.26 Relative Phase Difference Between Port 2 and Output Ports.

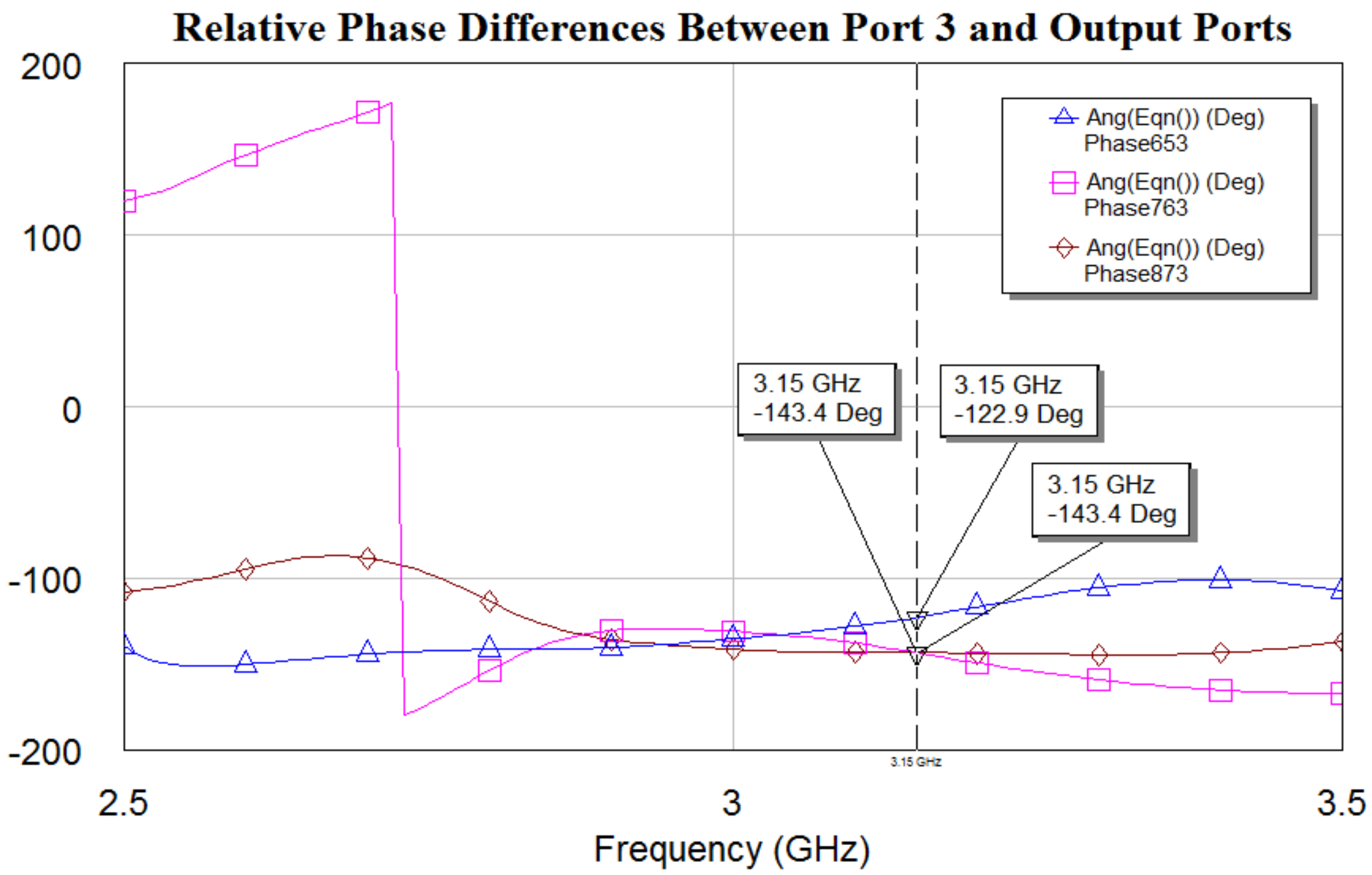

Figure 3.27 Relative Phase Difference Between Port 3 and Output Ports.

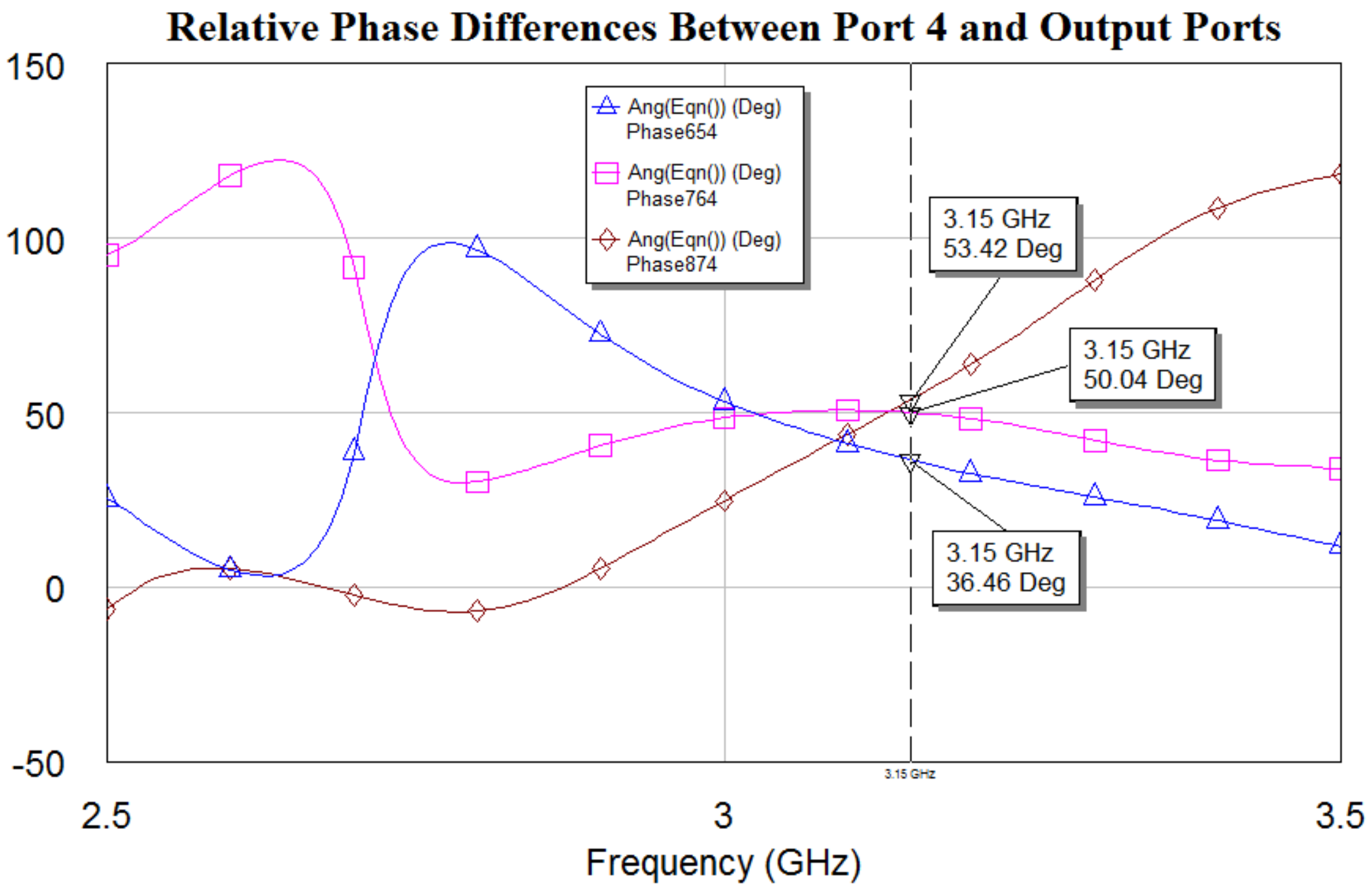

Figure 3.28 Relative Phase Difference Between Port 4 and Output Ports.

As it can be seen from Figures 3.25 and 3.26, the relative phase differences between port 1 and output ports and port 2 and output ports are almost kept constant. But there are still some variations in the phases. The relative phases of $S_{61}/S_{51}$, $S_{71}/S_{61}$, and $S_{81}/S_{71}$ should ideally be –45˚, but in this case, the errors of around –5˚ to –8˚ are observed that are acceptable for this operation frequency. The same thing is applied to port 2 relative phases, then this port's ideal relative phases for $S_{62}/S_{52}$, $S_{72}/S_{62}$, and $S_{82}/S_{72}$ should be 135˚ but in simulations some errors are still observed. Because the structure of the Butler Matrix has kept its symmetry, the ports 3 and 4 have the same values but with different ± signs (Figures 3.27 and 3.28). It should be mentioned that the relative phases are calculated using the output equations written in MWO environment.

## 3.6 PCB LAYOUT OF THE BUTLER MATRIX

The prototype is fabricated on FR4 substrate with the dielectric constant ($\varepsilon_r$) of 4.7, substrate thickness (H) of 0.8 mm, copper thickness (T) of 0.035 mm (bottom/top), and loss tangent of 0.01. This substrate is suitable for the frequency of 3.15 GHz and it is a low cost substrate and easy to fabricate. The layout of the BM PCB design is shown in Figure 3.29.

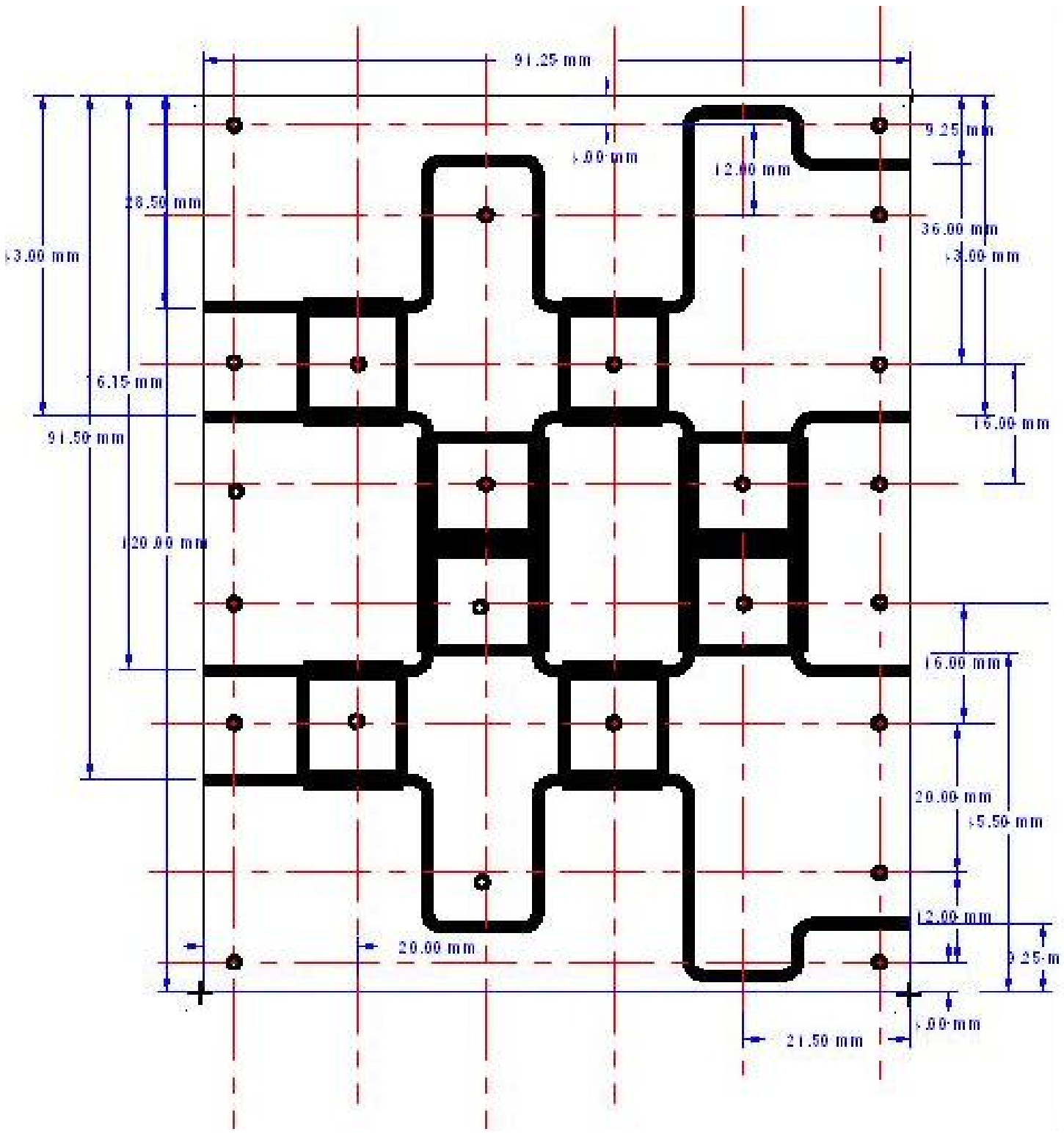

Figure 3.29 PCB Layout Design of the *4x4* Butler Matrix.

## 3.7 BM ARRAY FACTOR RADIATION PATTERNS

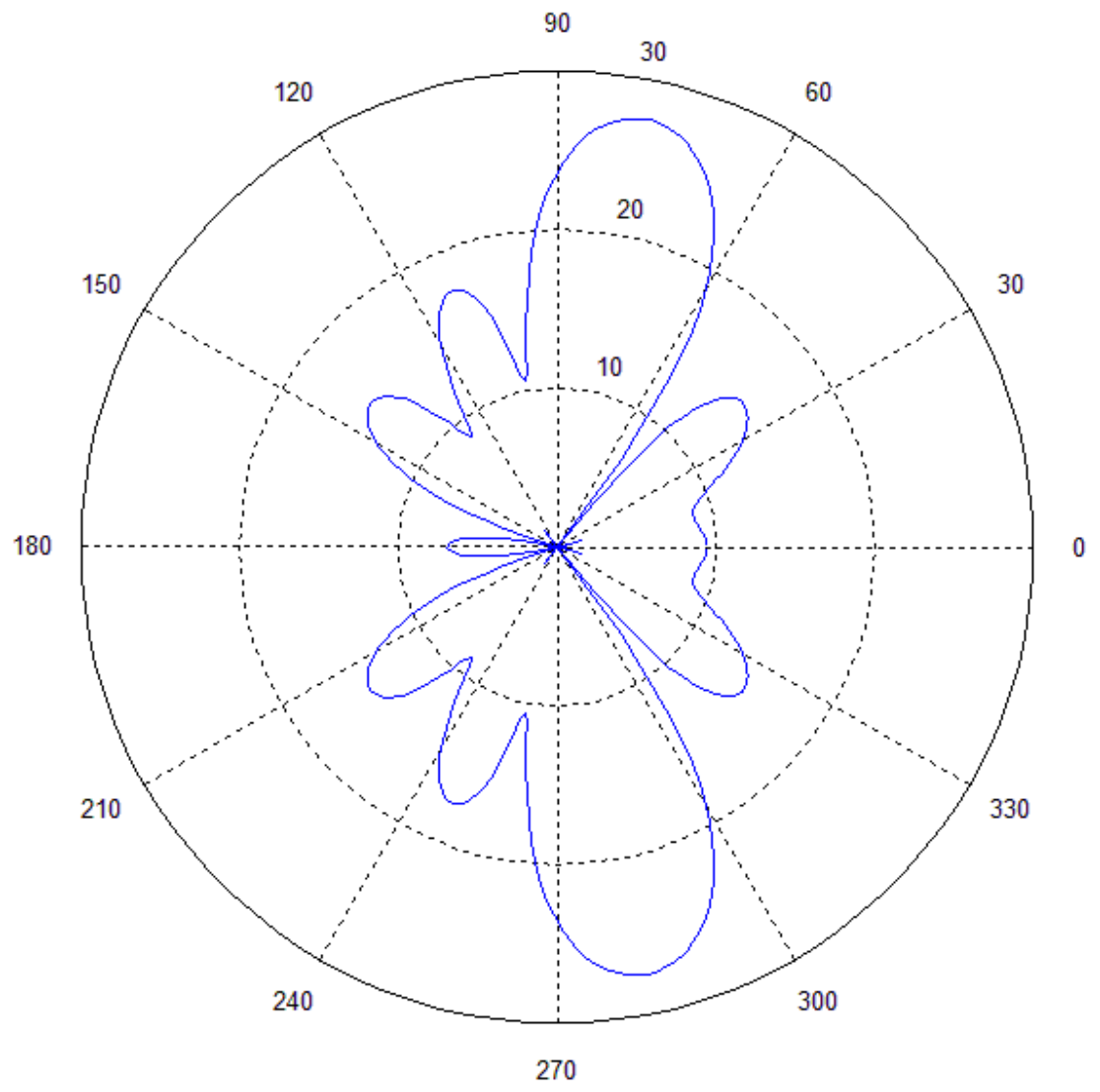

Figure 3.30 BM Simulated Array Factor Radiation Pattern of Port 1.

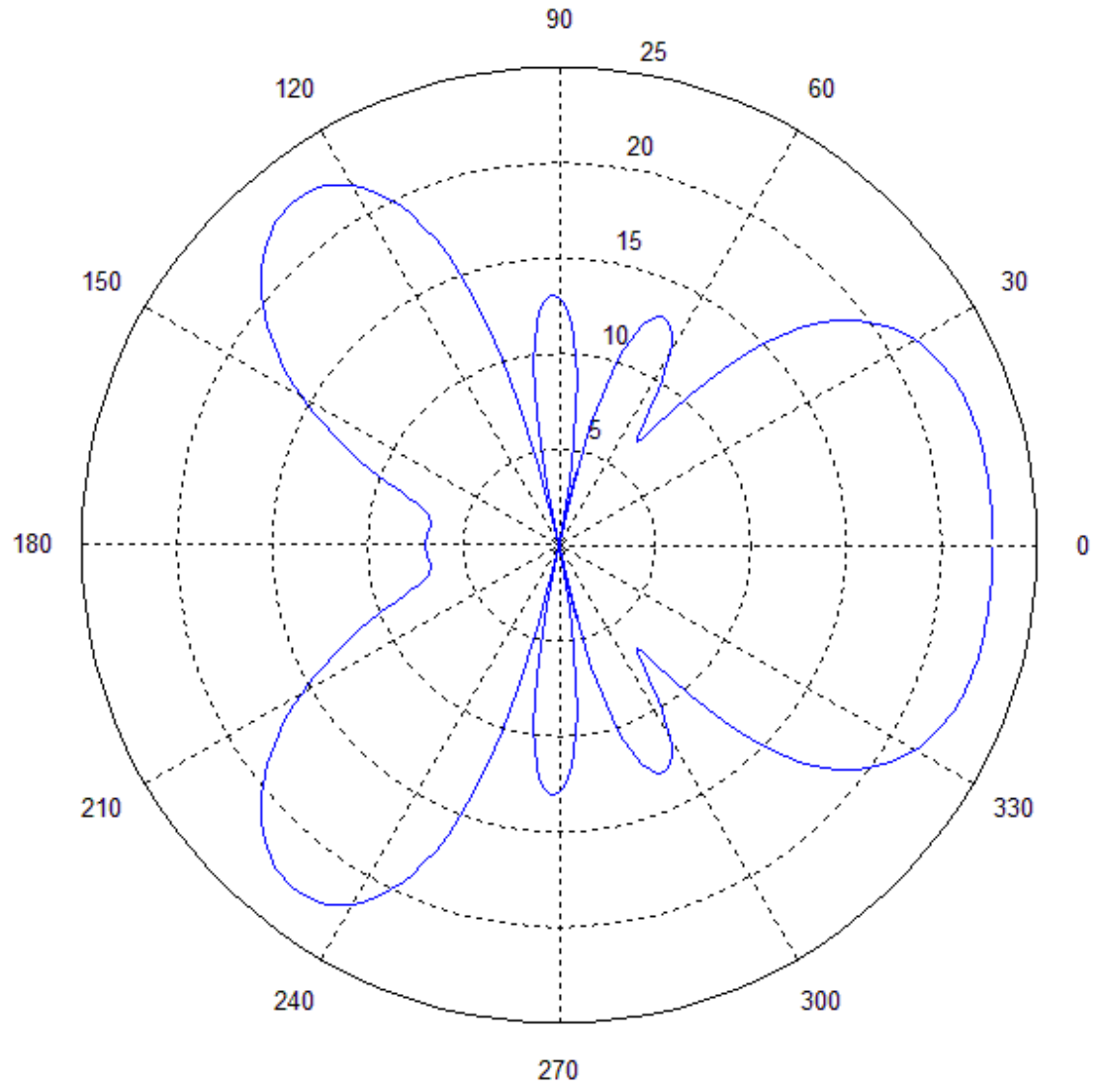

Figure 3.31 BM Simulated Array Factor Radiation Pattern of Port 2.

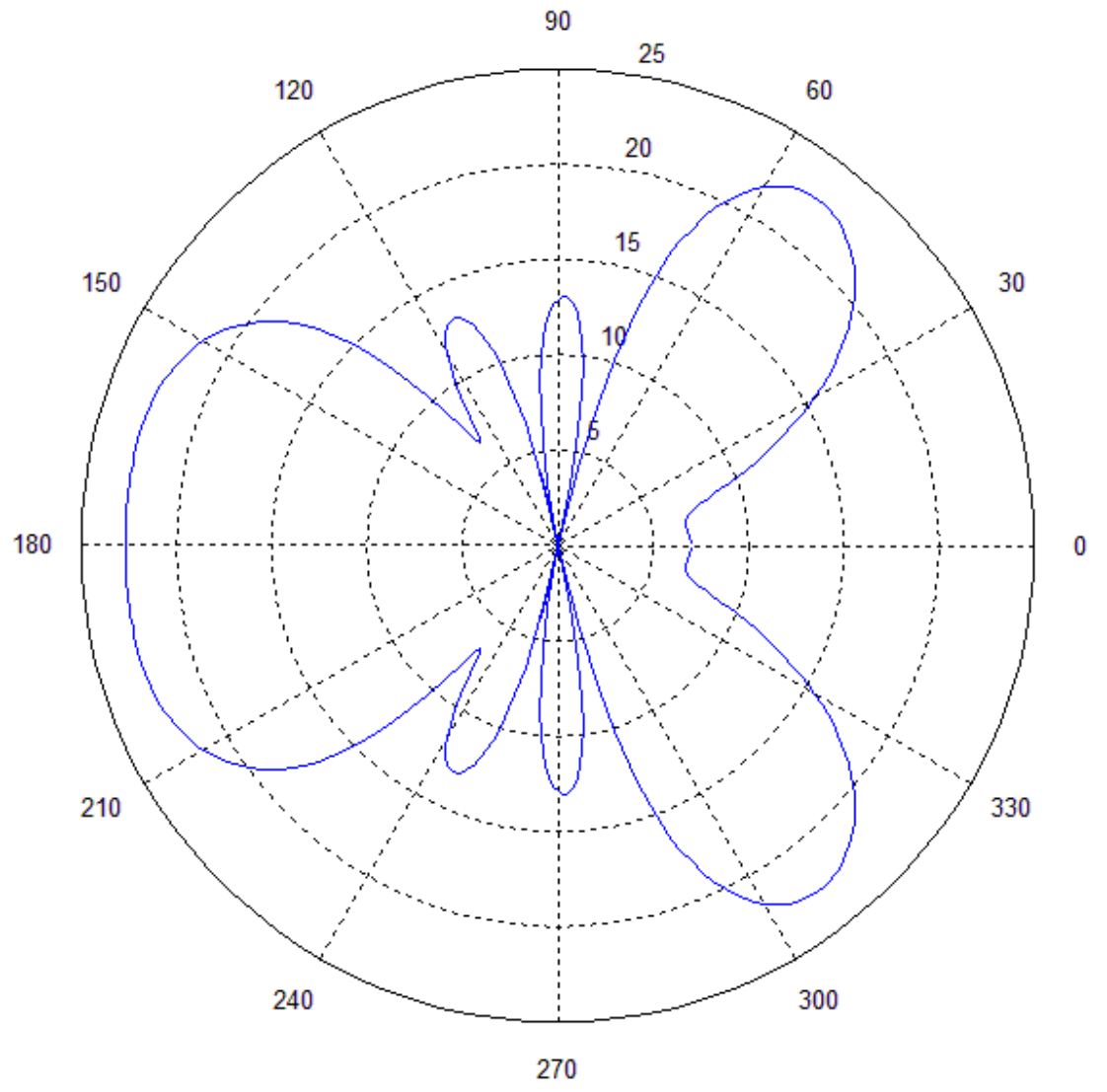

Figure 3.32 BM Simulated Array Factor Radiation Pattern of Port 3.

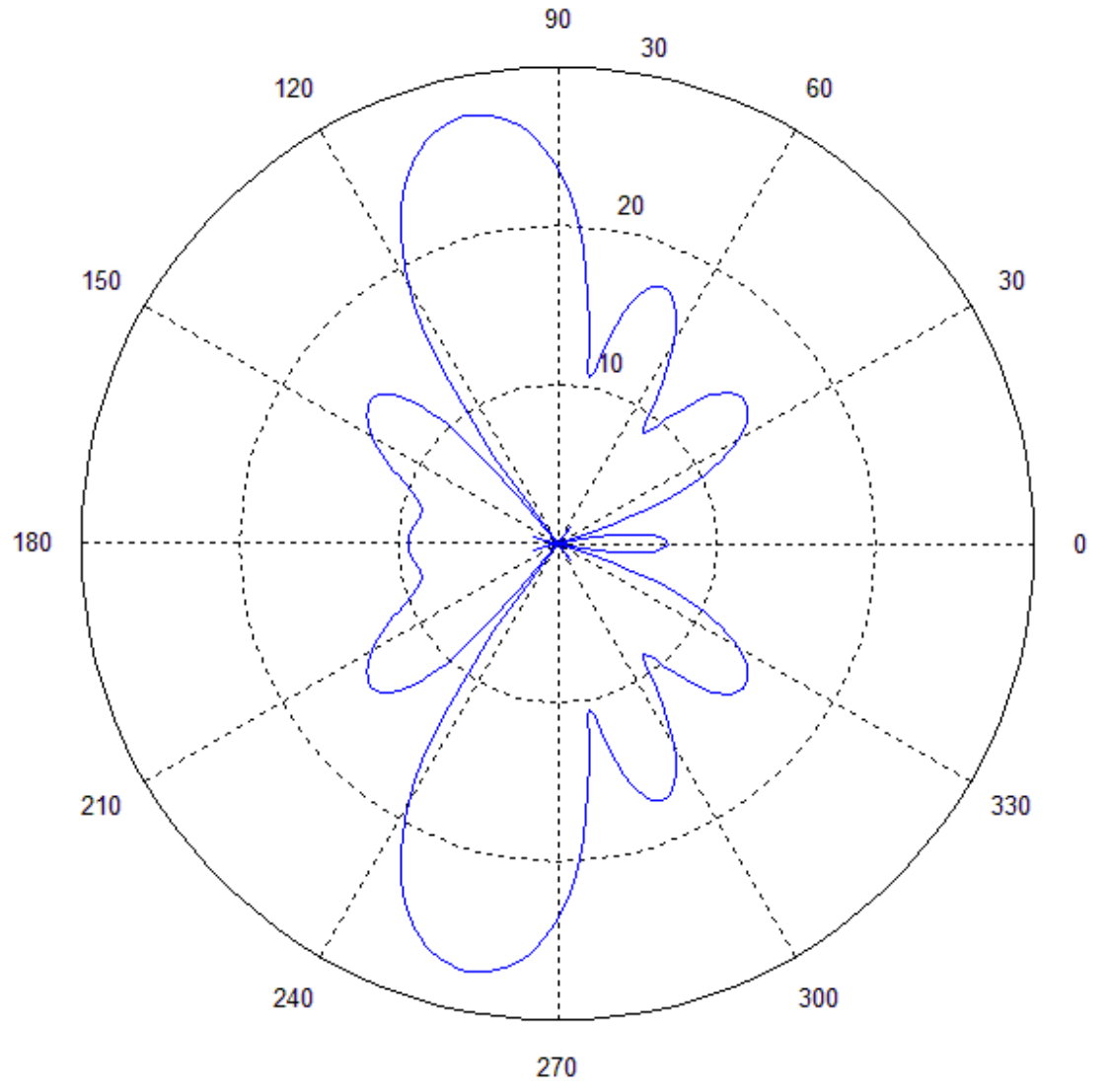

Figure 3.33 BM Simulated Array Factor Radiation Pattern of Port 4.

## 3.8 CASCADED BUTLER MATRICES

Suarez had proposed cascaded Butler Matrices for multibeam systems with circular array applications. The proposed system composed of circular array and two Butler Matrices as passive beamforming networks. It is indicated that the output from the first Butler Matrix produces several narrow and directional beams and by adding an extra Butler Matrix in the system, the number of directional beams can be increased [20]. As Figure 3.34 shows, the switched phase shifters are used to connect the two Butler Matrices.

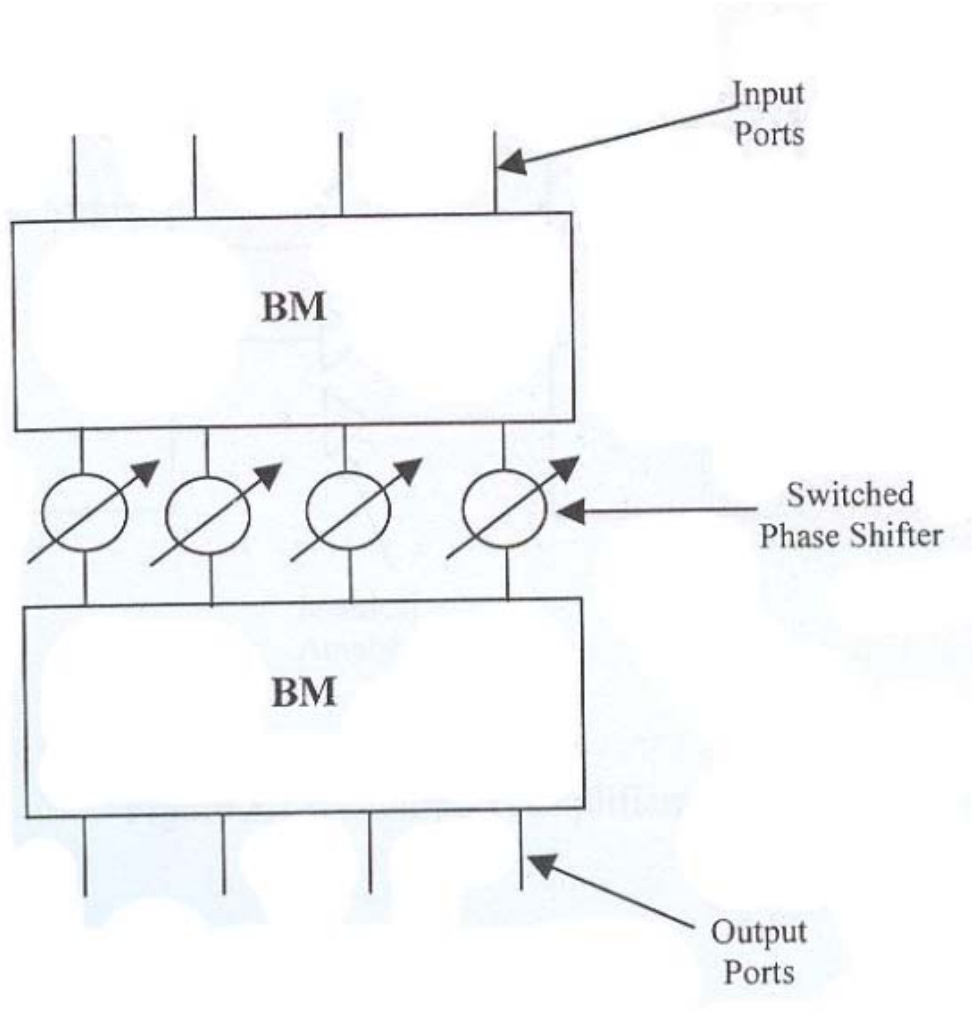

Figure 3.34 Cascaded Butler Matrices with Switched Shifters, taken from [4].

## 3.9 SUMMARY

This chapter covered the essential materials for designing the *4x4* Butler Matrix and its elements. The covered materials include an overview to the branch line coupler and its design procedure, 0dB crossover, phase shifters, along with their output characteristics. Then based on these elements, the final design of the *4x4* Butler Matrix was carried out and the output insertion and return losses, relative phase characteristics, and simulated array factor radiation patterns (for the phased array concept verification) were analysed. Also, the PCB (Printed Circuit Board) layout design of the *4x4* Butler Matrix and its substrate properties were shown. Finally the concept of the Cascaded Butler Matrices was discussed briefly.

# CHAPTER 4
# THE BUTLER MATRIX MEASUREMENTS

## 4.1 INTRODUCTION

This chapter will cover the sections regarding circuit implementation of the *4x4* Butler Matrix and its experimental results. The measured S–Parameters along with a comparison with the simulated results and the computed array factor radiation patterns will be shown.

## 4.2 BM EXPERIMENTAL RESULTS

Figure 4.1 indicates the fabricated *4x4* Butler Matrix board shielded with a metal box along with absorbing foam attached to the top lid of the inner box in order to reduce internal coupling and external interference effects in the shielded metal box. It has been tested using the network analyser having only one input port and one output port ($S_{21}$) measured at a time and all other ports are terminated using 50Ω terminations. In the following sections, the measured S–Parameters with their computed array factor radiation patterns are presented.

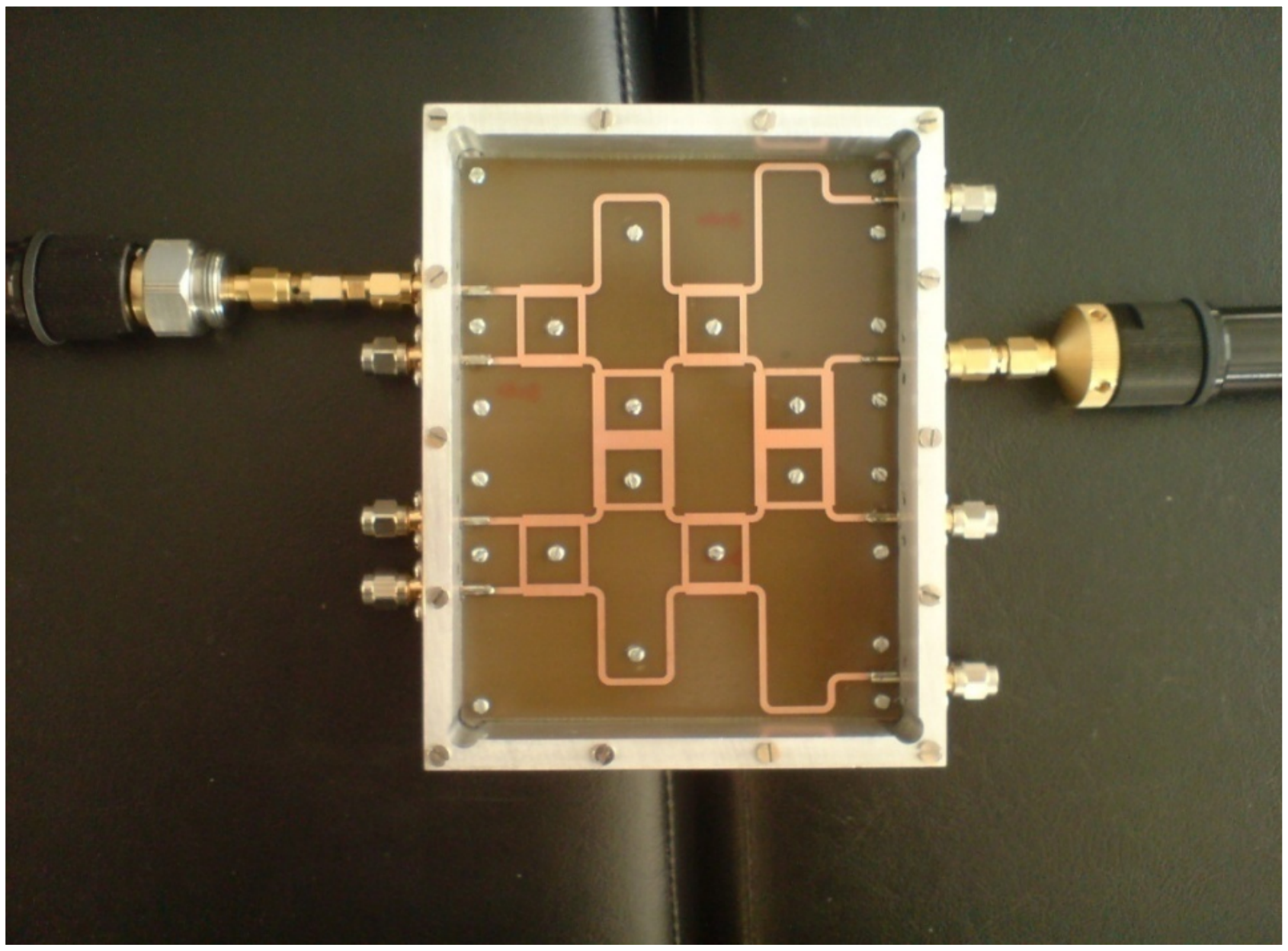

Figure 4.1 *4x4* Butler Matrix Circuit Configuration.

## 4.2.1 BM PORT 1 MEASUREMENTS

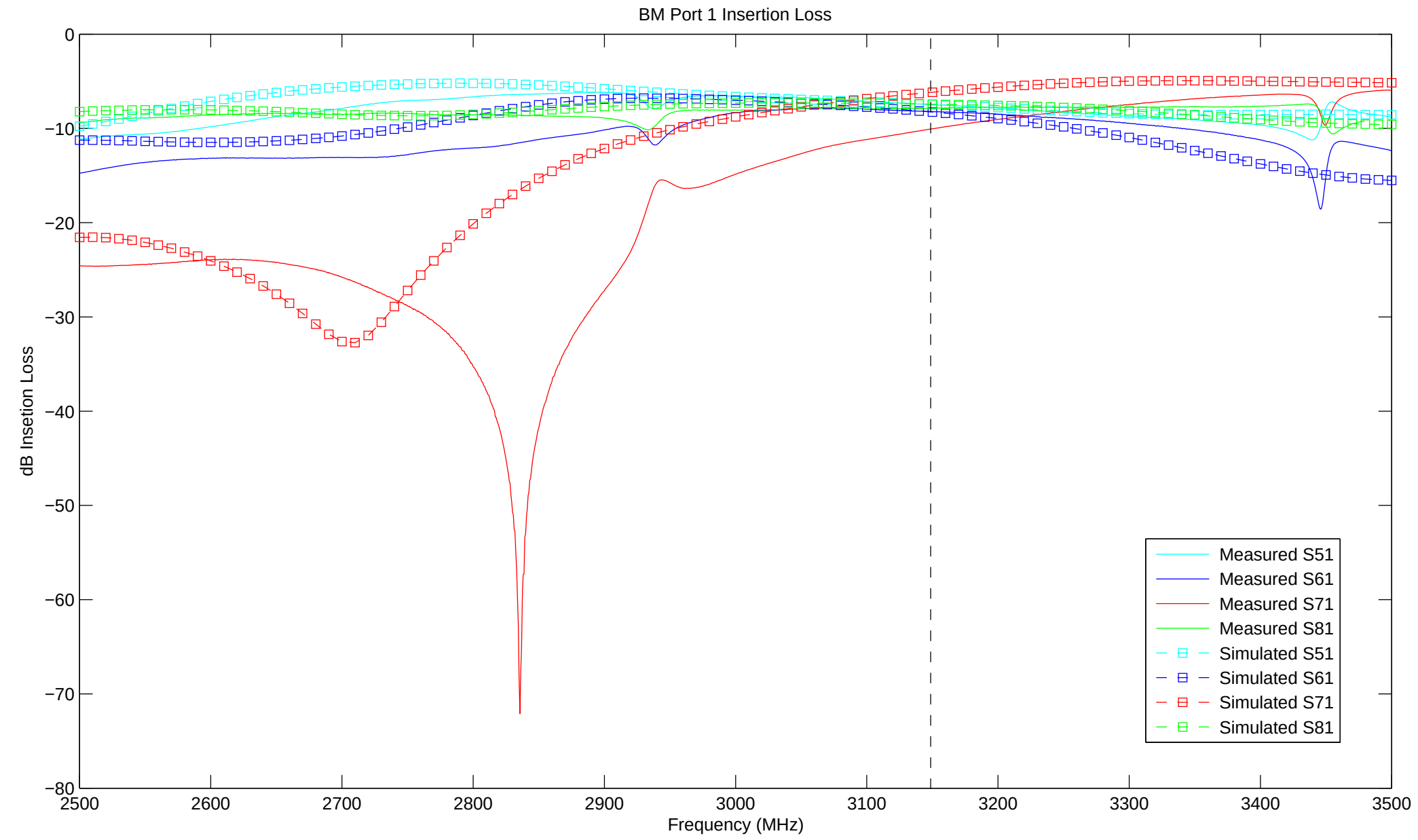

Figure 4.2 Measured and Simulated Insertion Loss of Port 1.

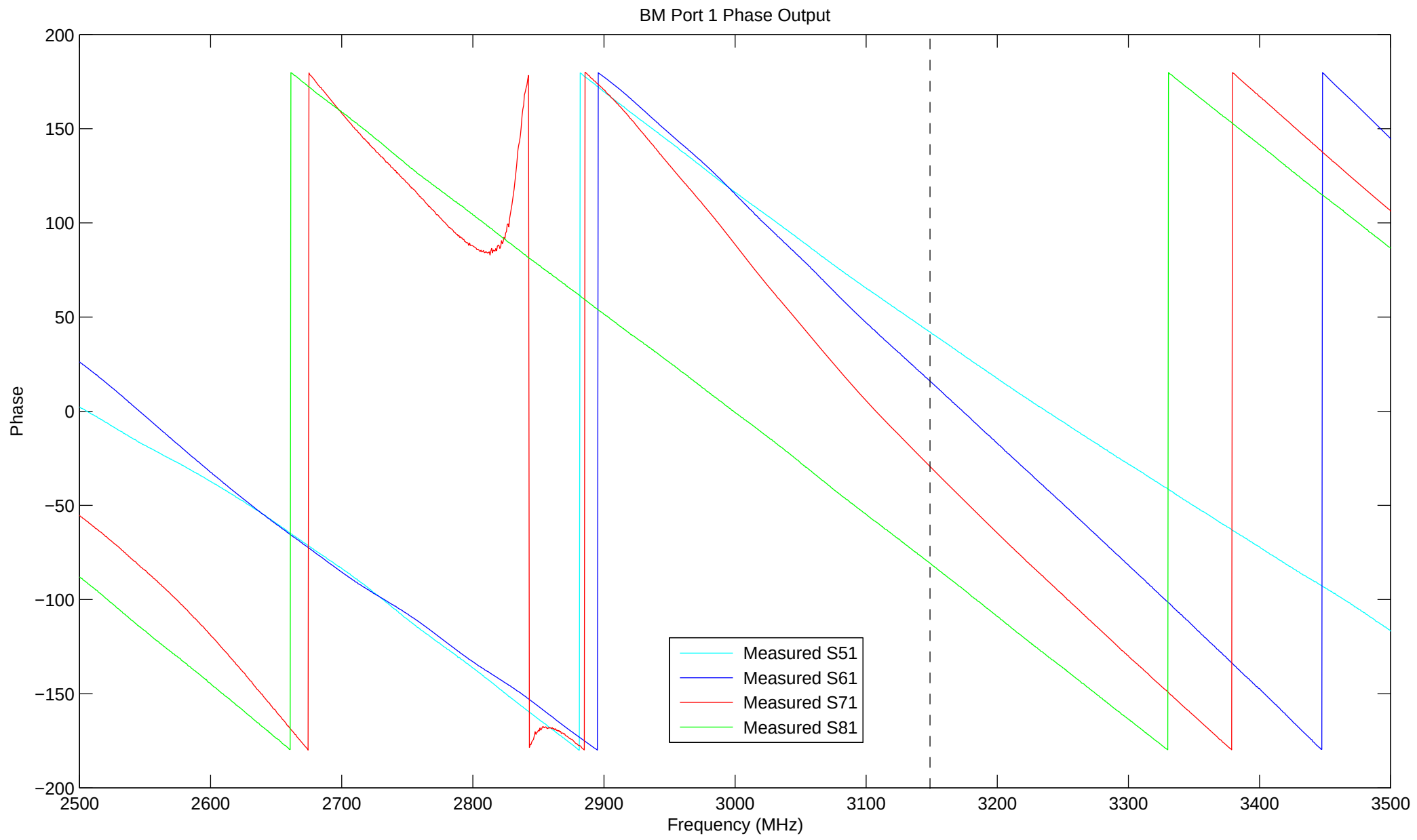

Figure 4.3 Measured Output Phase of Port 1.

## 4.2.2 BM PORT 2 MEASUREMENTS

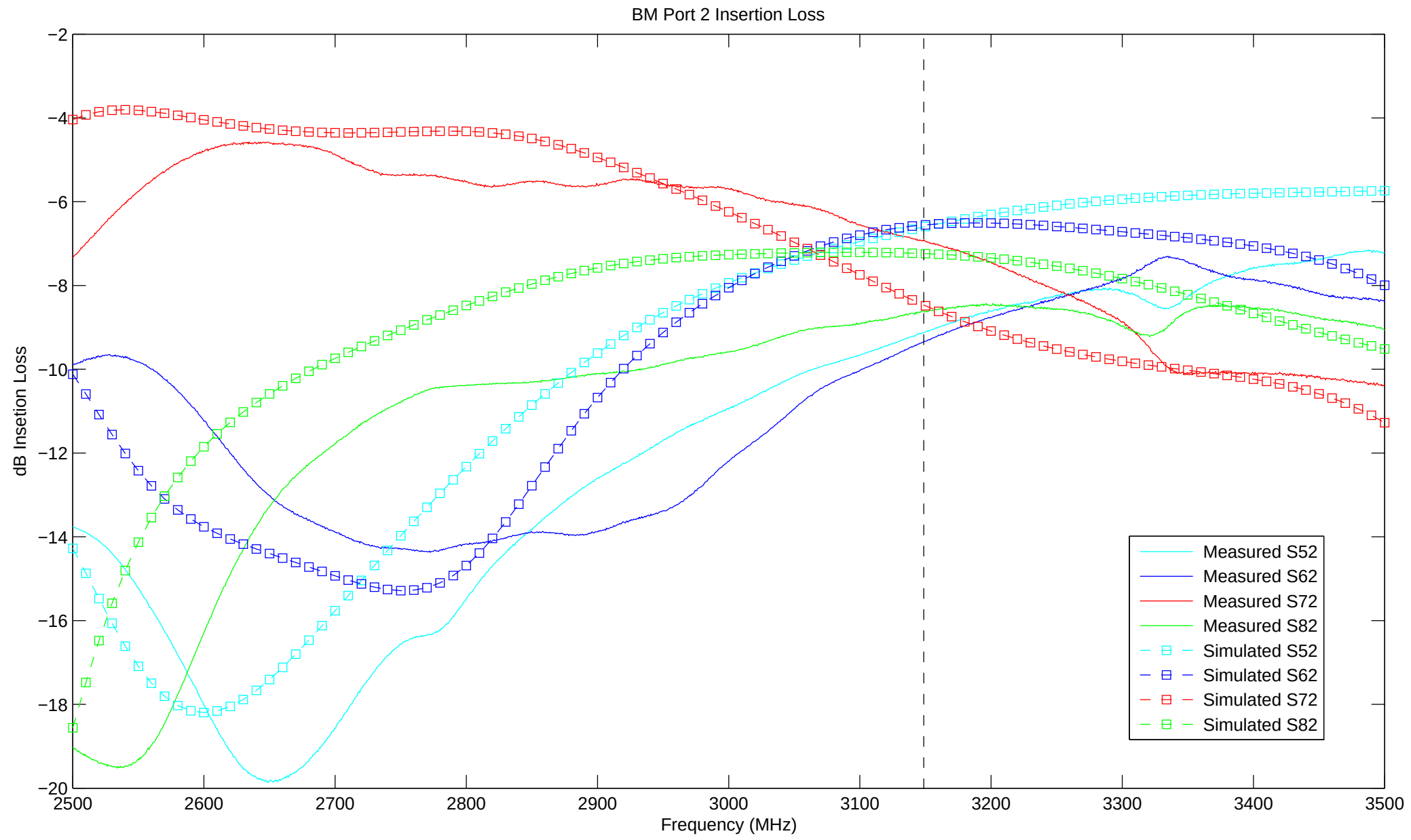

Figure 4.4 Measured and Simulated Insertion Loss of Port 2.

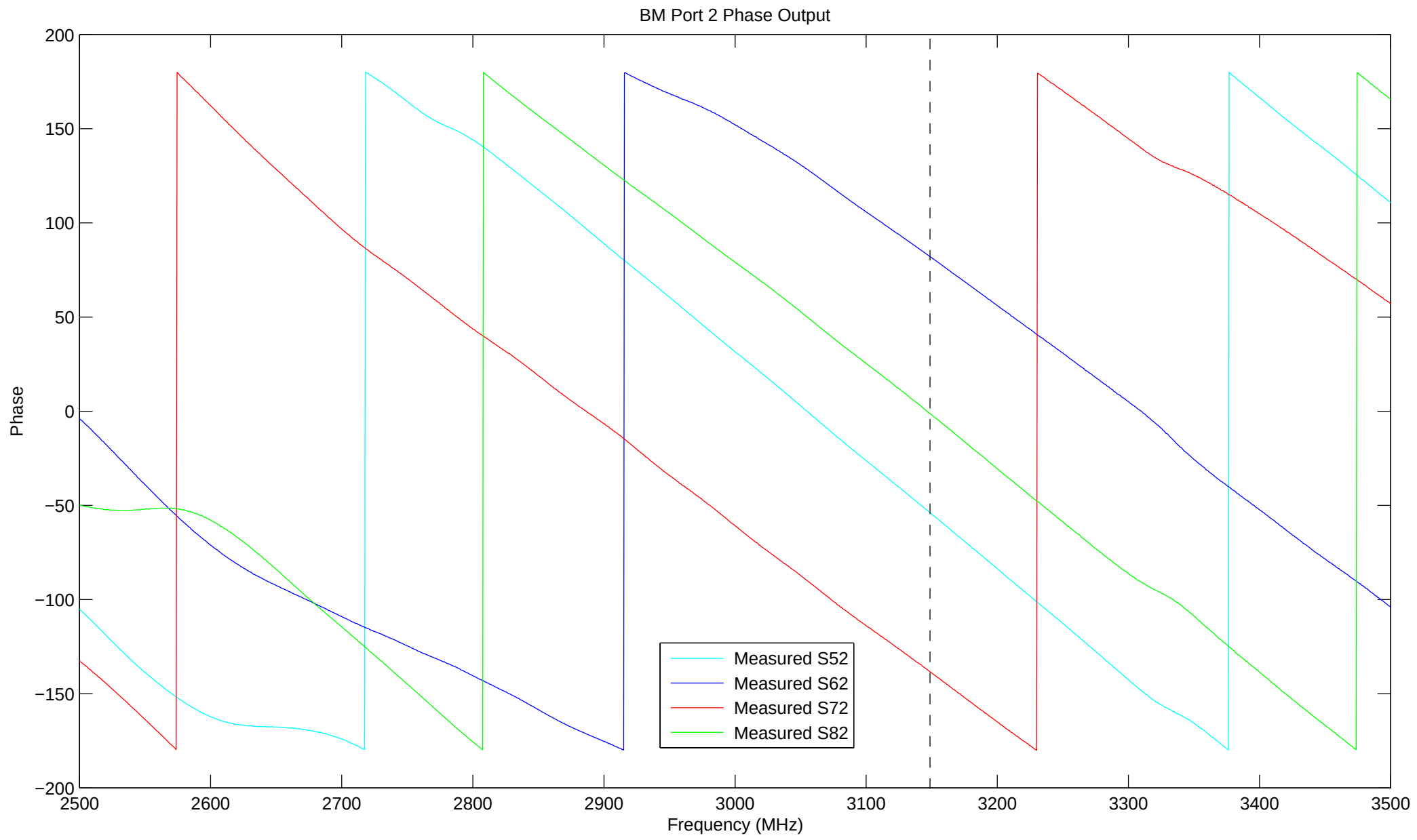

Figure 4.5 Measured Output Phase of Port 2.

### 4.2.3 BM ARRAY FACTOR RADIATION PATTERNS

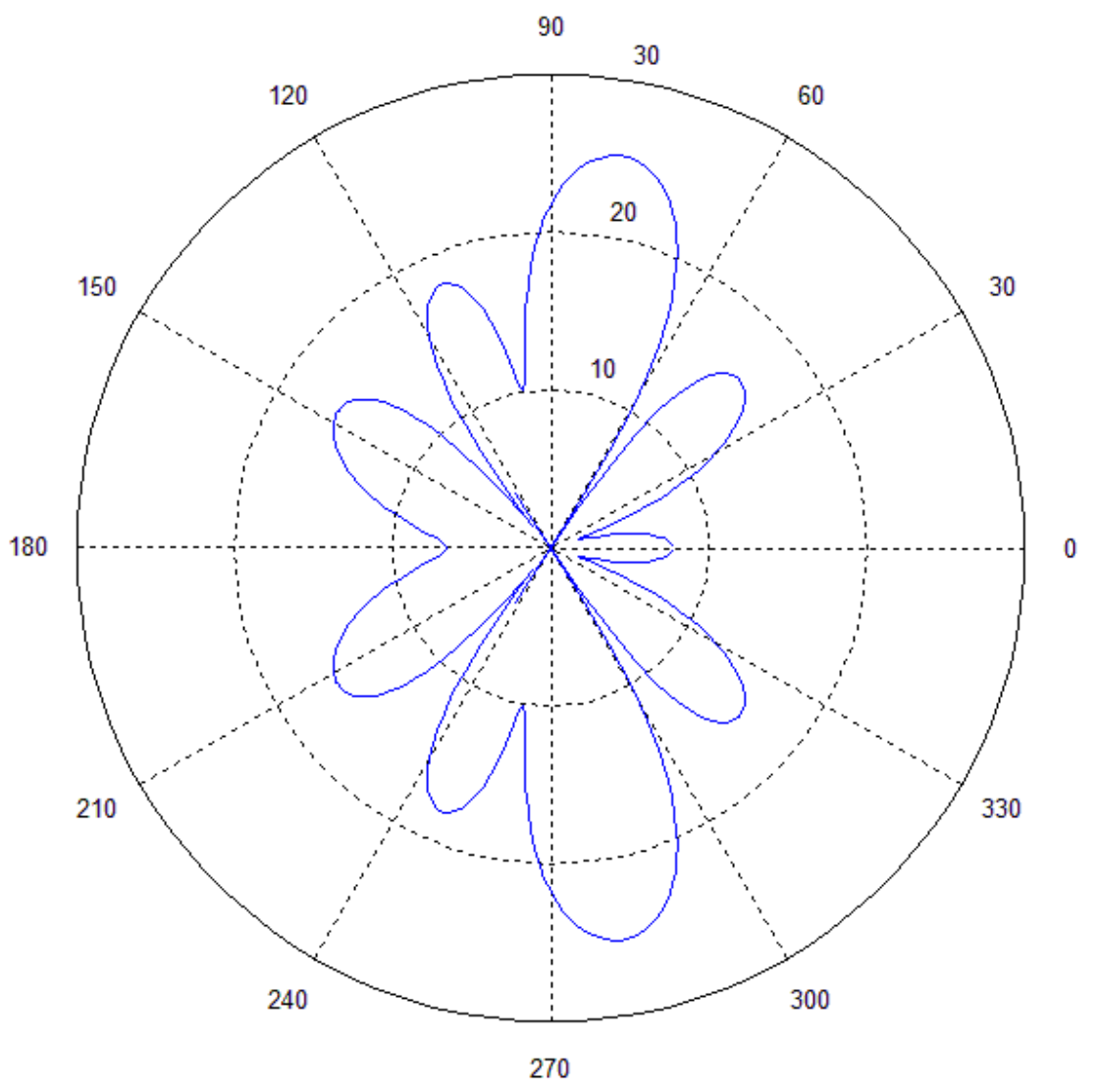

Figure 4.6 BM Measured Array Factor Radiation Pattern of Port 1.

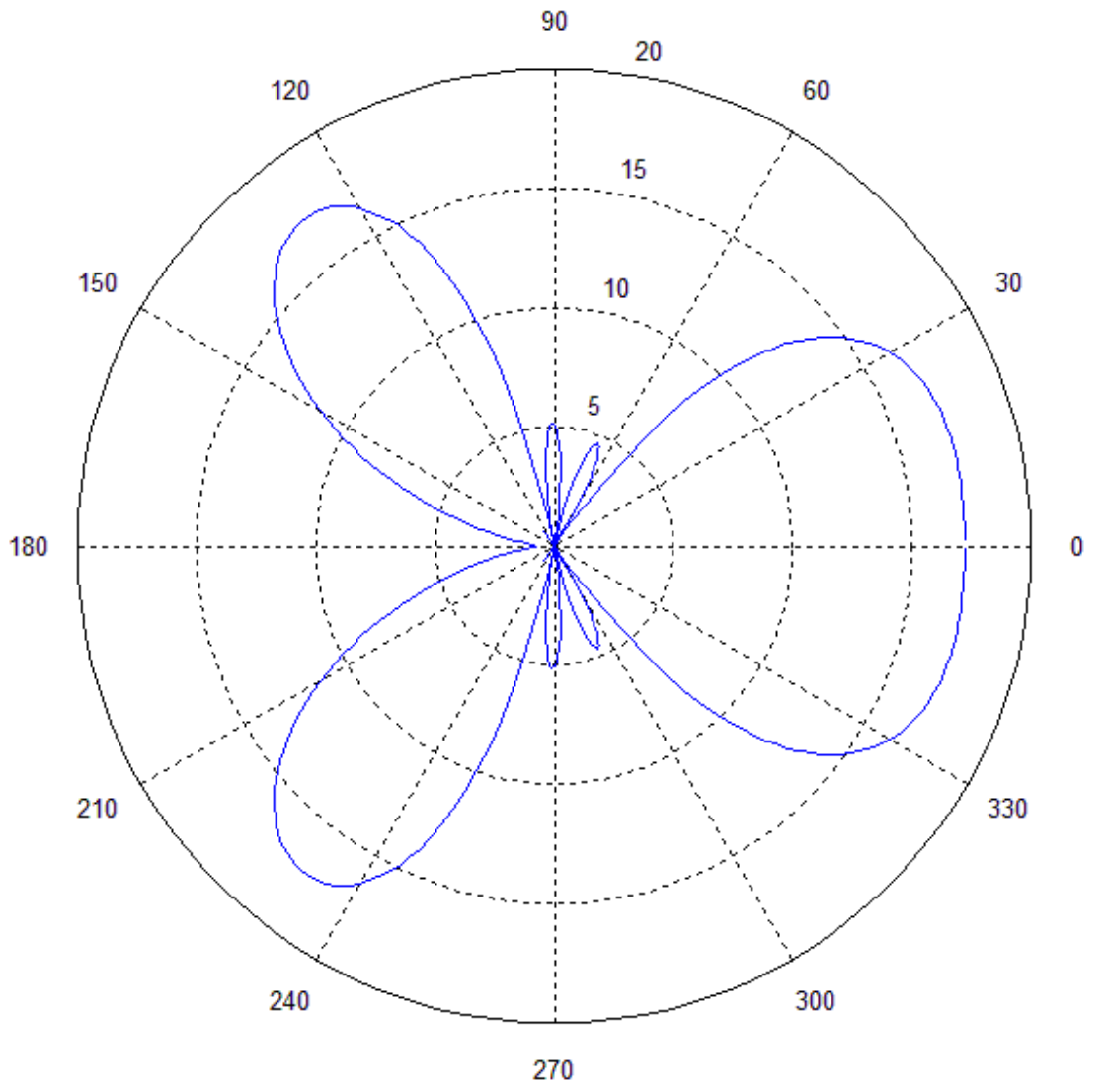

Figure 4.7 BM Measured Array Factor Radiation Pattern of Port 2.

As Figures 4.2 and 4.4 indicate, the measured insertion losses of the output ports 1 and 2 agree with their corresponding simulated values. Figures 4.3 and 4.5 also show the output phases of different generated beams having almost the same phase progression for ports 1 and 2 respectively.

Figures 4.6 and 4.7 show the array factor radiation patterns for ports 1 and 2 respectively, utilising the measured S–Parameter amplitudes and phases. These radiation patterns indicate beams formed in the expected directions having their main lobes showed strong identity with at least 10dB of isolation from the side lobes. Also, it is shown that despite the non–ideal performance of the *4x4* Butler Matrix, in terms of phase and amplitude distributions, it is still capable of forming well–defined beams suitable for the beam steering experiments by causing the main lobe to be directed in a certain direction. The high gain narrow–beam pattern can be used for long–distance communication while the broad–beam can be selected for shorter distance communication.

The differences between the simulations and measurements and slight distortion of the beam shape might be due to the employed FR4 board and fabrication process, non–uniformity of transmission line width, cross–coupling effects, and measurement process errors. This matrix can be used as a beamforming network for multibeam antenna arrays.

## 4.3 SUMMARY

This chapter covered the materials regarding the *4x4* Butler Matrix circuit implementation and measurements. The measured S–Parameters along with a comparison with the simulated results and measured array factor radiation patterns were shown and analysed.

# CHAPTER 5
# THE ROTMAN LENS SIMULATIONS

## 5.1 INTRODUCTION

This chapter will cover the definition, lens geometry, design approach, and simulation of the Rotman Lens as another type of multiple beamforming networks. The design and simulation of the *4x4* and *8x8* Rotman Lenses will be carried out and the output results will be shown.

## 5.2 ROTMAN LENS BRIEFLY

As it is mentioned earlier, microwave lenses are another well–known class of multiple beamforming networks. Rotman Lens is an attractive beamforming network due to its low cost, reliability, design simplicity, and wide–angle scanning capabilities. It avoids the complexities of phase shifters to steer a beam over wide angles and has proven itself to be a useful beamformer for Electronic Scanning Arrays (ESA). Because Rotman lens is a true time–delay device, it produces beam steering independent of frequency and is therefore capable of wide–band operation. Also, the cost of a Rotman Lens implemented on microwave material is primarily driven by the cost of the material itself and the price of photo etching [21].

A Rotman Lens is a parallel plate device used to feed an antenna array. It has a carefully chosen shape and appropriate length transmission lines in order to produce a wave–front across the output that is phased by the time delay in the signal transmission. Each input port will produce a distinct beam that is shifted in angle at the output [22].

The design of the lens is controlled by a series of equations that set the focal points and array positions. The inputs, during the design of the system, include the desired scan angle of the array, the working frequency, the desired number of beams and array elements, and the spacing of the elements [22]. The details of the equations needed for the lens geometry and its design procedure are provided in the next section.

## 5.3 LENS GEOMETRY AND DESIGN APPROACH

Figure 5.1 indicates the cross–section of a tri–focal Rotman Lens. One focal point $F_0$ is located on the central axis and two others $F_1$ and $F_2$ are located on either side on a circular focal arc. Contour $I_2$ is a straight line and defines the position of the radiating elements. $I_1$ is the inner contour of the lens and also called the array contour. The inner and outer contours are connected by TEM mode transmission line W(N). Two off axis focal points $F_1$ and $F_2$ are located on the focal arc and make angles $\alpha$ and $-\alpha$ with the *X*–axis. It is required that the lens be designed in such a way that the outgoing beam make angles $-\alpha$, 0, and $\alpha$ with the *X*–axis when feed are placed at $F_1$, $F_0$, and $F_2$ respectively [23]. Figure 5.2 also indicates the direction of outgoing beams for different input ports.

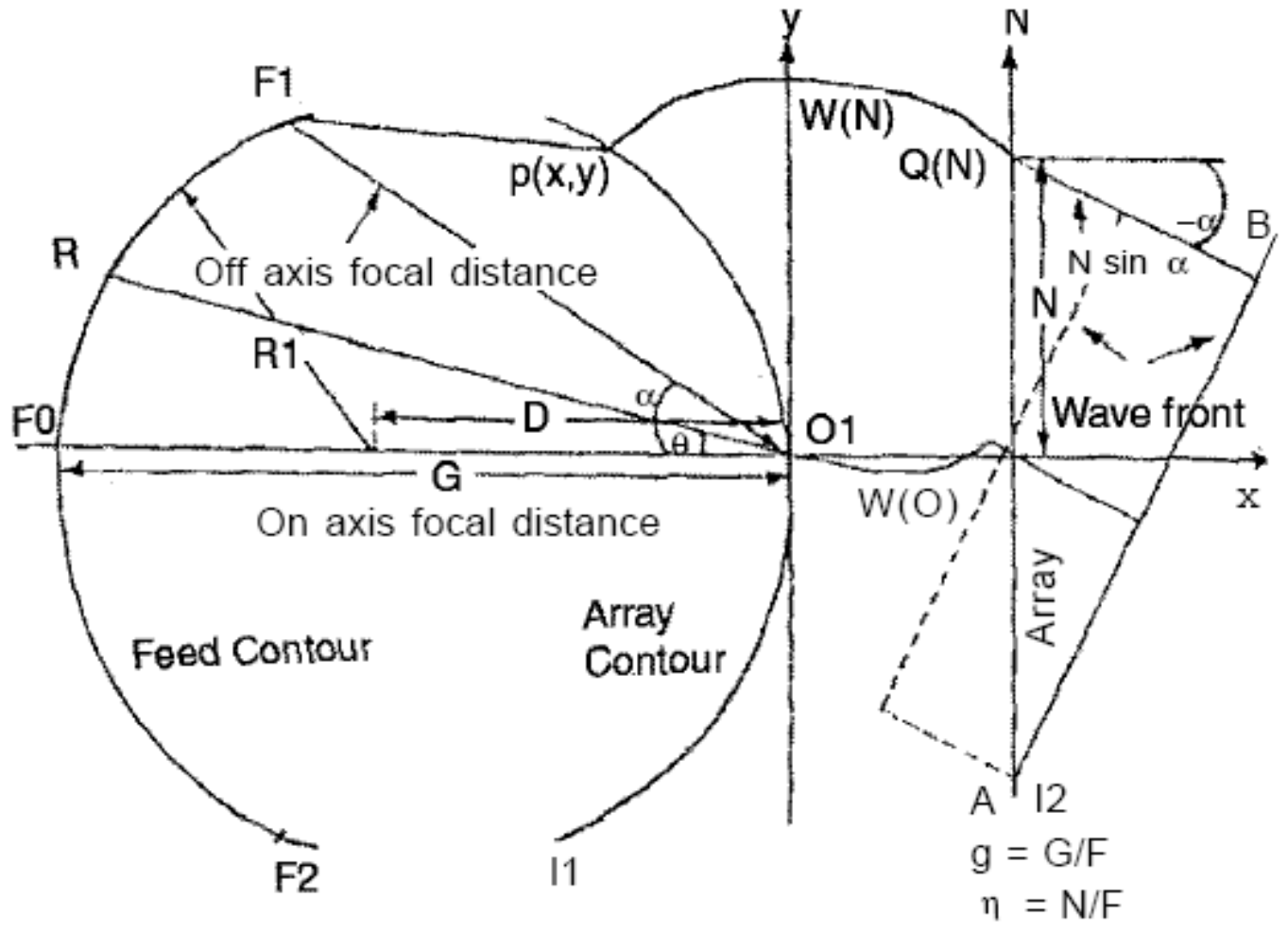

Figure 5.1 Configuration of a Tri–focal Rotman Lens, taken from [23].

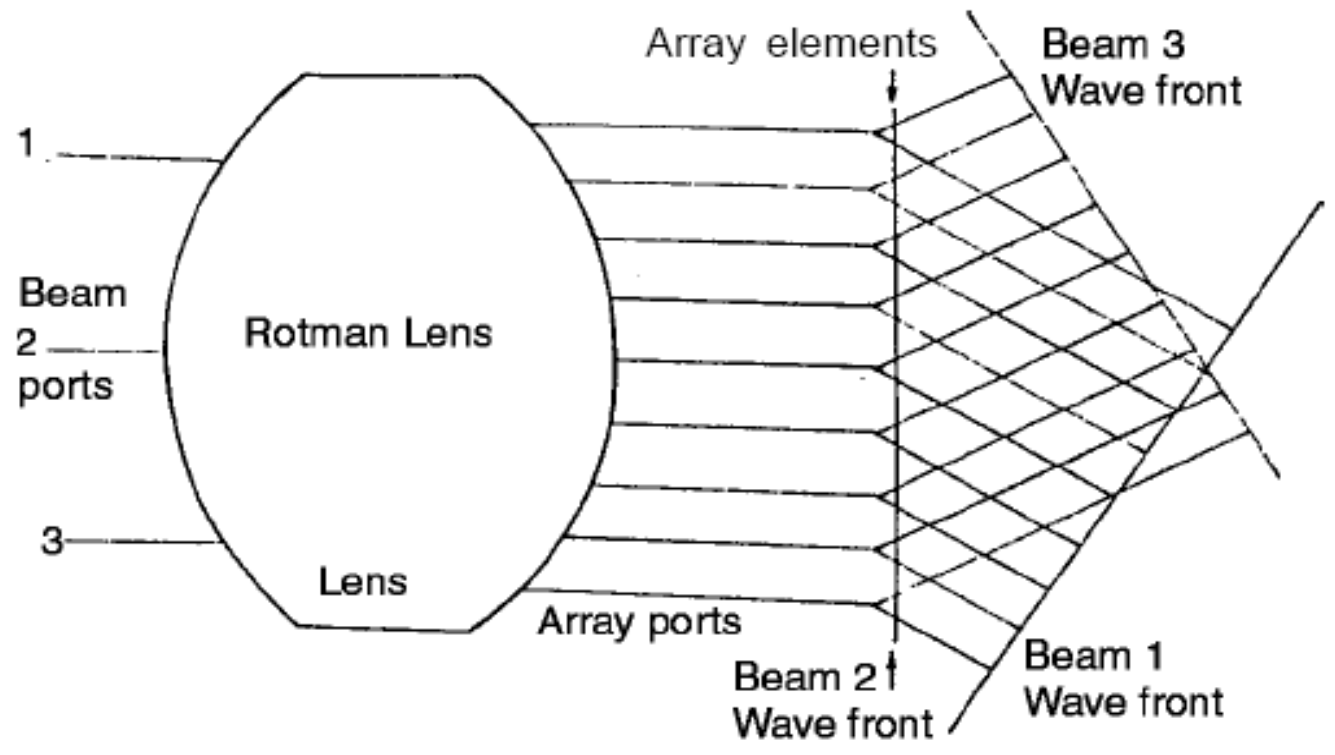

Figure 5.2 Direction of Outgoing Beams, taken from [23].

In designing the Rotman Lens, the following design parameters are taken into account:

i) Off–axis focal length $F$: The distance between off–axis focal point and mid–point of the array contour (distance $O_1F_1$ or $O_1F_2$ as defined in Figure 5.1).

ii) On–axis focal length $G$: The distance between on–axis focal point $F_0$ and mid–point of the array contour (distance $O_1F_0$ as defined in Figure 5.1).

iii) Antenna element spacing $d$: Radiating elements are located along the straight line $I_2$. Number of antenna elements and spacing between them determine the length of lens contour $I_2$.

iv) Focal angle $\alpha$: It determines the angular coverage provided by the lens. The lens can cover an angular area of $\pm\alpha$.

A ray originating from $F_1$ may reach the wave–front through a general point P(X,Y) on the inner contour $I_1$, transmission line W(N), and point Q(N) on the outer contour and then tracing a straight line at an angle $-\alpha$, terminate perpendicular to the wave–front AB at an angle $-\alpha$ to the broadside. Also, the ray from $F_1$ may reach the wave front from $F_1$ to point $O_1$ and then through W(0), terminate at the wave–front AB [23].

Inner contour $I_1$, and the transmission lines are designed based on the equations derived from the fact that at the wave–front all these rays must be in phase, independent of the path they travel. This requires that the total phase shift in traversing the path to reach the wave–front in each case be equal. The lens design equations are written by equating the path length from the focal points to the corresponding wave–front [24].

$$F_1P + \mathrm{W(N)} + N\sin\alpha = F + \mathrm{W(0)} \tag{5.1}$$

$$F_2P + \mathrm{W(N)} - N\sin\alpha = F + \mathrm{W(0)} \tag{5.2}$$

$$F_0P + \mathrm{W(N)} = G + \mathrm{W(0)} \tag{5.3}$$

where

$$(F_1P)^2 = (F\cos\alpha + X)^2 + (F\sin\alpha - Y)^2$$

$$(F_2P)^2 = (F\cos\alpha + X)^2 + (F\sin\alpha + Y)^2$$

$$(F_0P)^2 = (G + X)^2 + Y^2$$

Lens dimensions are normalised by off–axis focal length $F$ as follows:

$$\eta = N / F,\ g = G / F,\ x = X / F,\ y = Y / F, \text{ and } w = (\mathrm{W(n)} - \mathrm{W(0)}) / F$$

Let it be assumed that:

$$a = \cos\alpha \text{ and } b = \sin\alpha$$

Algebraic manipulation of above equations yield the following relations:

$$y = \eta\,(1 - w) \tag{5.4}$$

$$x^2 + y^2 + 2gx = w^2 - 2gw \tag{5.5}$$

$$a_0\, w^2 + b_0\, w + c_0 = 0 \tag{5.6}$$

where

$$a_0 = (1 - 1\eta^2 - (g - 1)^2) / (g - a)^2$$

$$b_0 = 2g\,(g - 1) / (g - a) - (g - 1)\, b^2\, \eta^2 / (g - a)^2 + 2\eta^2 - 2g$$

$$c_0 = gb^2\eta^2 / (g - a) - b^4\eta^4 / 4(g - a)^2 - \eta^2$$

For fixed value of design parameters $\alpha$ and $g$, $w$ can be calculated as a function of $\eta$. These values of w and $\eta$, give $x$ and $y$. This procedure completes the lens design.

In principle, the lenses with the wide–angle scanning capabilities are wideband systems since their design is based on the different optical path lengths. However, their band–width is limited by the circuit elements used in realising the lens such as the array elements and their properties, connecting transmission lines, and the array geometries and mutual coupling effects. Band–width for a given lens must be carefully defined in terms of array performance [24].

According to Singhal, different types of focal arcs are suggested like parabolic, hyperbolic, elliptical, and straight line. Having investigated the effects of design parameters on the shape of the lens and the phase error, it is suggested that the lens obtained with elliptical focal arc is more compact compared to circular focal arc lens. In design of the *4x4* and *8x8* Rotman Lenses, the shapes of focal arc are considered as circular and elliptical respectively.

In elliptical focal arc lens, the equations to design the array contour and transmission lines will be exactly the same as circular. The only difference will be in the focal arc [25]. The focal arc is given by the following equation:

$$X^2 / a^2 + Y^2 / b^2 = 1 \tag{5.7}$$

where

$$b^2 = a^2 (1 - e^2)$$

$$a^2 = F^2 \{(1 - e^2 \cos^2\alpha) / (1 - e^2)\}$$

$e$ is the eccentricity of the elliptical focal arc. The centre of the elliptical focal arc is located at the origin (0,0). $F$ is the off–axis focal length, and $\alpha$ is the focal angle [25].

## 5.4 EFFECTS ON THE SHAPE

The lens shape determines the mutual coupling between ports, multiple scattering between ports, and spillover losses[2]. Figure 5.3 shows the effect of focal angle $\alpha$ for an elliptical focal arc lens. As the value of $\alpha$ increases, the array contour closes (its curvature increases), and the feed contour opens. The height of the array contour is a function of lens aperture $\eta$. As $\eta$ increases, the height the array contour increases up to a certain range. The value of $\eta$ should be chosen in a way in order to equalize the height both contours. Equal heights of both contours are required to couple the maximum power from the feed contour to array elements [25].

Figure 5.4 indicates the dependence of the lens contour $g$. As $g$ increases, the array contour opens, the feed contour closes, and the gap between the feed contour and array contour also increases. A small value of *g* gives a compact lens configuration hence it has fewer spillover losses.

[2] The coupling of energy from the beam ports to array ports due to sidewall reflection and vice versa.

The off–axis focal length has no effect on the shape of the lens and it only changes the dimensions of the lens [25].

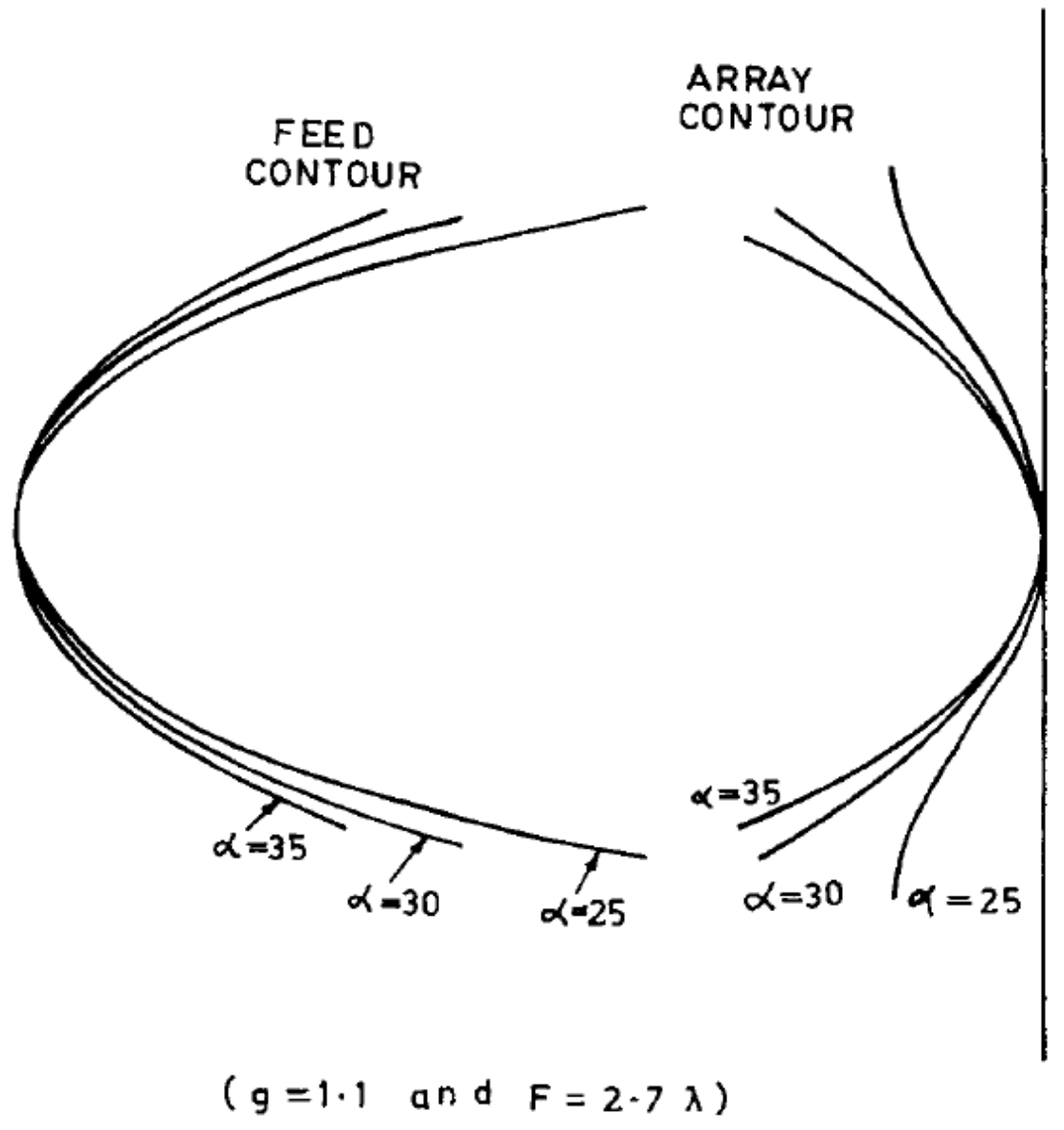

Figure 5.3 Effect of $\alpha$ on the Shape of the Lens, taken from [25].

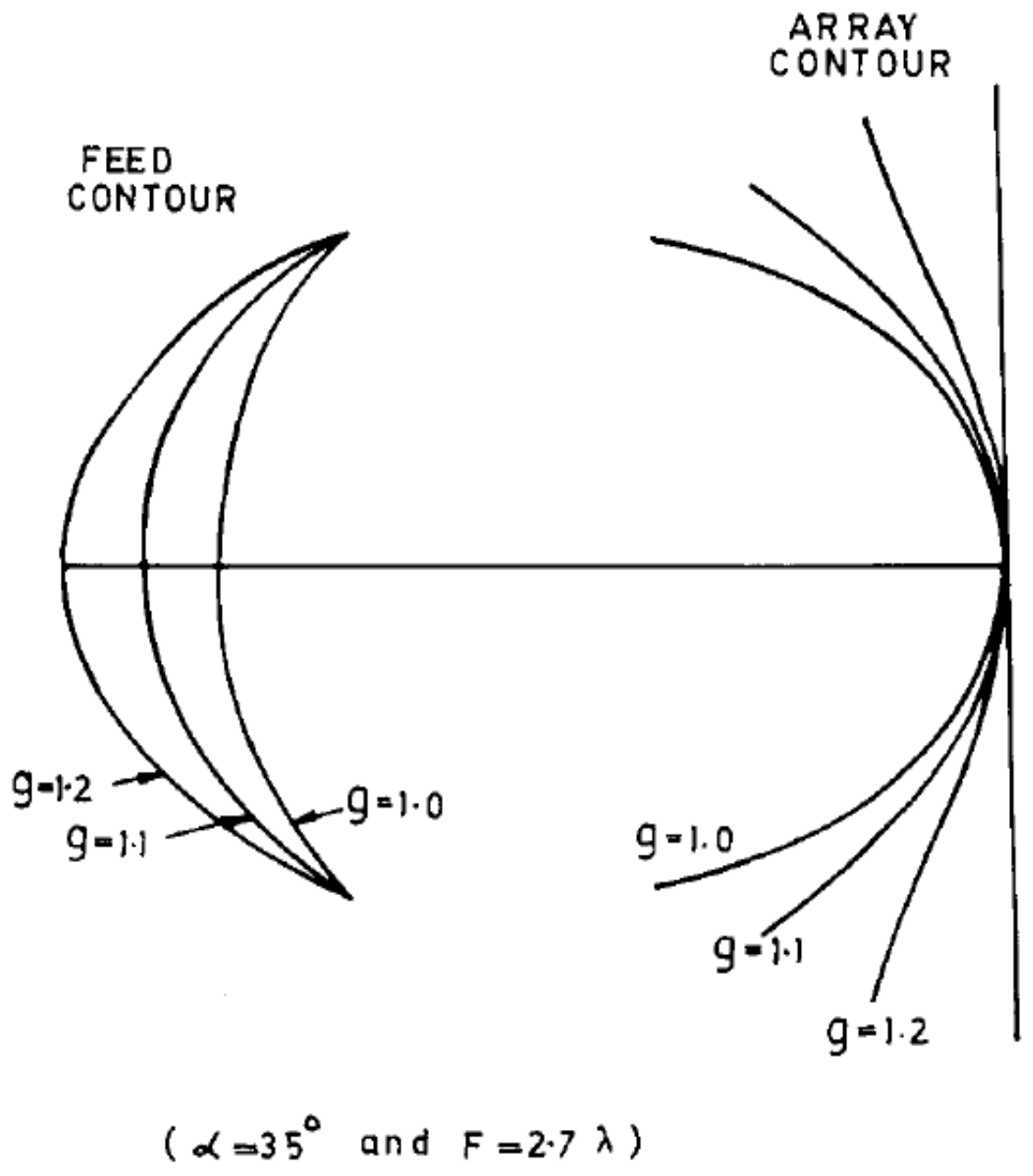

Figure 5.4: Effect of *g* on the Shape of the Lens, taken from [25].

## 5.5 EFFECTS ON THE PHASE ERROR

When a feed point is placed at any one of the focal points, the corresponding wave–front has no phase error. When the feed is displaced from these focal points, the corresponding wave–front will have a phase error. However, for wide–angle scanning, the lens must be focused at all intermediate points along the focal arc [25].

Let a feed be located at point $R$ in Figure 5.1 on the focal arc. Line $RO_1$ makes an angle $\theta$ with the central axis. Then the direction of the outgoing beam should be an angle of $-\theta$. Let $R_a$ and $R_b$ be the path lengths from the feed position to the wave–front when the ray is passing through $O_1$ and P(X,Y) respectively. The phase error is given by [24]:

$$\delta L = R_a - R_b \tag{5.8}$$

The phase error is a function of lens aperture $N$ and scan angle $\theta$ and it increases with scan angle and lens aperture.

## 5.6 LENS COUPLING

To calculate performance of the microwave lens, the coupling between ports using aperture theory is approximated and a uniform distribution to the port aperture is implied. These port radiation patterns are used to compute the direct path and singly reflected path propagation from port to port as shown in Figure 5.5.

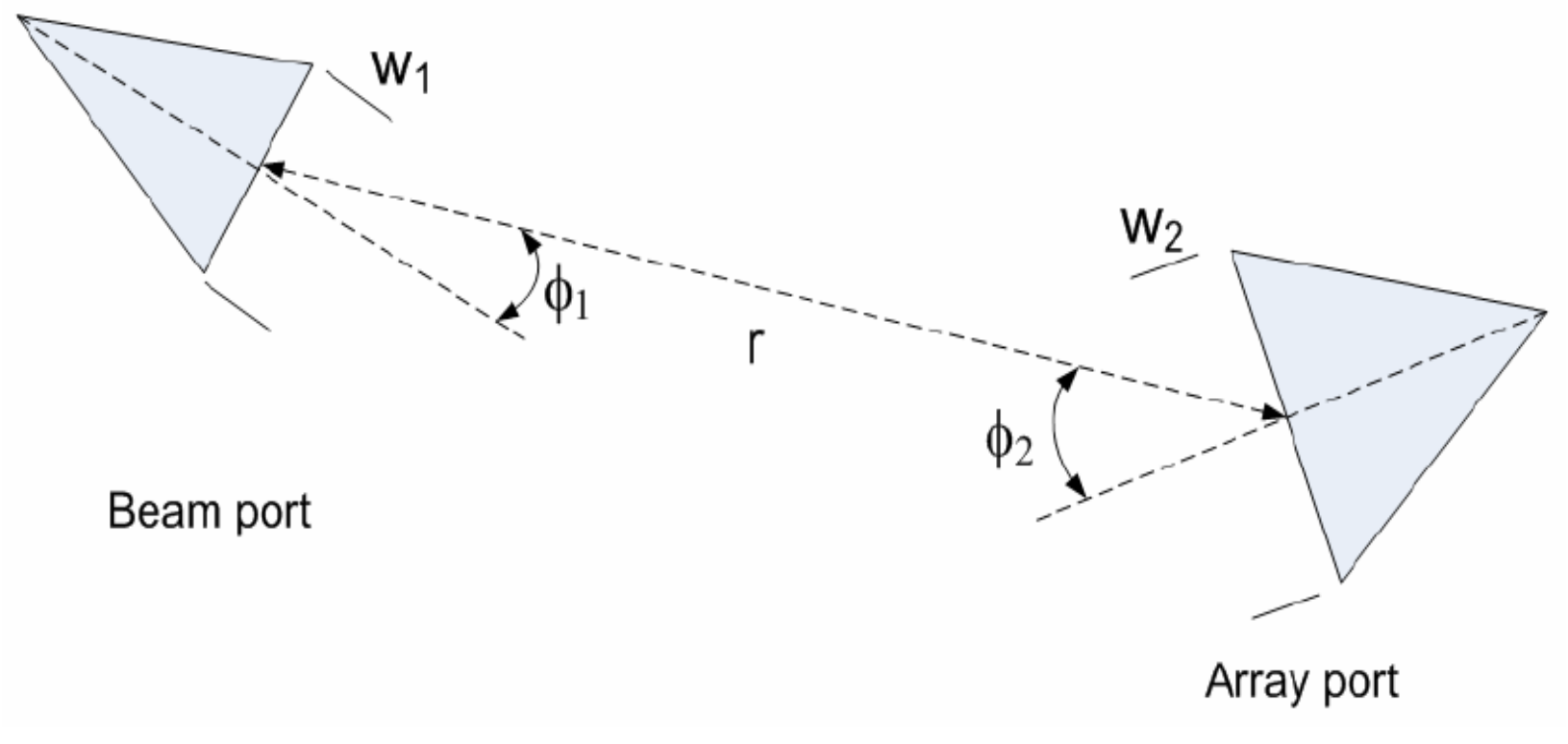

Figure 5.5 Ray–based Coupling Model.

The port to port coupling is then approximated as the following:

$$S_{ij} = \frac{\sin x_i}{x_i} \frac{\sin x_j}{x_j} \sqrt{\frac{w_i w_j}{\lambda r}} e^{-j(kr+\frac{\pi}{4})} \tag{5.9}$$

where

$$x_i = (kw_i / 2) \sin\varphi_i$$

A similar calculation is done for reflected paths but is done in two stages, the incident ray is calculated and then reflected according the properties of either the absorber boundary or dummy port boundary, and then the reflected ray–path to the receiving port is calculated.

## 5.7 FACTORS LIMITING SCAN ANGLE

The following factors cause the limitation of the scan angle capability as one of the lens key performance factors [8]:

i) Array grating lobes limitation that is caused by insufficiently filled aperture.
ii) Array blindness limitation that is caused by mutual coupling.
iii) Pattern degradation limitation (reduced gain, increased beam–width, and sidelobe level) that is caused by reduced effective aperture size with scan, phase, and amplitude errors.
iv) Spillover that is caused by insufficient objective or feed aperture size.

## 5.8 ROTMAN LENS DESIGNER SOFTWARE

Remcom's Rotman Lens Designer (RLD) is a computer–aided software package for design, synthesis, and analysis of the Rotman Lenses and their variants. It is based on Geometrical Optics (GO) combined with the Rotman Lens design equations. It is designed for rapid development and analysis of Rotman Lenses given several physical and electrical input parameters. RLD generates the proper lens contour, transmission line geometry, absorptive port (dummy port) geometry, provides an approximate analysis of performance, and generates geometry files for import into Remcom's XFDTD software for further analysis or Gerber photo plotter file exportation. It is capable of providing the following results:

i) Calculation of the lens contour and transmission line geometry.

ii) Element locations along the lens contour.

iii) Insertion loss per beam port.

iv) Amplitude and phase distributions of the array ports versus frequency.

v) Phase errors.

The analysis is based on geometry and geometrical optics and therefore assumes:

i) Radiative leaks are not accounted for.

ii) Transmission line and material dispersion is negligible.

iii) Dummy load effectiveness is assumed ideal.

## 5.9 4X4 ROTMAN LENS SIMULATIONS

As Figure 5.6 indicates, a realistic Rotman Lens is designed and simulated. The design parameters are based on those used in previous sections. The *4x4* Rotman Lens is designed to have 4 beam ports, 4 array ports suitable for a 4–element antenna array, a scan angle of ±50° at a centre frequency of 3.15 GHz, and the element spacing of 47 mm. Also, the prototype is assumed to be FR4 with the dielectric constant ($\varepsilon_r$) of 4.7, substrate thickness (H) of 0.8 mm, and loss tangent of 0.01.

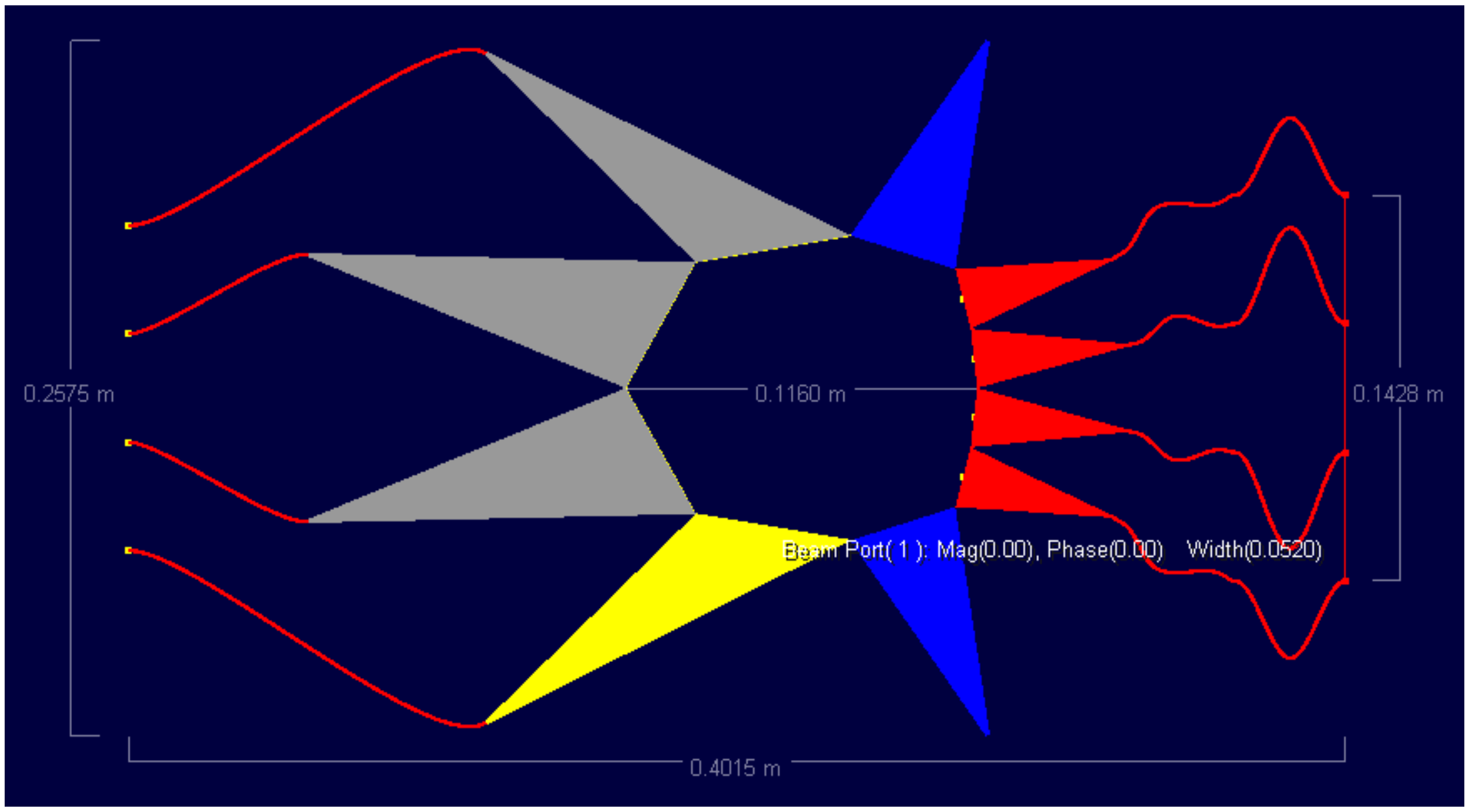

Figure 5.6 Simulated *4x4* Rotman Lens.

The above lens has a circular curvature on the beam port side and 2 dummy ports to absorb energy and reduce reflections. As Figure 5.7 shows, in order to obtain better performance, the beam and array ports are adjusted so that each line is pointing toward the centre of the lens on the opposite side rather than being normal to the lens surface. This is called *port pointing* that improves the response of the outer beams [22].

The lens has routed lines that connect the array contour (inner lens contour) to the linear array (outer lens contour) for each corresponding element. The geometry of the transmission line routing is adjusted in a way to ensure no overlapping, proper spacing between lines, proper curvature, and maintaining overall physical length requirement.

To obtain the desired performance, the lens requires to be tuned in terms of phase error or the array factor (beam pattern). The tuning involves adjusting the focal ratio ($g$) of the lens that will minimise the error reported by the phase error. This factor determines the curvature and focus of the lens, and if not adjusted accurately, will produce a messy beam. In this case, a final design with well–focused beams was reached with a focal ratio of 1.2970. This procedure can be seen in Figure 5.8.

Figures 5.9 and 5.10 indicate the beam to array coupling amplitude (the array ports distribution from a given beam or set of beams) and beam to array spillover coupling amplitude (the coupling of energy from the beam ports to array ports due to sidewall reflection) respectively. These figures show how the amplitude distribution along the array contour is much more uniform with beam port pointing enabled. Depending on the shape of the lens, this can be very useful to minimise spillover. Similar behavior can be seen for certain lens geometries using array port pointing.

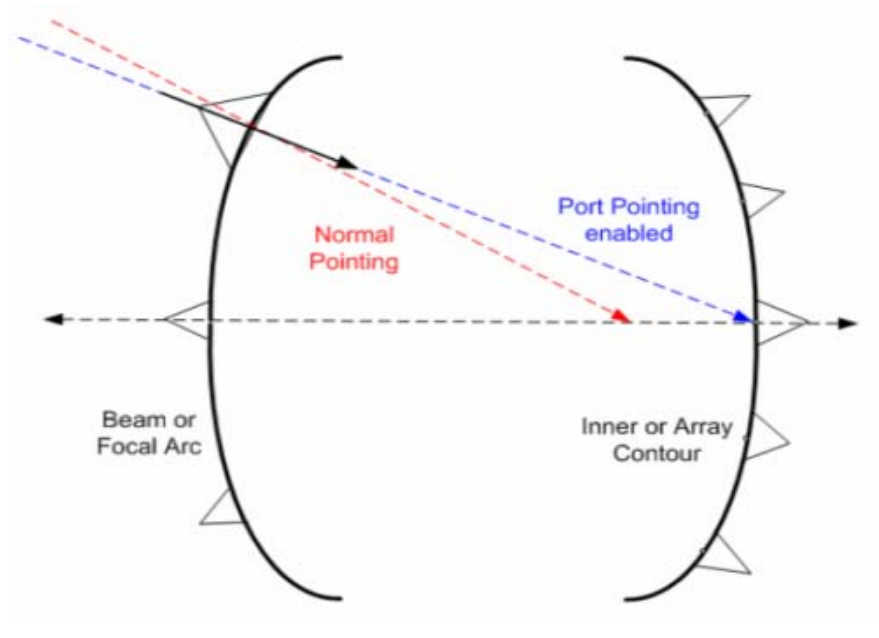

Figure 5.7 Port Pointing Concept.

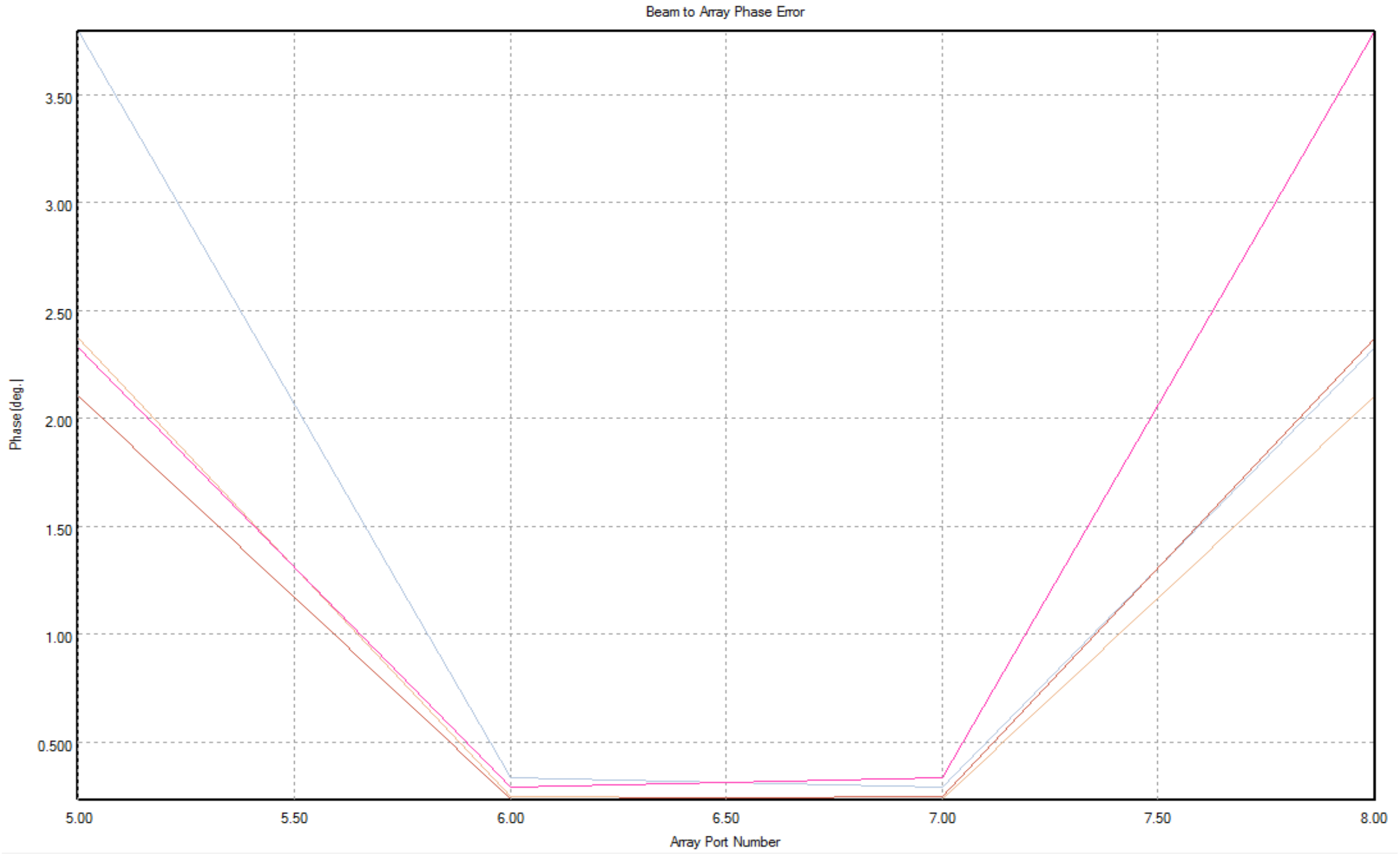

Figure 5.8 *4x4* RL Beam to Array Phase Error.

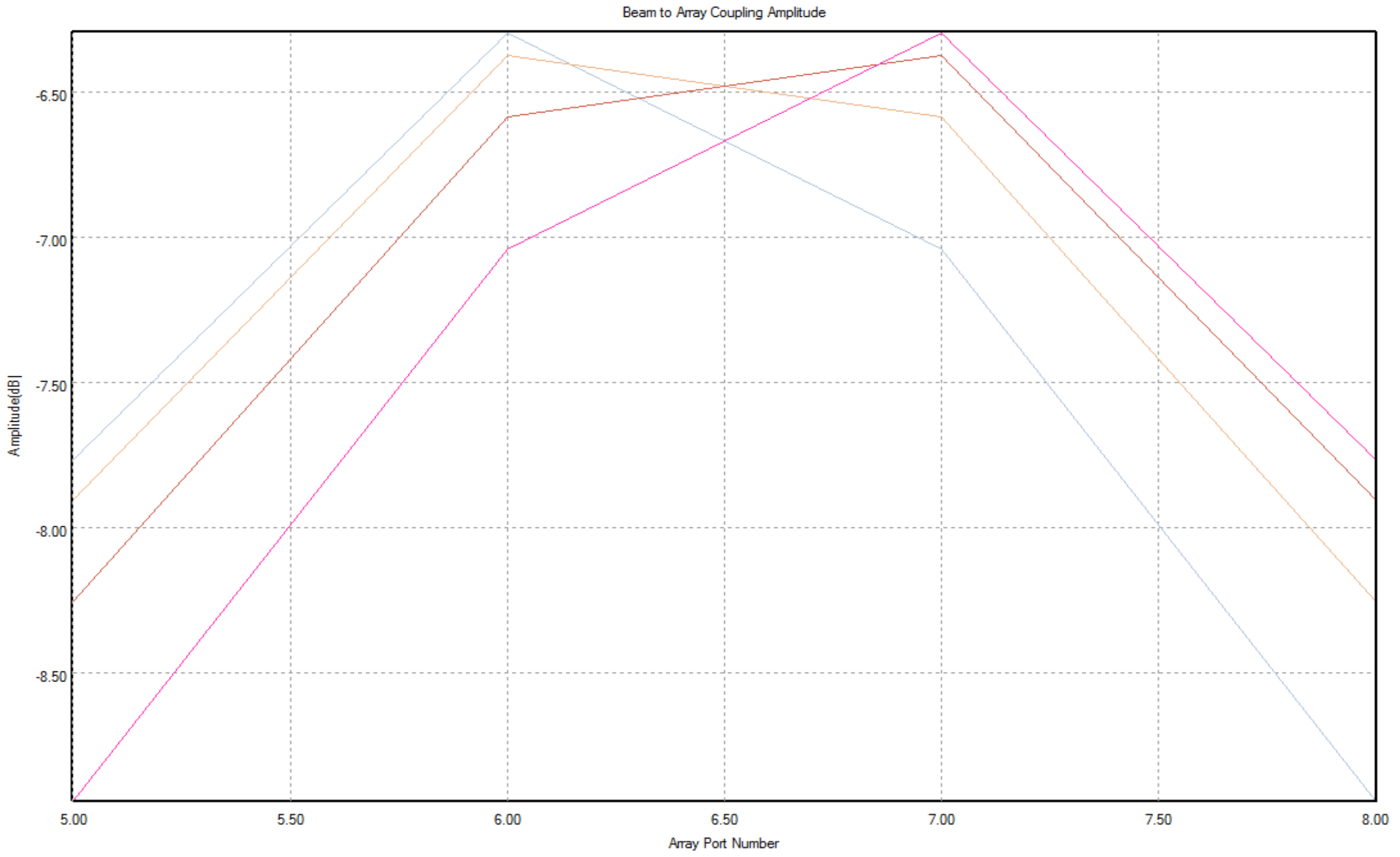

Figure 5.9 *4x4* RL Beam to Array Coupling Amplitude.

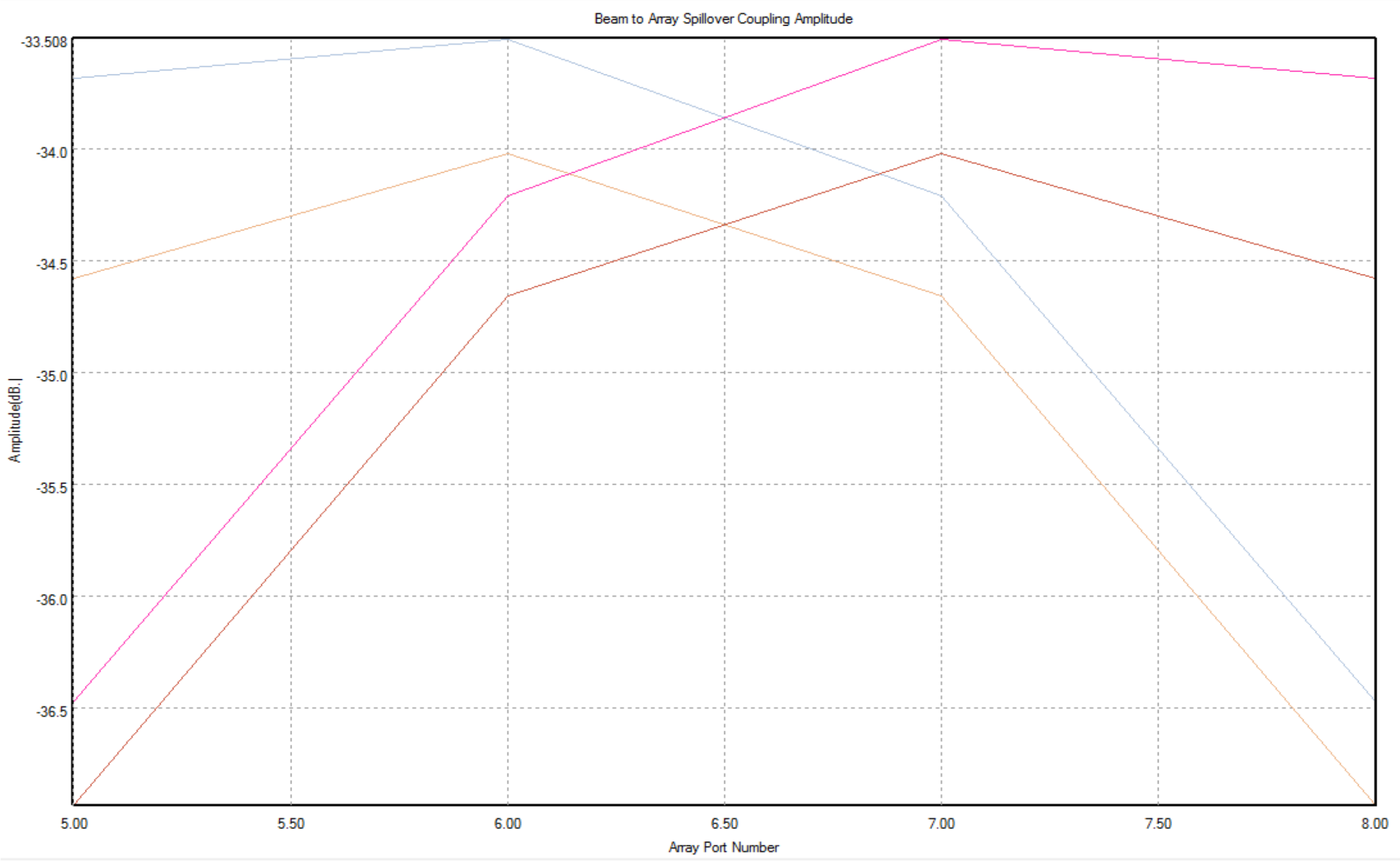

Figure 5.10 *4x4* RL Beam to Array Spillover Coupling Amplitude.

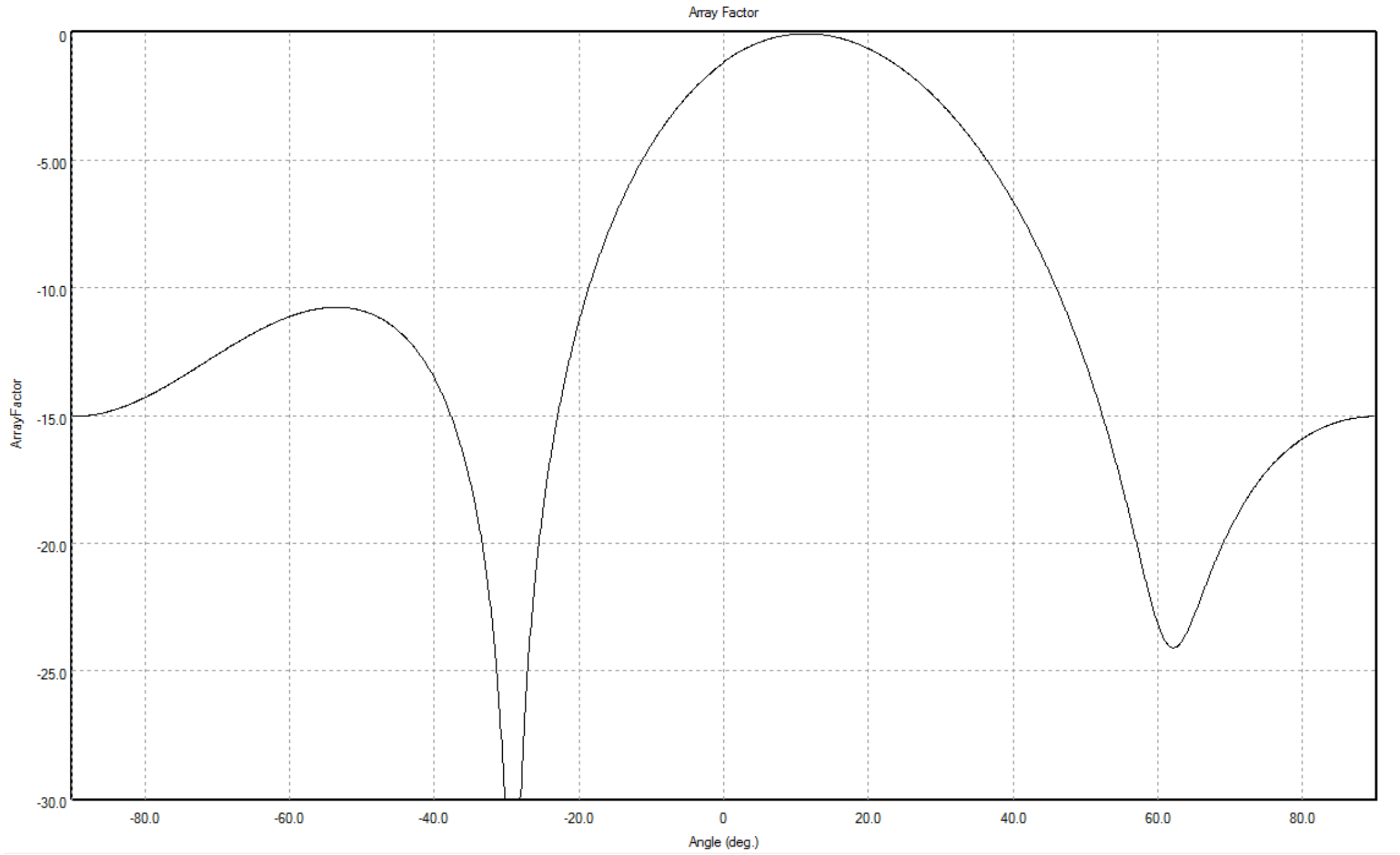

Figure 5.11 *4x4* RL Array Factor Radiation Pattern of Port 2.

## 5.10 8X8 ROTMAN LENS SIMULATIONS

The same design procedure as the *4x4* Rotman Lens is carried out for the design and simulation of the *8x8* Rotman Lens. In this case, the lens is designed to have 8 beam ports, 8 array ports suitable for an 8–element antenna array, a scan angle of ±50° at a centre frequency of 6.3 GHz ($\frac{1}{10}$ th of the automotive communication systems frequency, 63–64 GHz), and the element spacing of 28 mm.

The prototype is fabricated on Taconic TLC–30 substrate with the dielectric constant ($\varepsilon_r$) of 3.0, substrate thickness (H) of 1.3 mm, and loss tangent of 0.003. The lens has an elliptical curvature on the beam port side and dummy ports are replaced with the absorber sidewalls to introduce a novelty in the lens structure, reduce the size, and increase the performance of the lens. Also, the focal ratio ($g$) is adjusted to 1.2670 to minimise the phase error and produce well–focused beams. The lens simulation, the PCB layout design, and the beam to array phase error can be seen in Figures 5.12, 5.13, and 5.14 respectively.

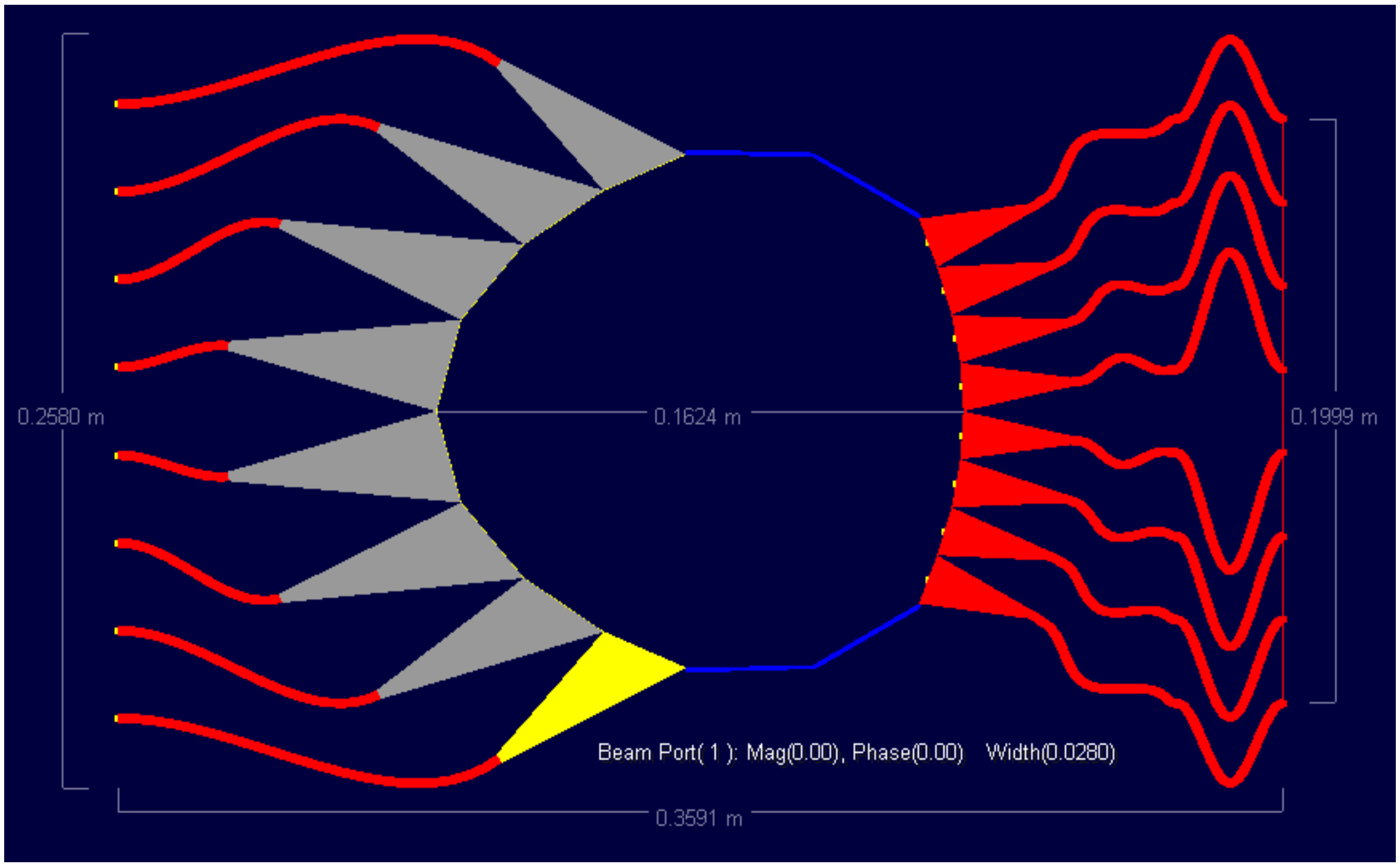

Figure 5.12 Simulated *8x8* Rotman Lens.

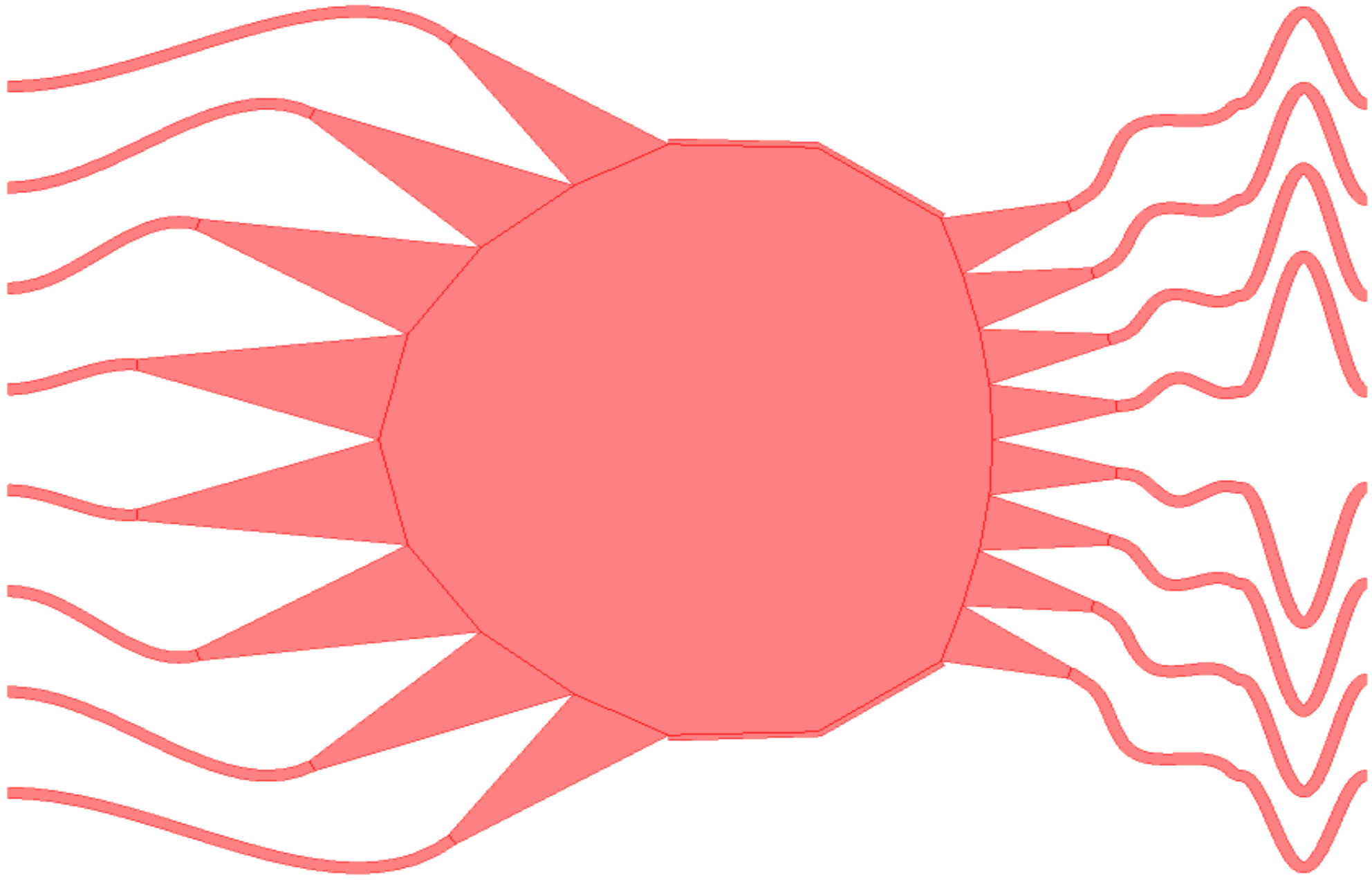

Figure 5.13 *8x8* Rotman Lens 2D Layout.

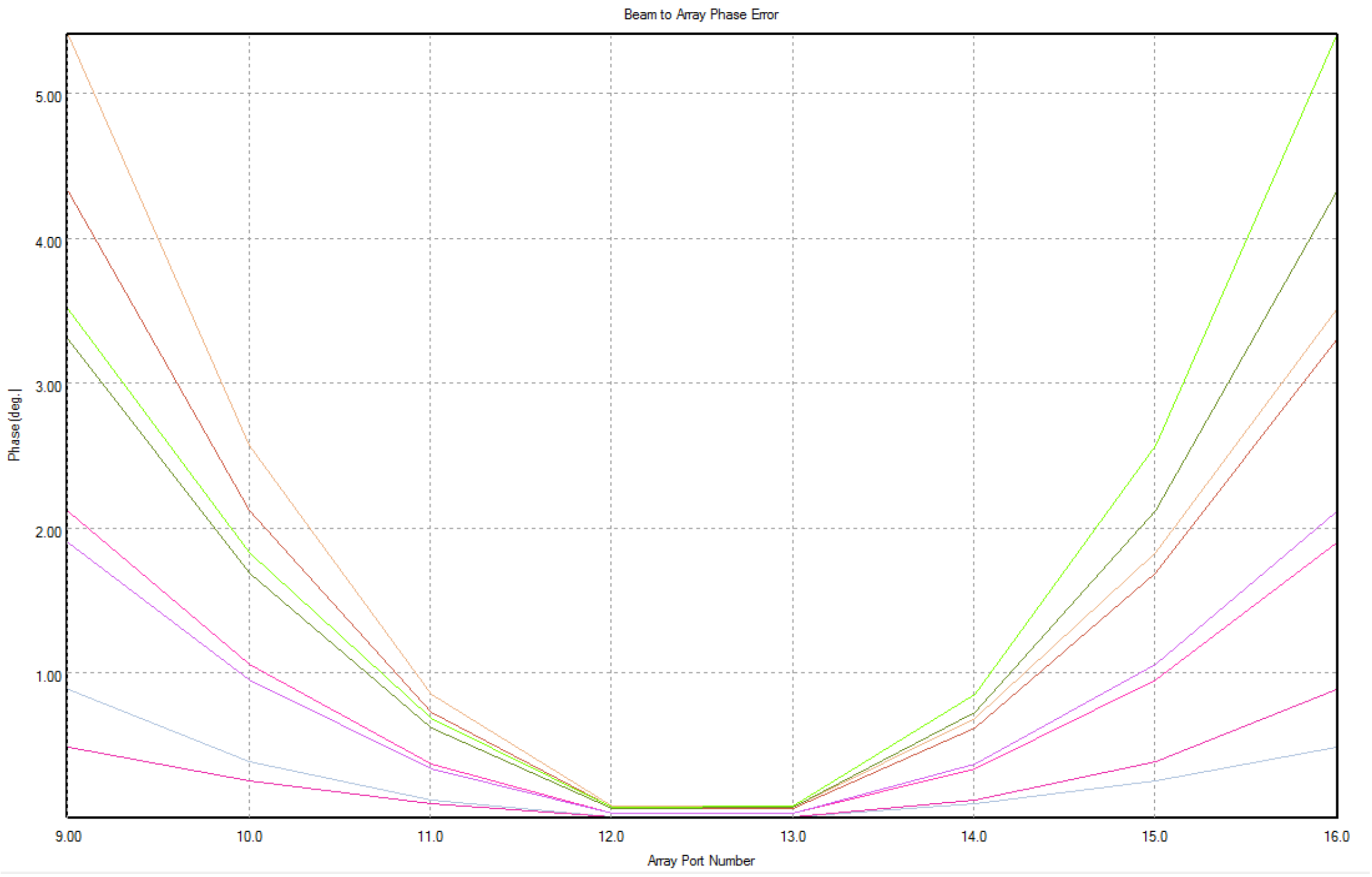

Figure 5.14 *8x8* RL Beam to Array Phase Error.

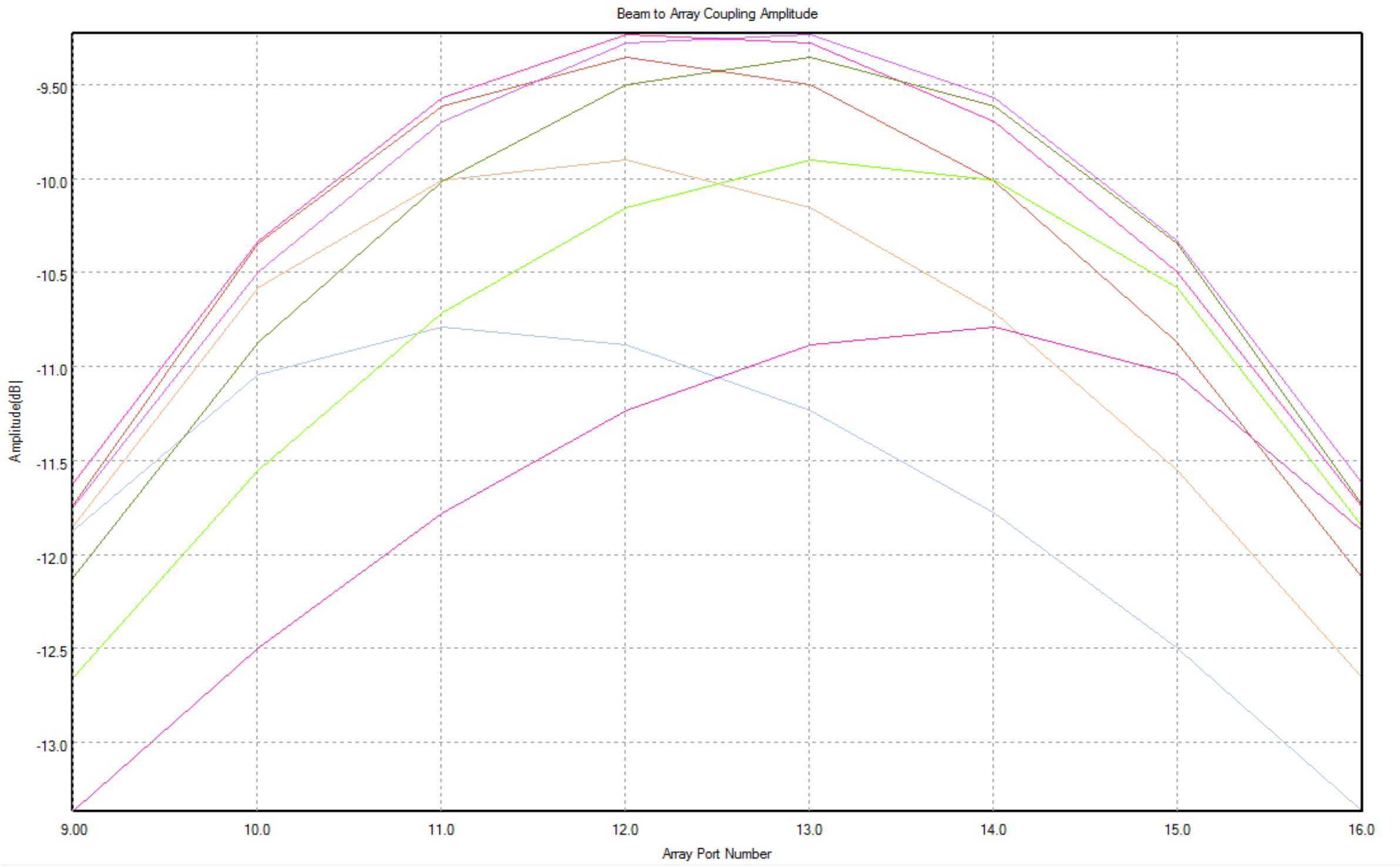

Figure 5.15 *8x8* RL Beam to Array Coupling Amplitude.

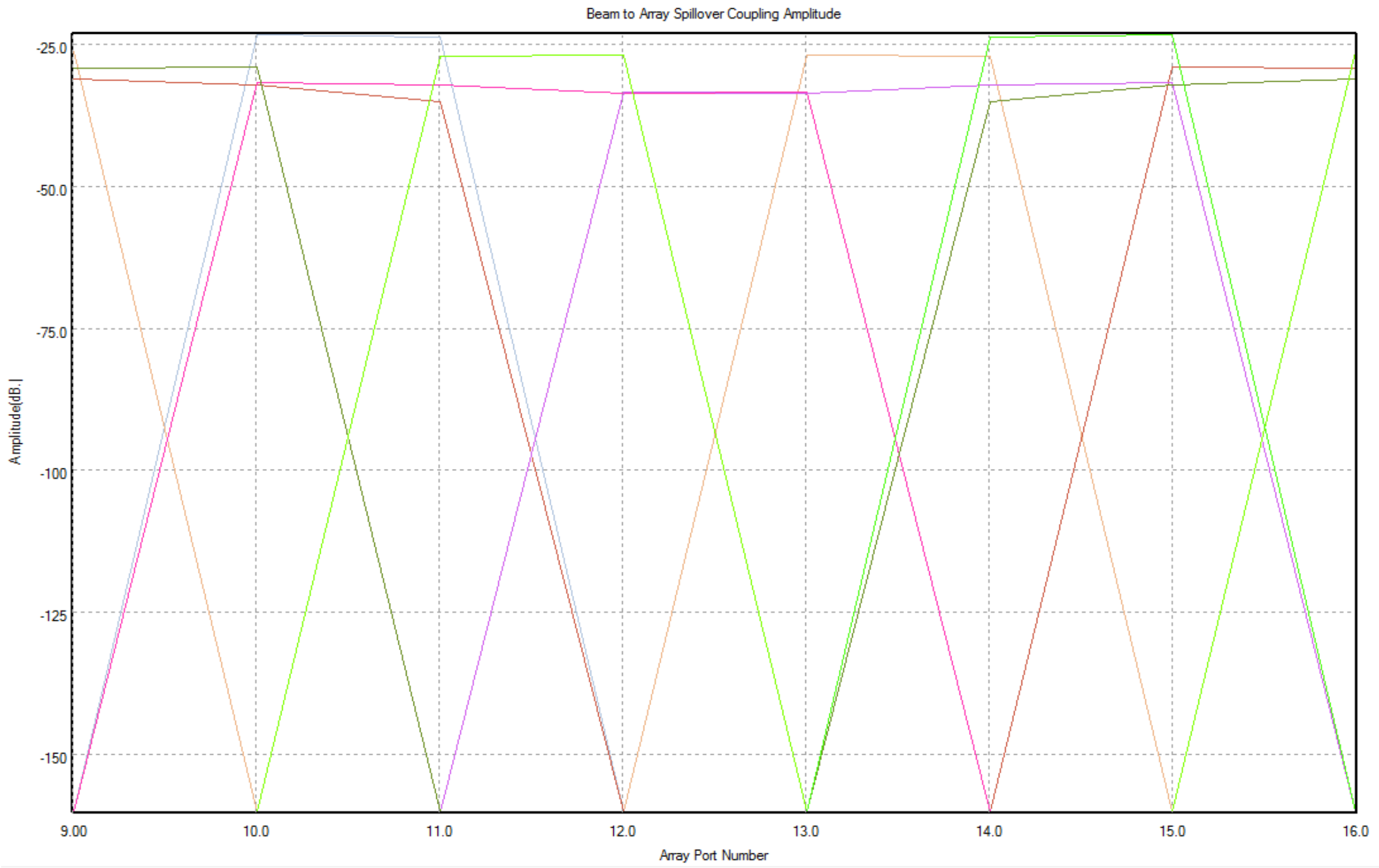

Figure 5.16 *8x8* RL Beam to Array Spillover Coupling Amplitude.

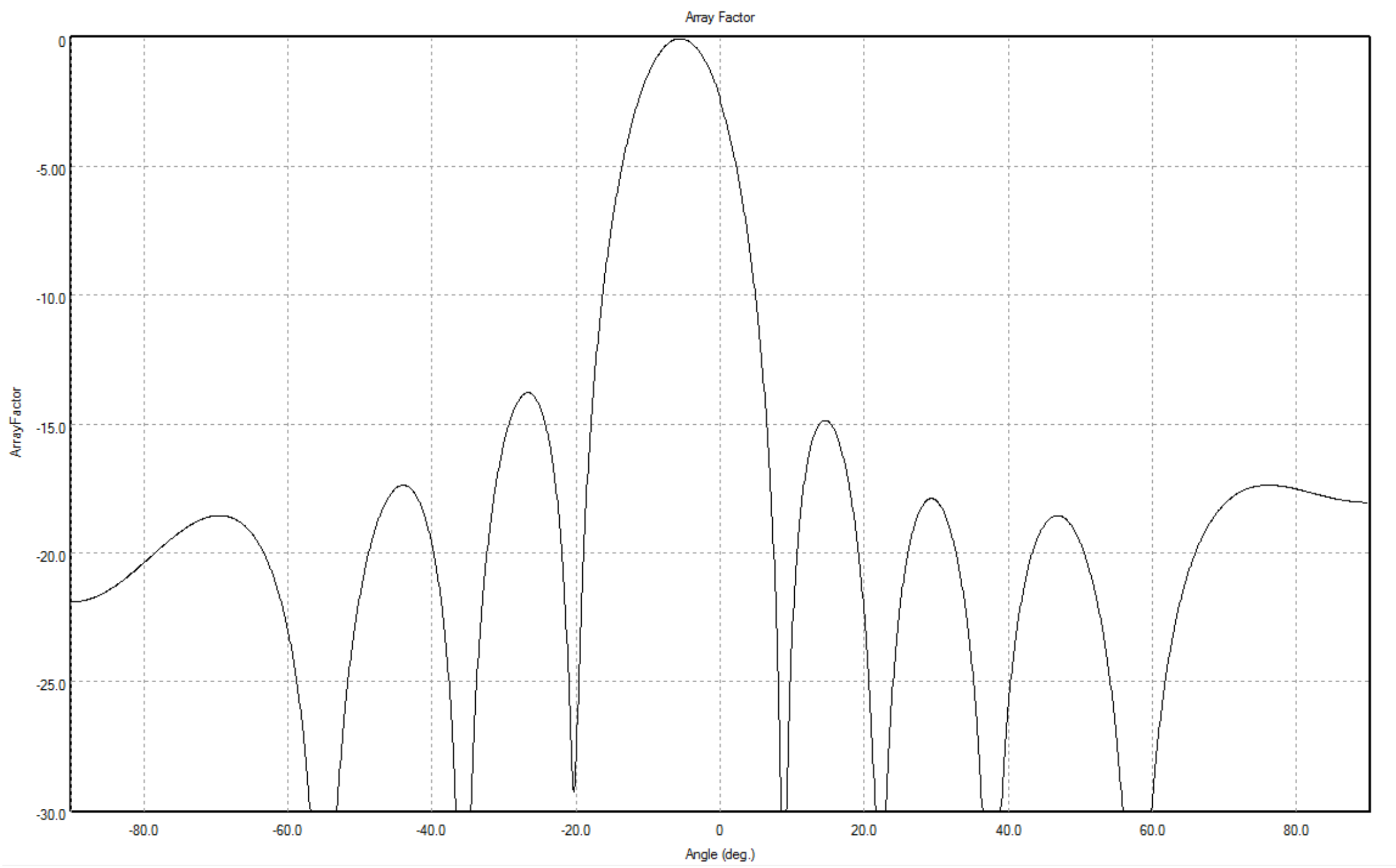

Figure 5.17 *8x8* RL Array Factor Radiation Pattern of Port 5.

## 5.11 SUMMARY

This chapter covered the definition, lens geometry, design equations, phase error, and coupling of the Rotman Lens. The simulations regarding design of the *4x4* and *8x8* Rotman Lenses were carried out and their output characteristics including beam to array phase errors, beam to array coupling amplitudes, beam to array spillover amplitudes, and array factor radiation patterns were shown. Apparently the simulated results for the *4x4* Rotman Lens agreed with the simulated results of the *4x4* Butler Matrix in terms of coupling amplitude and array factor radiation pattern. The simulated results for the novel *8x8* Rotman Lens indicated the expected outcomes and its main lobe of the array factor radiation pattern is more than 10dB greater than the side lobes. Also, the beam to array phase error for both of the lenses were minimised and distributed over the output ports (Figures 5.8 and 5.14).

# CHAPTER 6
# THE ROTMAN LENS MEASUREMENTS

## 6.1 INTRODUCTION

This chapter will cover the sections regarding circuit implementation of the *8x8* Rotman Lens and its experimental results. The measured array factor radiation patterns corresponding each port and relative phase differences of the output ports will be shown.

## 6.2 RL EXPERIMENTAL RESULTS

Figure 6.1 indicates the fabricated *8x8* Rotman Lens board being tested using the network analyser. Only one input port and one output port ($S_{21}$) can be measured at a time and other 14 ports are terminated using 50Ω terminations. In the following sections, the measured array factor radiation patterns for each port and relative phase differences between each port and output ports are presented and discussed.

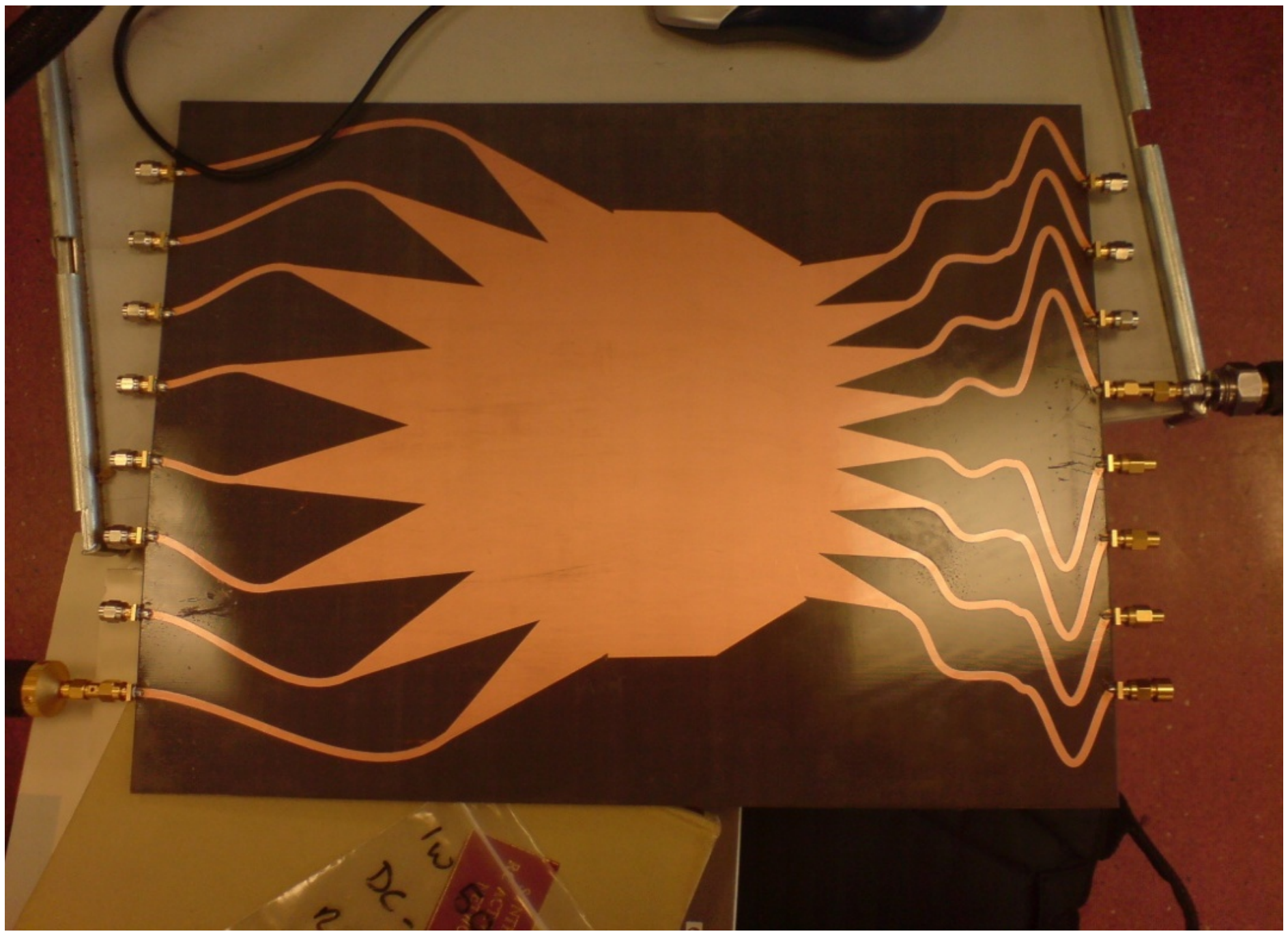

Figure 6.1 *8x8* Rotman Lens Circuit Configuration.

### 6.2.1 RL ARRAY FACTOR RADIATION PATTERNS

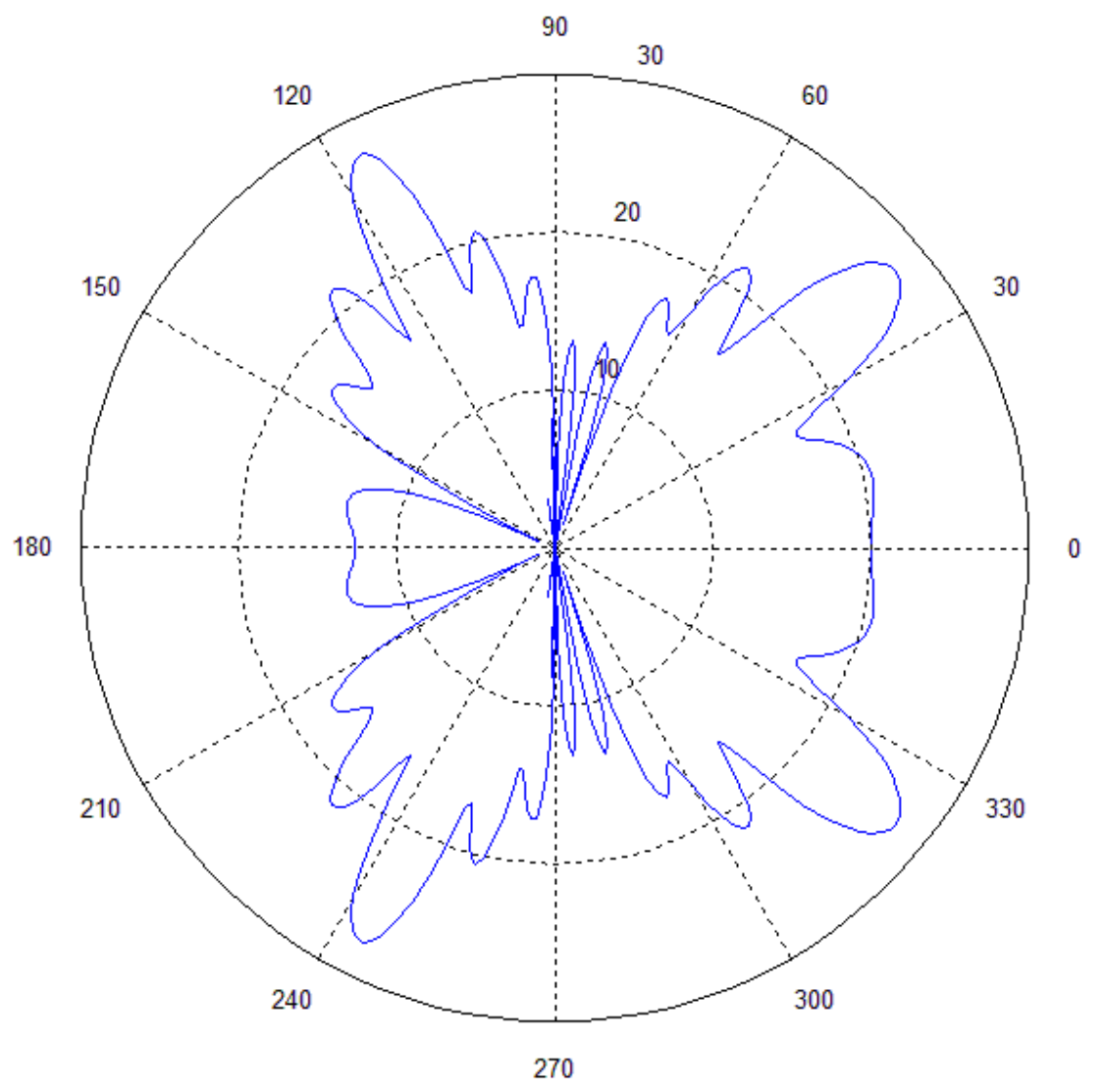

Figure 6.2 RL Measured Array Factor Radiation Pattern of Port 1.

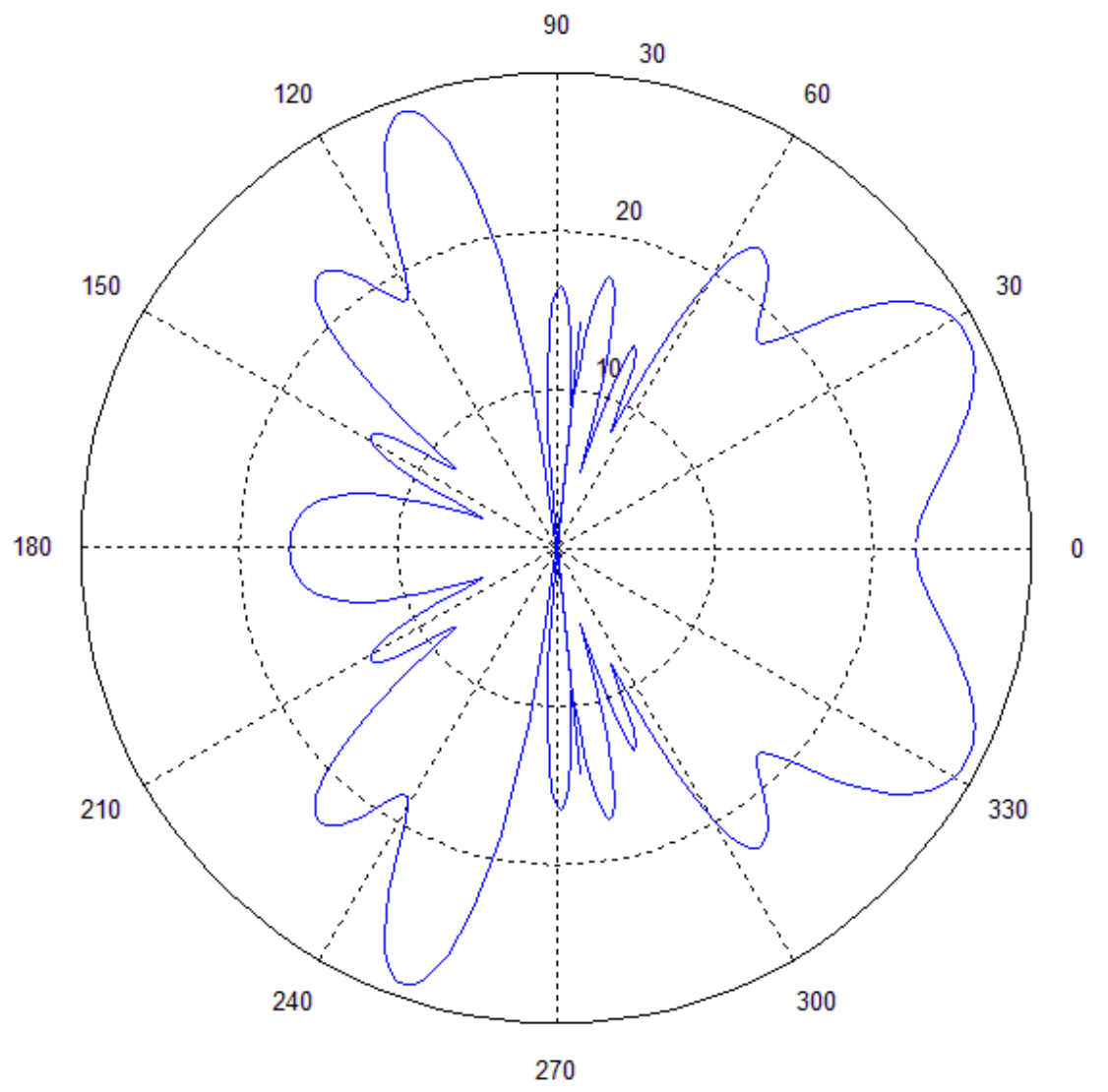

Figure 6.3 RL Measured Array Factor Radiation Pattern of Port 2.

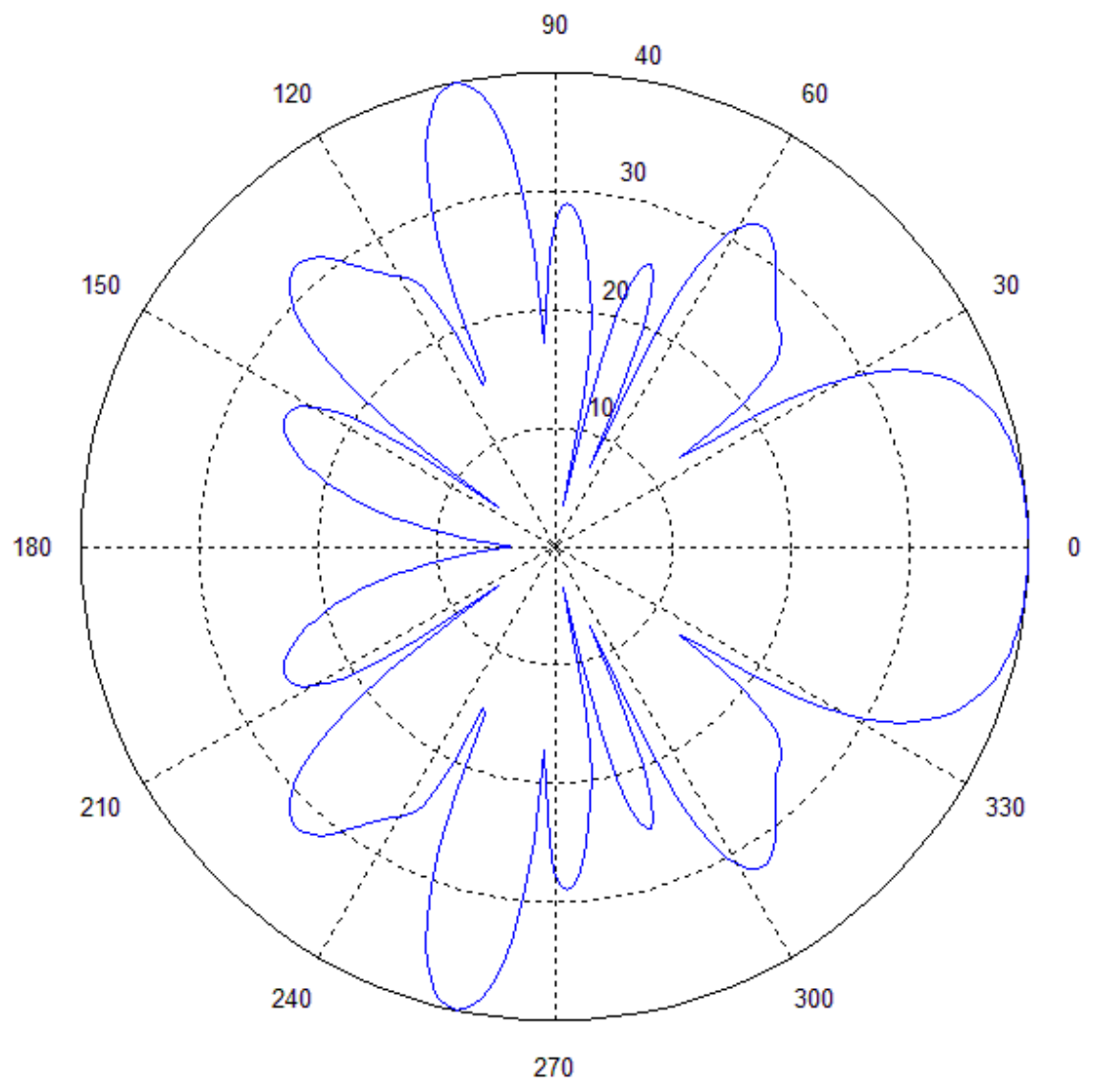

Figure 6.4 RL Measured Array Factor Radiation Pattern of Port 3.

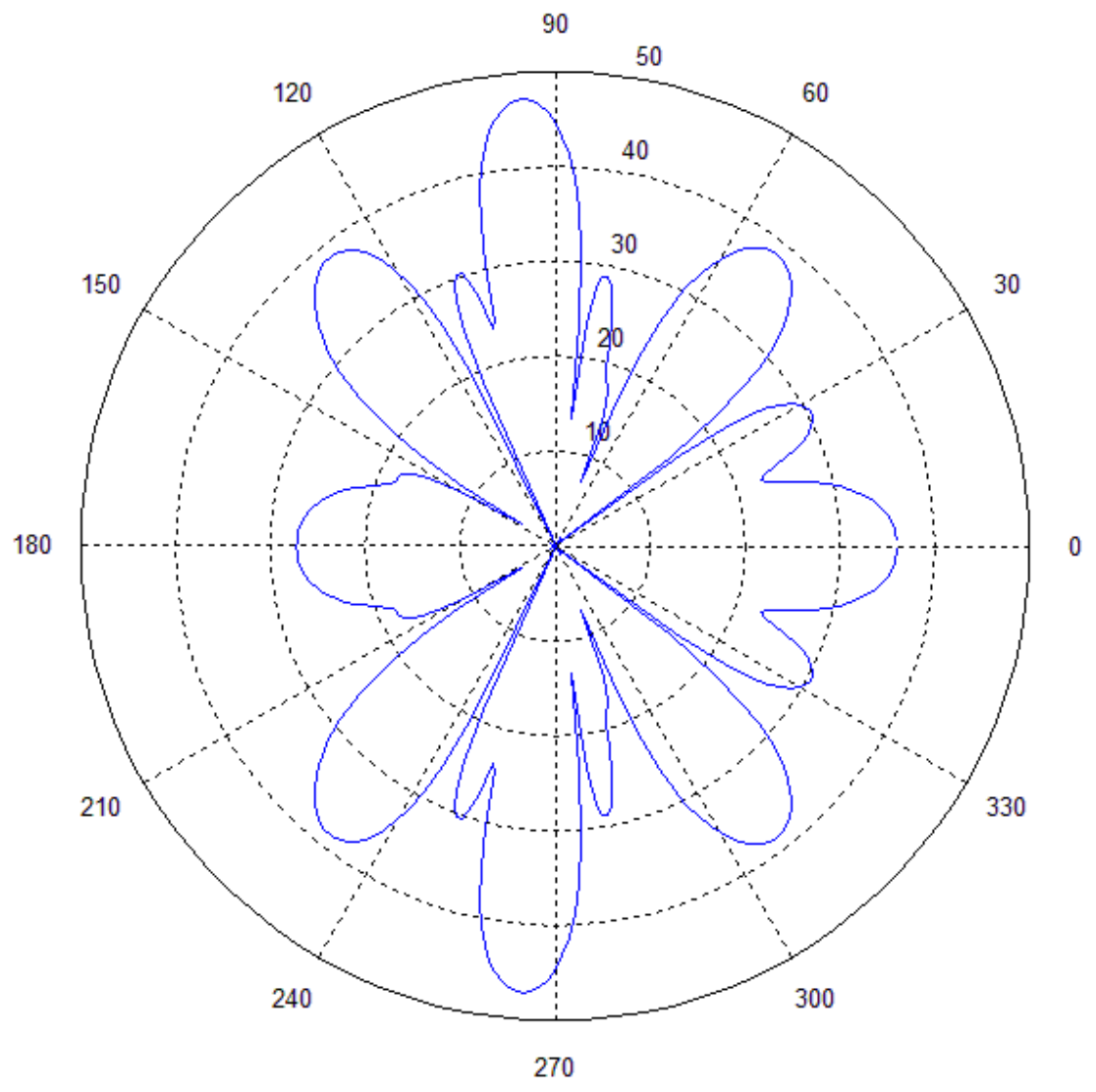

Figure 6.5 RL Measured Array Factor Radiation Pattern of Port 4.

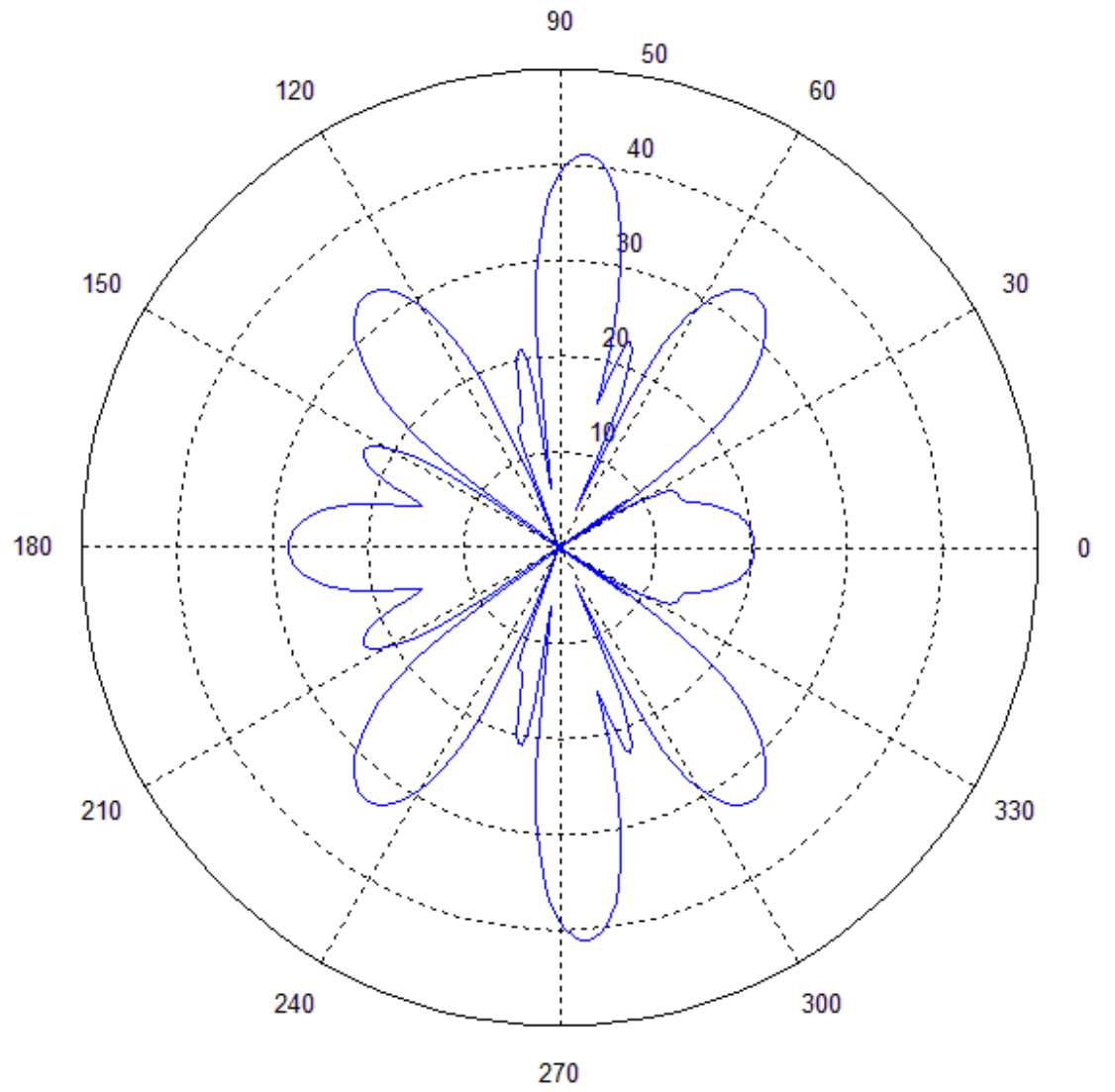

Figure 6.6 RL Measured Array Factor Radiation Pattern of Port 5.

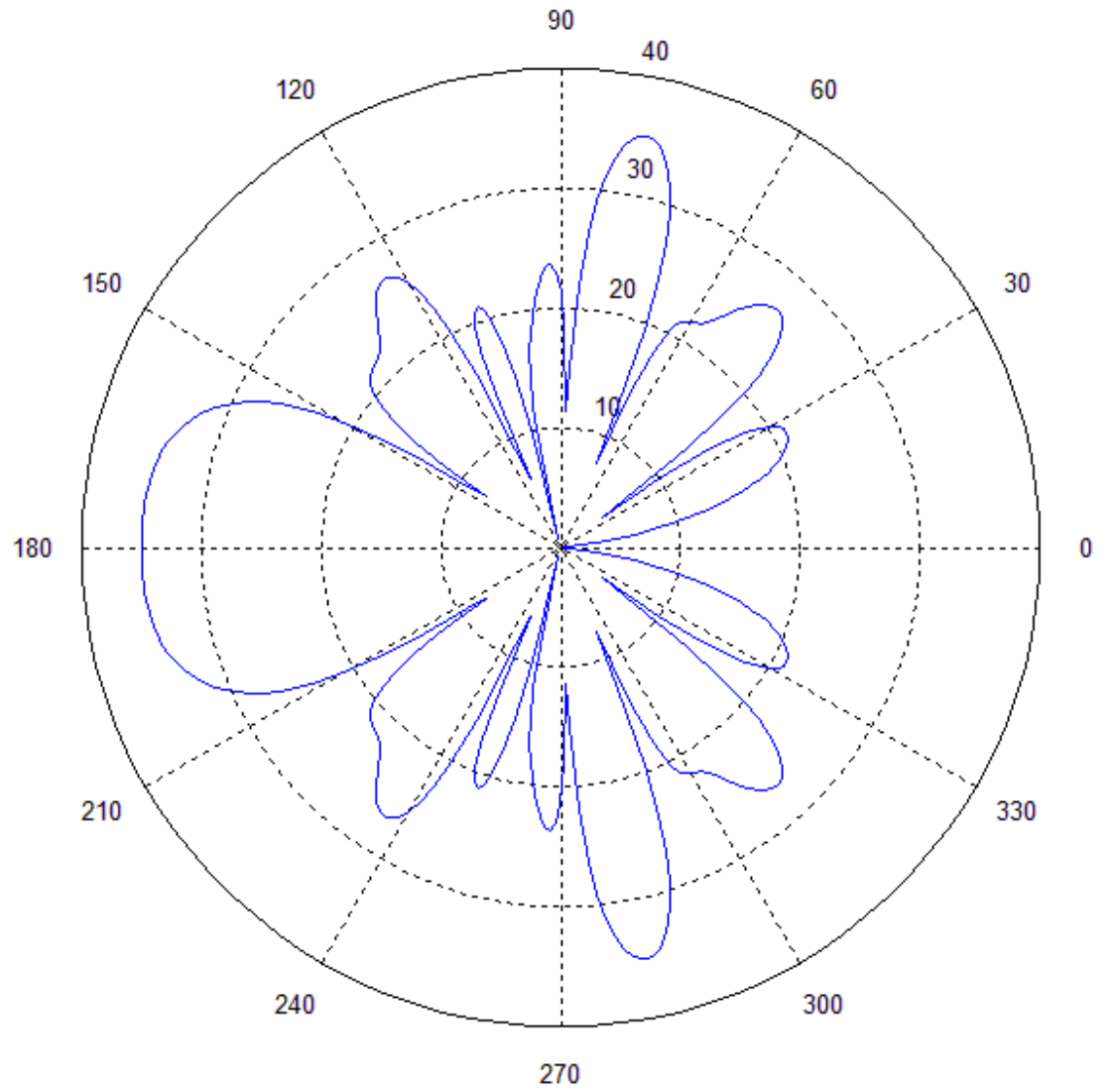

Figure 6.7 RL Measured Array Factor Radiation Pattern of Port 6.

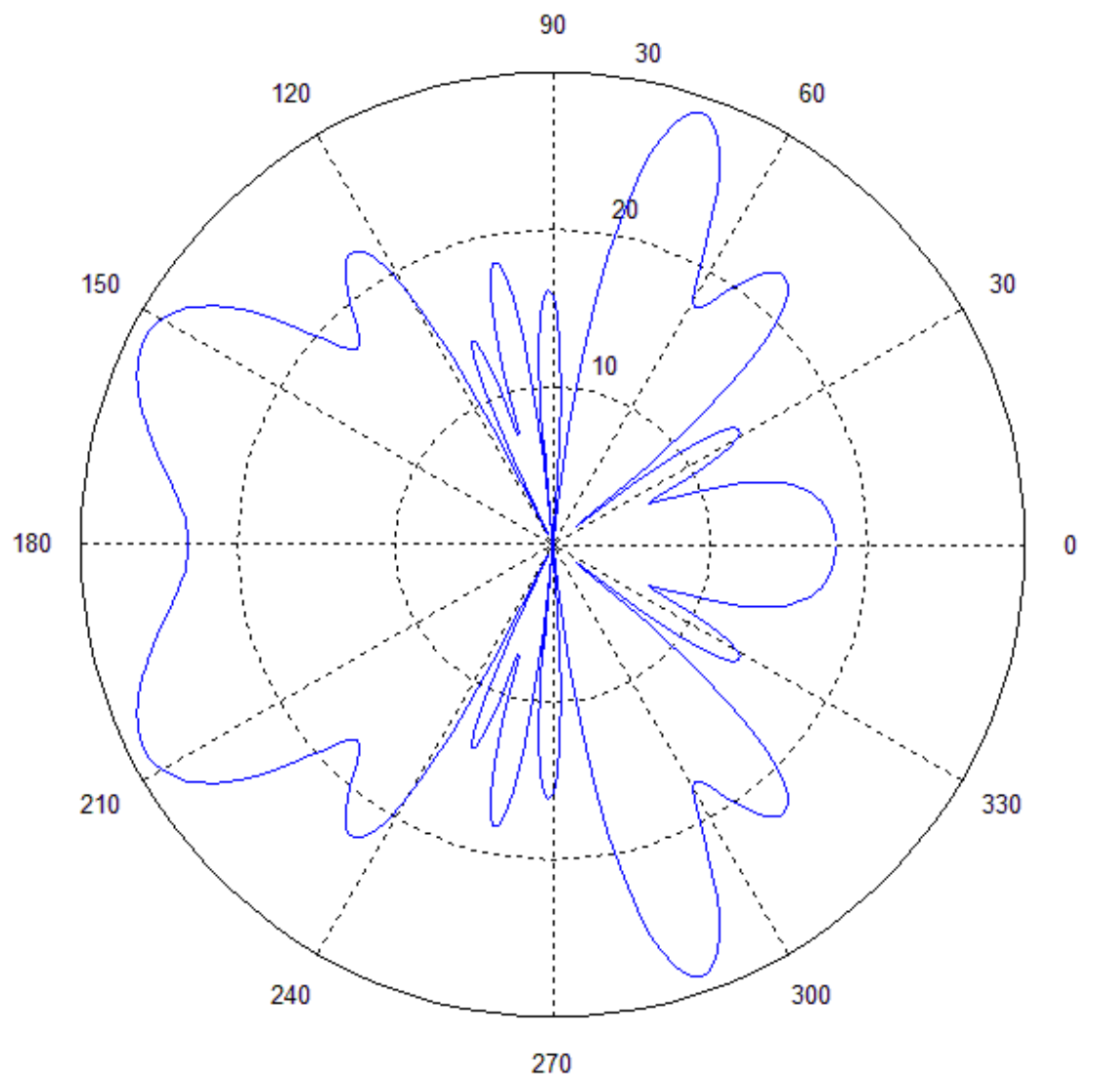

Figure 6.8 RL Measured Array Factor Radiation Pattern of Port 7.

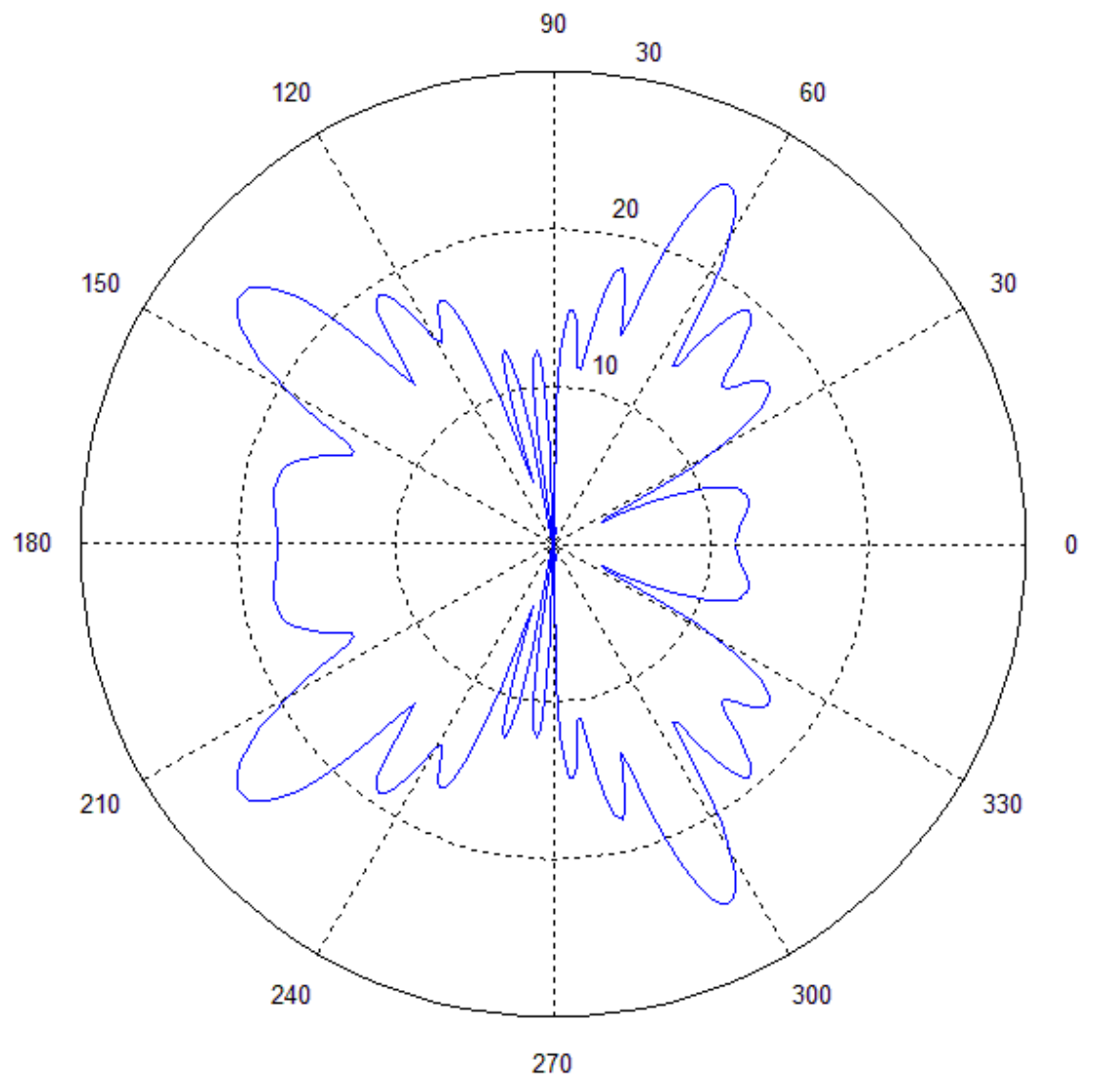

Figure 6.9 RL Measured Array Factor Radiation Pattern of Port 8.

As Figures 6.2 to 6.9 indicate, the array factor radiation patterns for ports 1 to 8 have their beams formed in the expected directions having their main lobes showed strong identity with at least 10dB of isolation from the side lobes.

Also, it is shown that despite the non–ideal performance of the *8x8* Rotman Lens, in terms of phase and amplitude distributions, it is still capable of forming well–defined beams suitable for the beam steering experiments by causing the main lobe to be directed in certain directions for ports 1 to 8.

The difference in beam shape between the measured radiation patterns and simulated array factor is mainly in the nulls between the beams, which are not deep enough as the measured results because of small phase and amplitude deviations, the cross–coupling effects that are not taken into account in the simulated array factor radiation pattern, fabrication process errors, non–uniformity of transmission line width, and measurement process errors.

### 6.2.2 RL OUTPUT PORTS RELATIVE PHASE DIFFERENCES

Figures 6.10 to 6.17 show the relative phase differences between each port and the output ports. In ideal case, the output ports should have a constant phase progression among themselves, hence their relative phase differences would be constant and the figures should indicate a uniform distribution over the output ports.

However, because of the amplitude and phase variations caused by the errors mentioned in previous section, the relative phase differences are non–uniform and sharply fluctuated. But as Figures 6.10, 6.16, and 6.17 indicate, the ports 1, 7, and 8 tend to follow the ideal case behaviour.

By eliminating the mentioned errors and improving the system performance in term of achieving high gain narrow–beams with desired directions in array factor radiation patterns, the relative phase differences will have a uniform distribution. Hence, the system can generate different beams in right directions having their side lobes minimised.

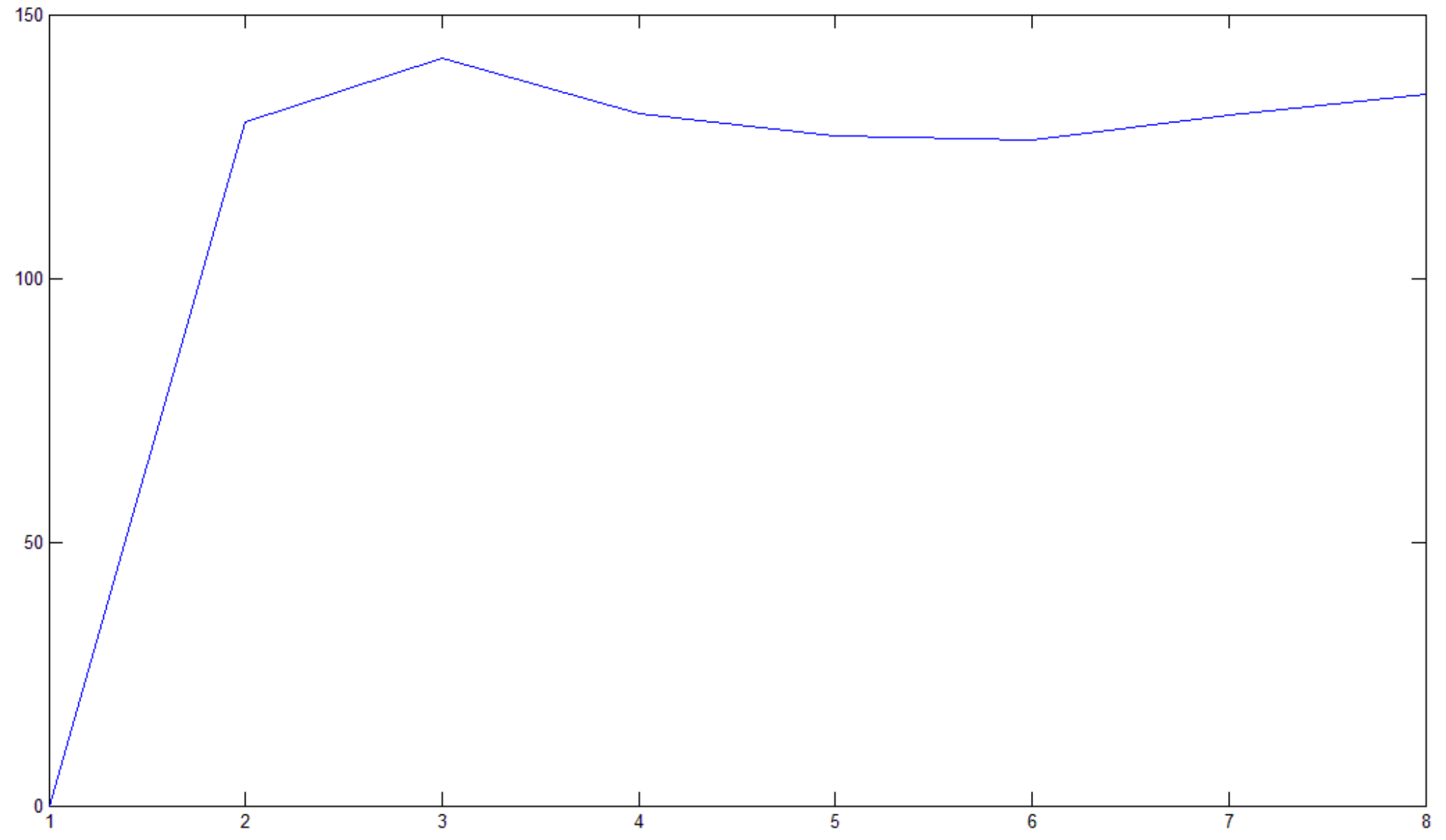

Figure 6.10 RL Measured Relative Phase Difference Between Port 1 and Output Ports.

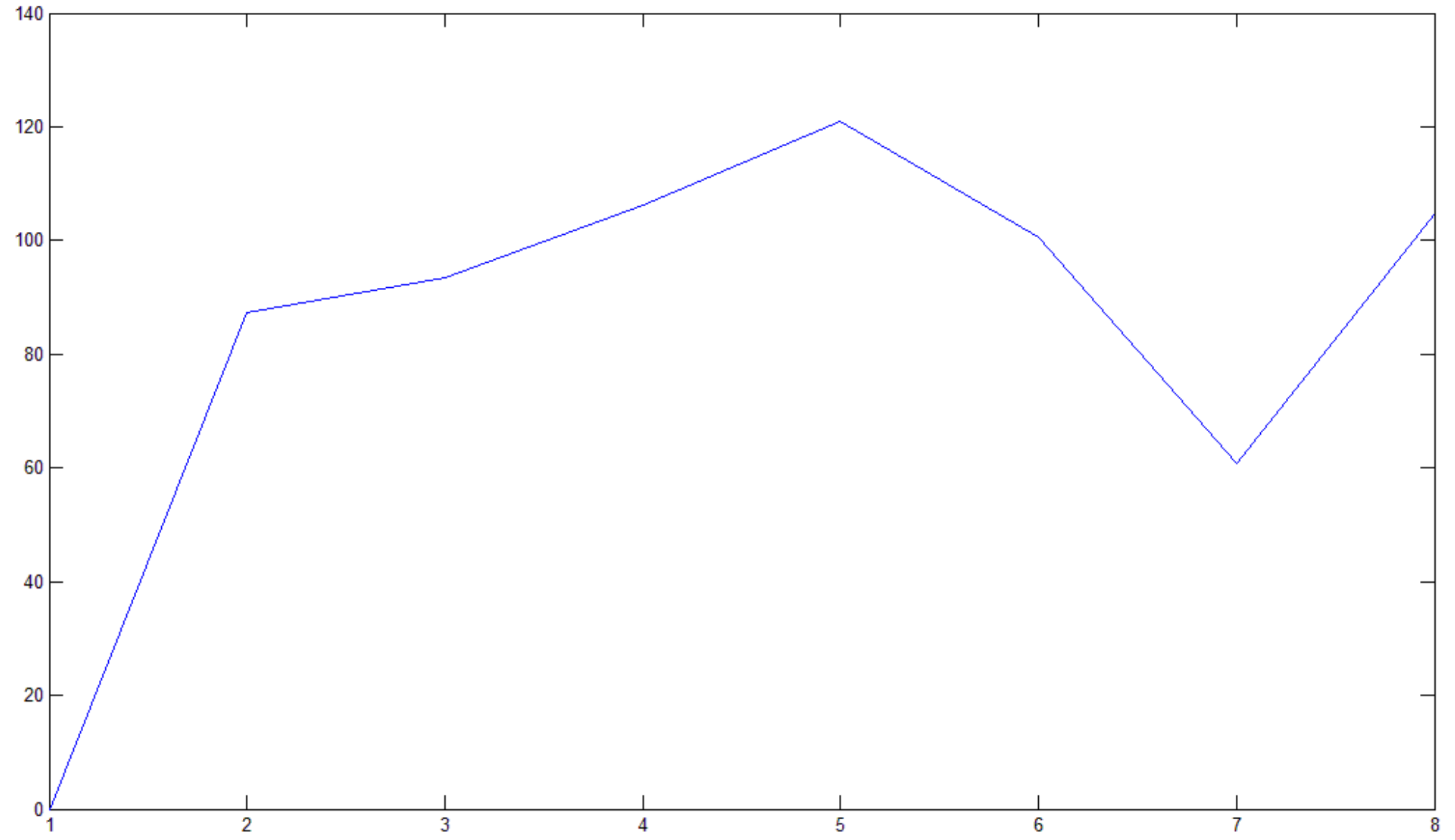

Figure 6.11 RL Measured Relative Phase Difference Between Port 2 and Output Ports.

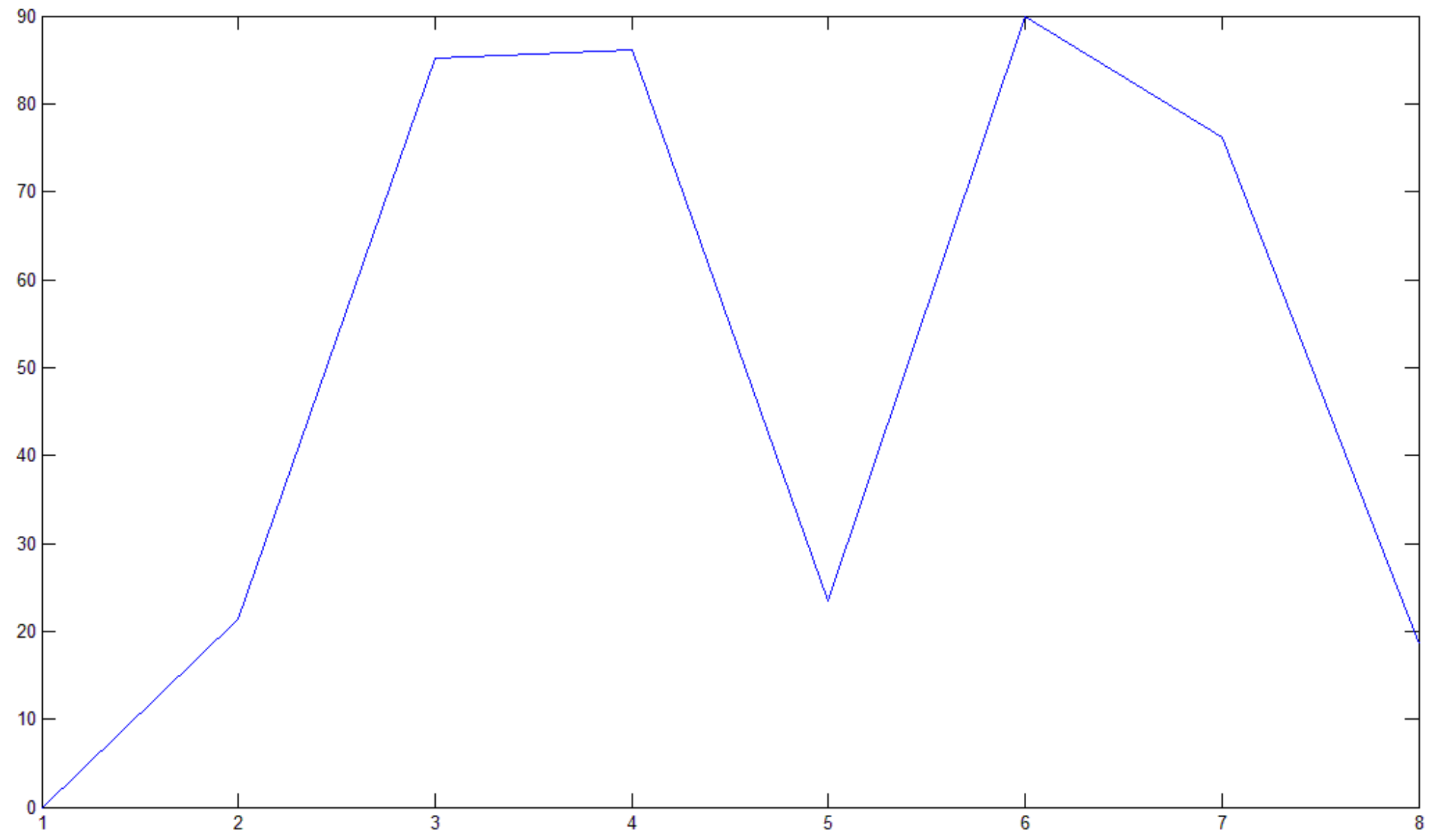

Figure 6.12 RL Measured Relative Phase Difference Between Port 3 and Output Ports.

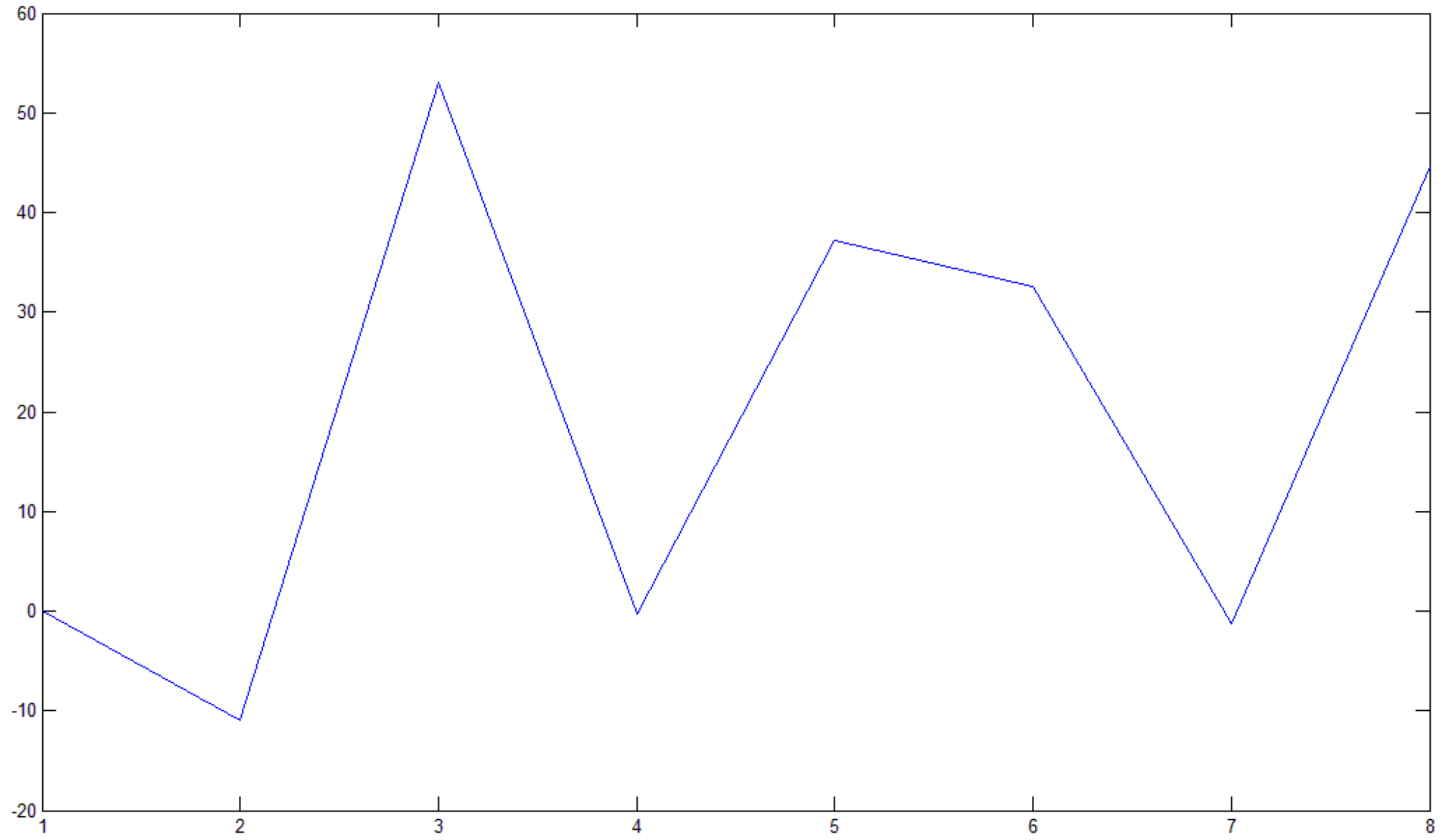

Figure 6.13 RL Measured Relative Phase Difference Between Port 4 and Output Ports.

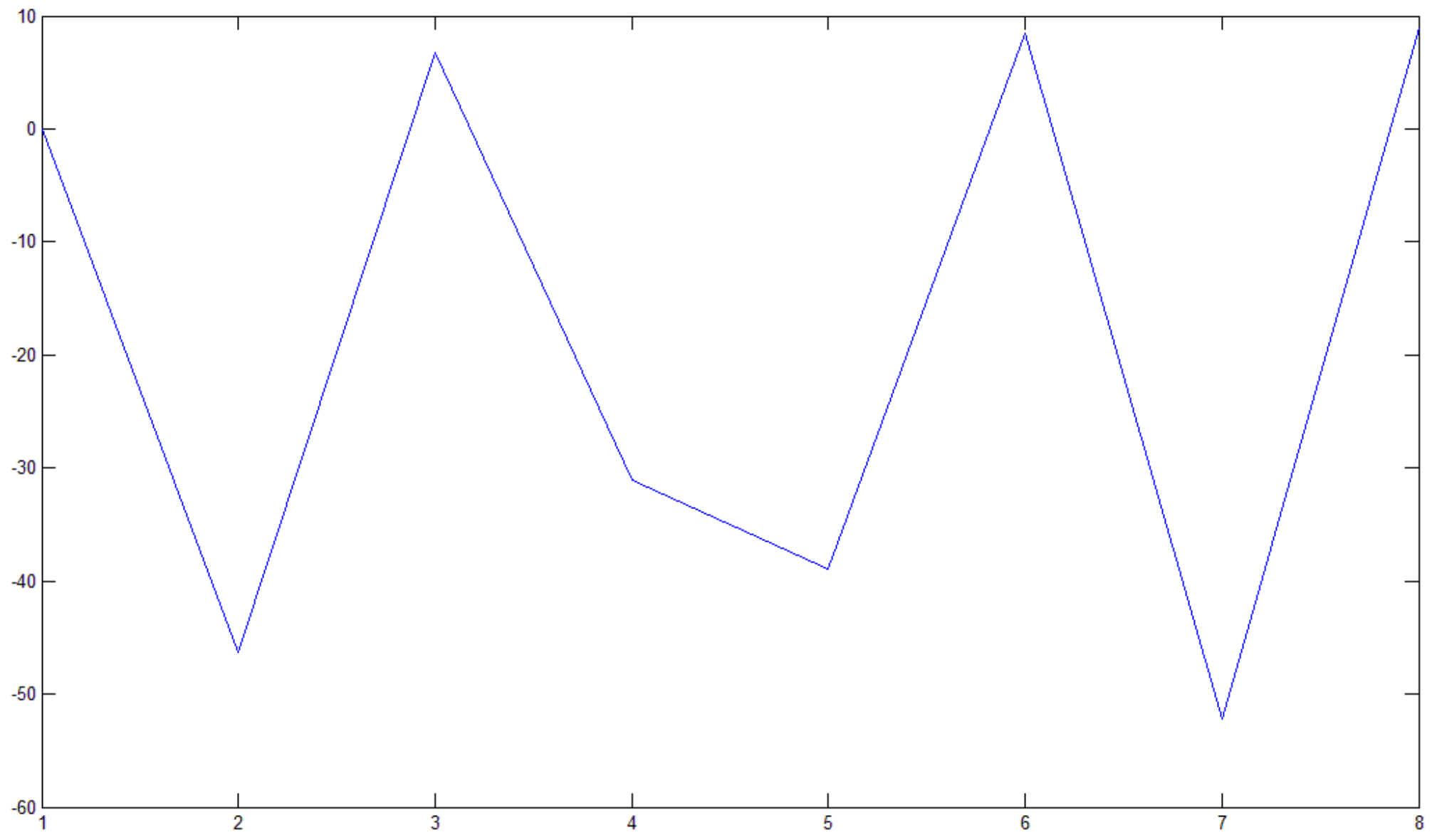

Figure 6.14 RL Measured Relative Phase Difference Between Port 5 and Output Ports.

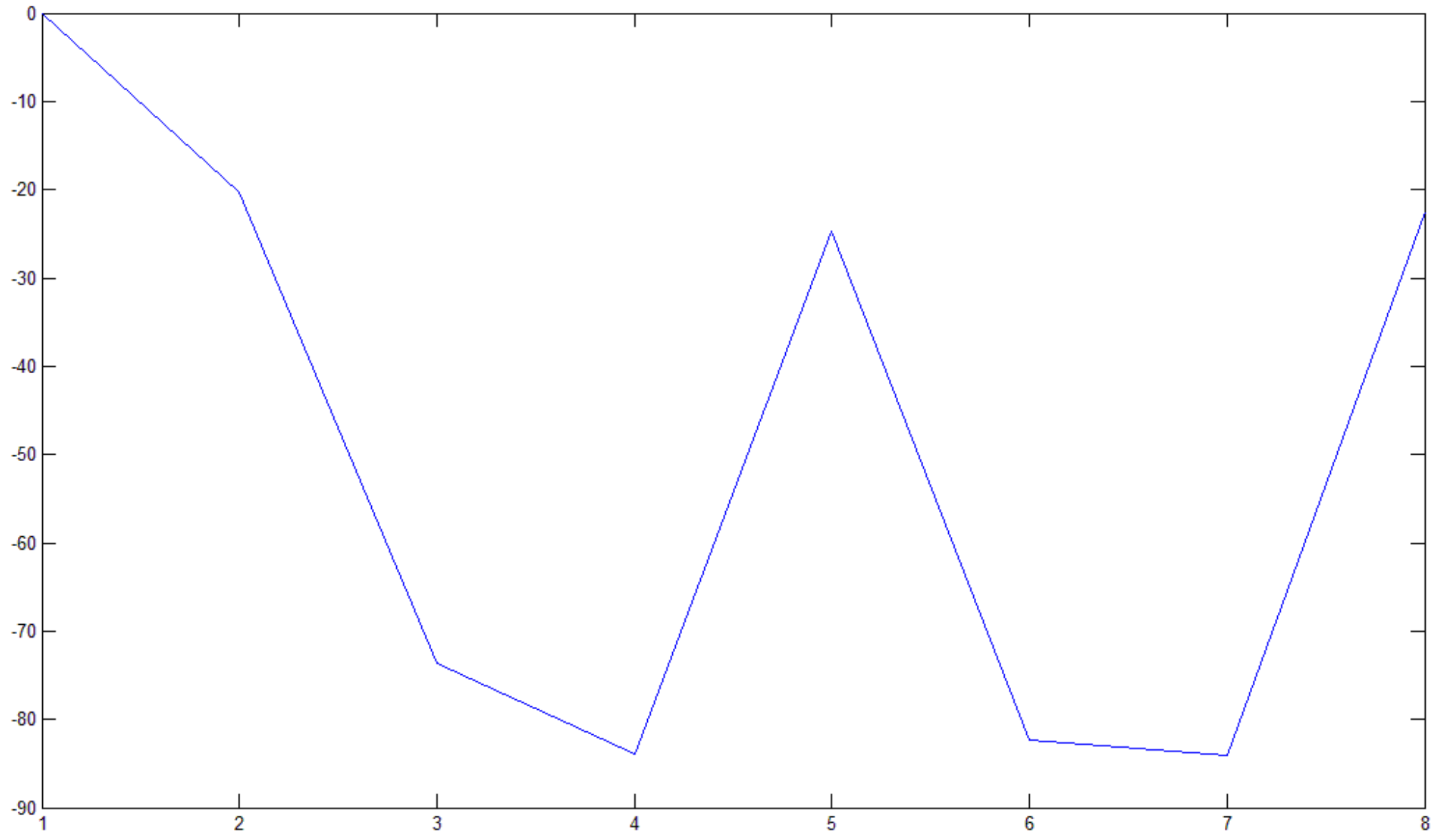

Figure 6.15 RL Measured Relative Phase Difference Between Port 6 and Output Ports.

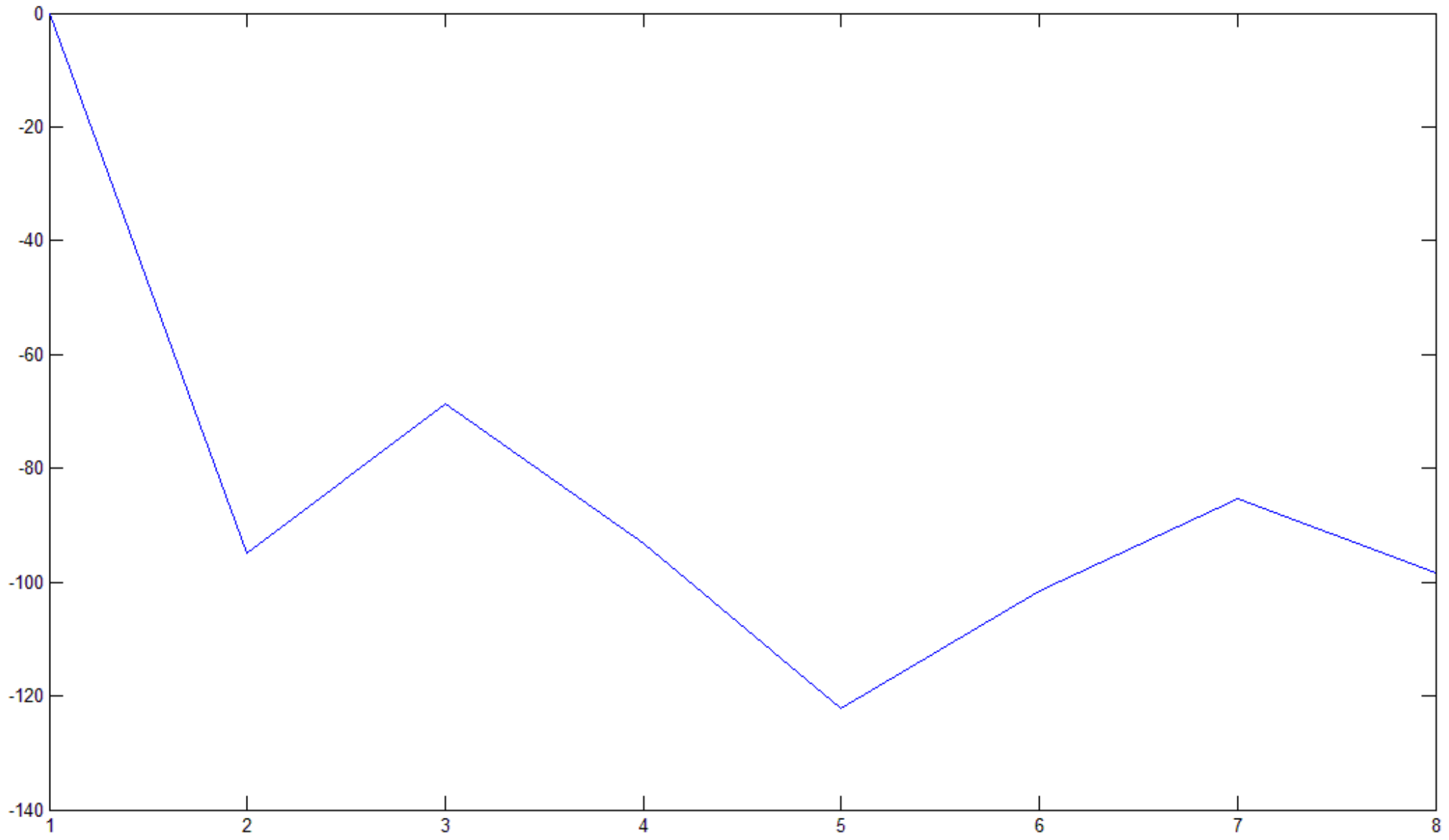

Figure 6.16 RL Measured Relative Phase Difference Between Port 7 and Output Ports.

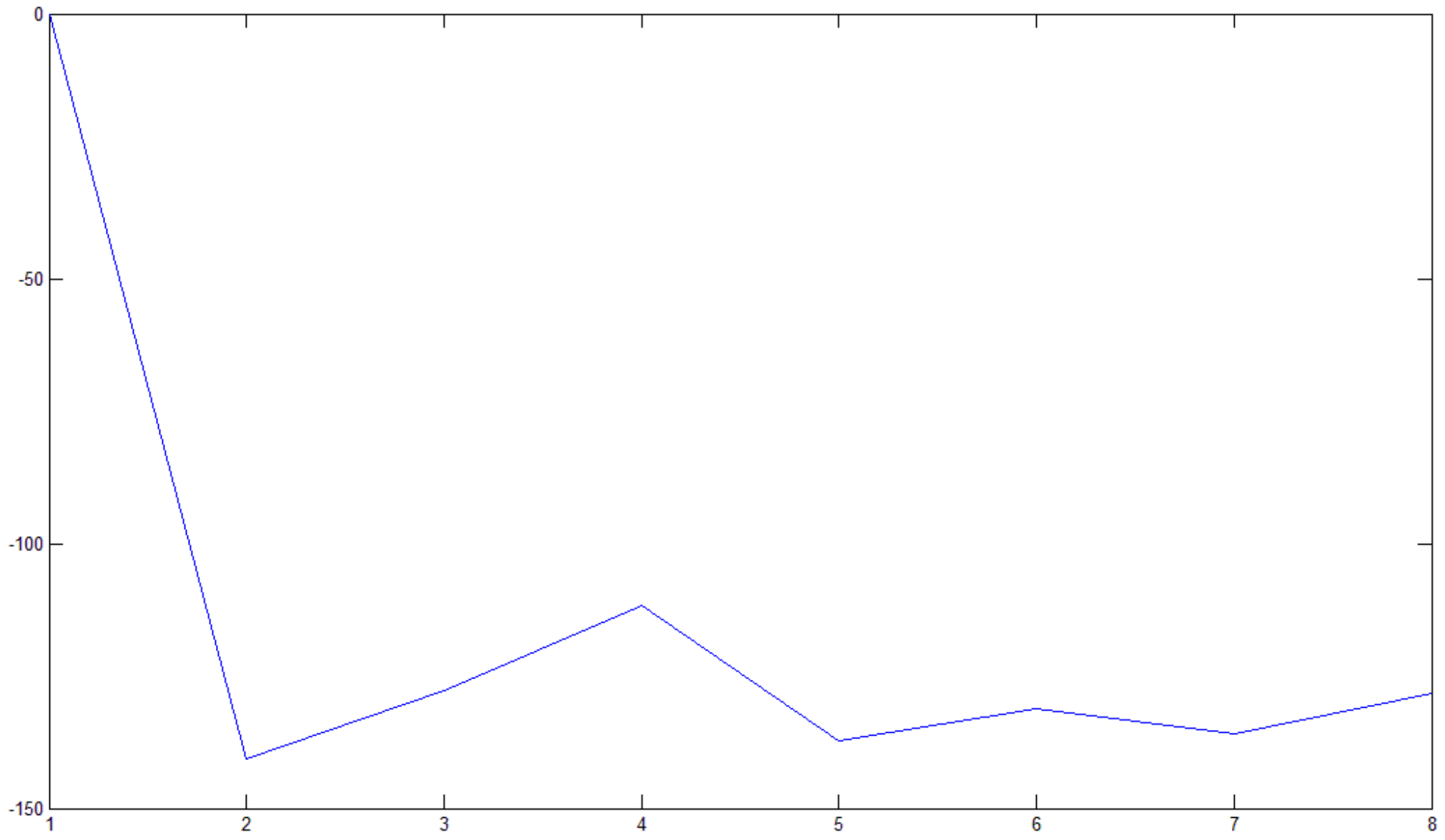

Figure 6.17 RL Measured Relative Phase Difference Between Port 8 and Output Ports.

## 6.3 SUMMARY

This chapter covered the materials regarding the novel *8x8* Rotman Lens circuit implementation and measurements. The measured array factor radiation patterns for each port and also output relative phase differences were shown and analysed. The array factor radiation patterns showed the beam steering functionality of the *8x8* Rotman Lens, although some of the patterns did not satisfy the desired beam directivity and low–level side lobes. This problem could be improved by eliminating the occurred errors mentioned in section 6.2.1 and/or using a shielded metal box with absorbing foams attached to the inner walls and top lid to reduce the interferences. The network analyser has to be calibrated correctly before use, because there are errors which are caused by the cables and connections. Also, during soldering process, a slight error is introduced which is primarily to do with the length of the pin. This pin has to lie on top of the copper track to avoid mismatch and also it should be perpendicular to edge of the transmission line.

# CHAPTER 7
# CONCLUSIONS AND FUTURE WORK

## 7.1 CONCLUSIONS

Researches on ITS and vehicular telematics such as MILTRANS and FATCAT projects were carried out to enhance the safety and efficiency of road transportation related to IVC and RVC systems. Hence, safer and more efficient system of trunk roads and highways can be achieved to the benefit of road users.

In this dissertation, summary designs of the system and results to support the ITS and vehicular telematics projects specifically related to RF beamforming techniques using Butler Matrix and Rotman Lens are presented. The contributions from this comprehensive research project are summarised in the following sections.

### 7.1.1 DESIGN, SIMULATION, FABRICATION, AND MEASUREMENT OF A 4X4 BUTLER MATRIX IN 3.15 GHZ

A *4x4* Butler Matrix beamforming network has been designed, simulated, fabricated, and measured in 3.15 GHz. Its different elements including the branch line coupler, 0dB crossover, and phase shifters have been designed and simulated separately and their output characteristics have been shown. Also, the matrix S–Parameters and computed array factor radiation patterns have been indicated and analysed. A comparison between the simulated results and measured results has been carried out to verify the beam steering concept and demonstrate the functionality of the *4x4* Butler Matrix.

### 7.1.2 DESIGN AND SIMULATION OF A 4X4 ROTMAN LENS IN 3.15 GHZ

A *4x4* Rotman Lens beamforming network has been designed and simulated in 3.15 GHz. Its output characteristics including beam to array phase error, beam to array coupling amplitude, beam to array spillover coupling amplitude, and array factor radiation pattern along with the lens geometry and design equations have been shown and analysed.

### 7.1.3 DESIGN, SIMULATION, FABRICATION, AND MEASUREMENT OF AN 8X8 ROTMAN LENS IN 6.3 GHZ

A novel *8x8* Rotman Lens beamforming network has been designed, simulated, fabricated, and measured in 6.3 GHz. Its simulated output characteristics including beam to array phase error, beam to array coupling amplitude, beam to array spillover coupling amplitude, and array factor radiation pattern have been indicated. Also, the lens measured S–Parameters and calculated array factor radiation patterns along with relative phase differences between each port and the output ports have been shown and analysed. Finally based on the analysis of radiation patterns, the concept of beam steering has been verified and functionality of the *8x8* Rotman Lens has been demonstrated.

## 7.2 FUTURE WORK

In the following, there are some suggested further improvements that can be carried out for both Butler Matrix and Rotman Lens beamforming networks:

i) Investigation of evolutionary computing and genetic algorithms applications, as potential tools of optimisation, in size reduction and performance improvement of the *4x4* Butler Matrix and *8x8* Rotman Lens.

ii) Design and fabrication of a *4x4* Butler Matrix using Lange Coupler, as key element of the matrix, and comparison of results in terms of array factor radiation patterns and space utilisation with the existing matrix.

iii) Integration of microstrip patch antennas with the *4x4* Butler Matrix and *8x8* Rotman Lens and analysis of antenna patterns and beam steering concept in both systems. The proposed systems can further be extended to wideband structures to support 63 GHz and 77 GHz operation frequencies for vehicle communication systems and collision avoidance radars respectively.

iv) Investigation of the cascaded Butler Matrices concept in RF front–end systems and vehicular telematics and further implementation of the system into MMIC (Monolithic Microwave Integrated Circuits) chips.

# APPENDIX A

## – MATLAB Code Listing –

### Calculation of Array Factor Radiation Pattern and Relative Phase Difference

```
clear all
clc

S=[
06300.8000000    -16.92 -178.73
06300.8000000    -12.51  134.96
06300.8000000    -15.03  141.73
06300.8000000    -10.60  110.60
06300.8000000    -14.23   71.72
06300.8000000    -10.99   80.19
06300.8000000    -14.54   28.04
06300.8000000    -16.63   36.94
];

% Element Numbers
N = 8;

% Element Spacing
d = 0.82;

% Theta Zero Direction
% 90 Degree for Braodside, 0 Degree for Endfire.
theta_zero = 90;

AF=zeros(1,360);

for theta=1:360

    % Change Degree to Radian
    deg2radq(theta)=(theta*pi)/180;

    % Array Factor Calculation
    for n=0:N-1
        AF(theta)= AF(theta)+((10^(S(n+1,2)/10)*exp(i*((n*2*pi*d*(cos(deg2radq(theta))-
cos(theta_zero*pi/180))+deg2rad(S(n+1,3)))))));
    end
    AF(theta) = 20*log10(abs(AF(theta)));

end

% Relative Phase Difference Calculation
for n=2:8
    R(n)=(10^(S(n,2)/10)*exp(i*deg2rad(S(n,3))))/(10^(S(n-1,2)/10)*exp(i*deg2rad(S(n-1,3))));
end

% Plot the Array Factor and Relative Phase
polar(deg2radq,50+AF);
figure
plot(AF);
figure
plot(((angle(R))*180)/pi);
```

# APPENDIX B

## – Dimensions of Metal Box for the *4x4* Butler Matrix Housing –

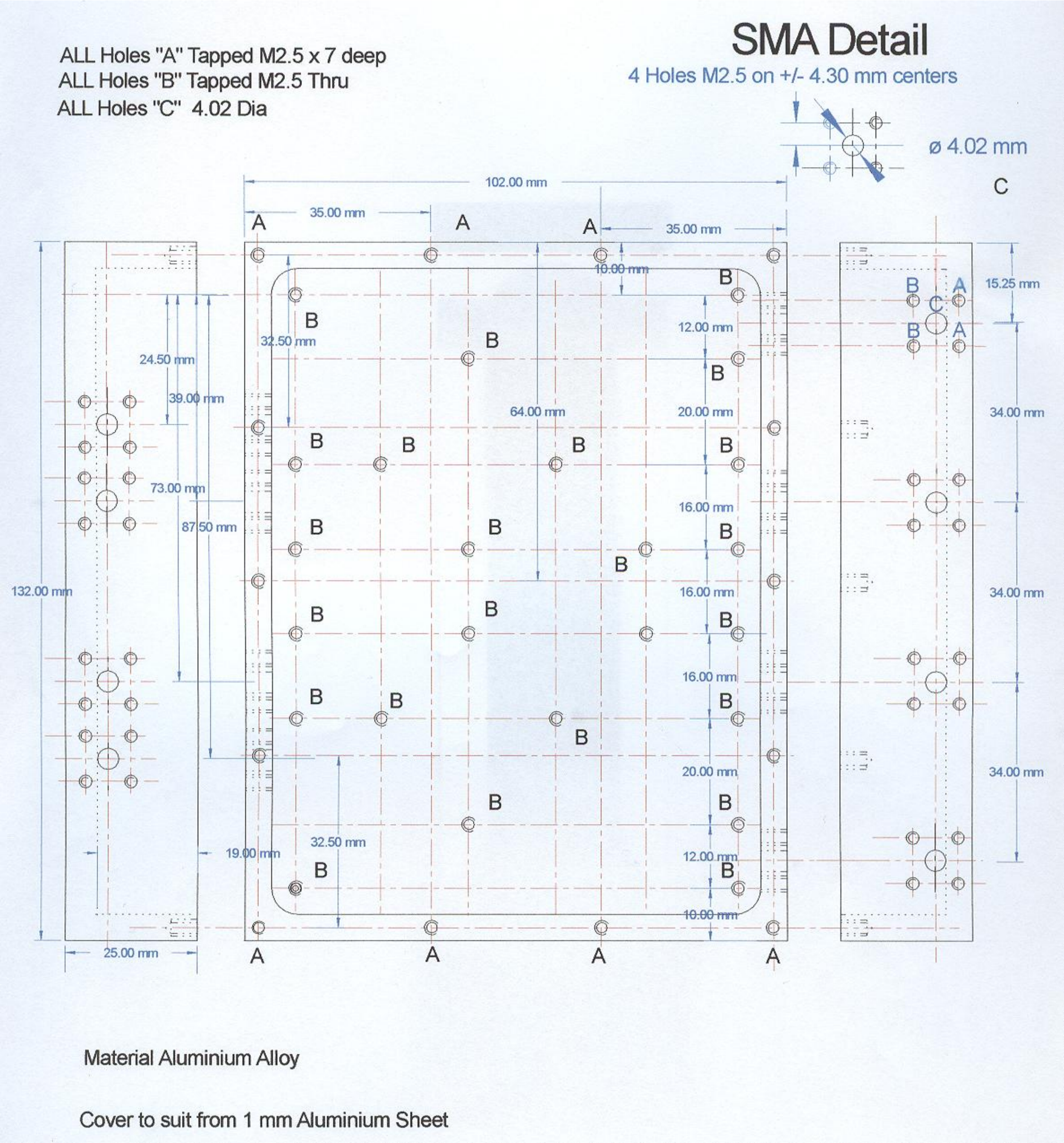

UNIVERSITY OF BIRMINGHAM

# APPENDIX C

**– Data Sheet of Taconic TLC–30 Substrate for the *8x8* Rotman Lens –**

**(Next Page)**

TACONIC
*Finding a better way*

# TLC Low Cost RF Substrate

TLC laminates are engineered to provide a cost effective substrate suitable for a wide range of microwave applications. TLC laminates offer superior electrical performance compared to thermoset laminates (e.g. FR-4, PPO, BT, polyimide and cyanate ester). TLC's construction also provides exceptional mechanical stability.

TLC laminates can be sheared, drilled, milled and plated using standard methods for PTFE/woven fiberglass materials. The laminates are dimensionally stable and exhibit virtually no moisture absorption during fabrication.

Taconic is a world leader in RF laminates and high speed digital materials, offering a wide range of high frequency laminates and prepregs. These advanced materials are used in the fabrication of antennas, multilayer RF and high speed digital boards, interconnections and devices.

## Benefits & Applications:

- **Low cost**
- **Tightly controlled Dk**
- **Low Df**
- **Excellent dimensional stability**
- **High flexural strength**
- **UL 94 V-0 rating**

- **LNBs**
- **Power amplifiers**
- **PCS/PCN large format antennas**
- **Passive components**

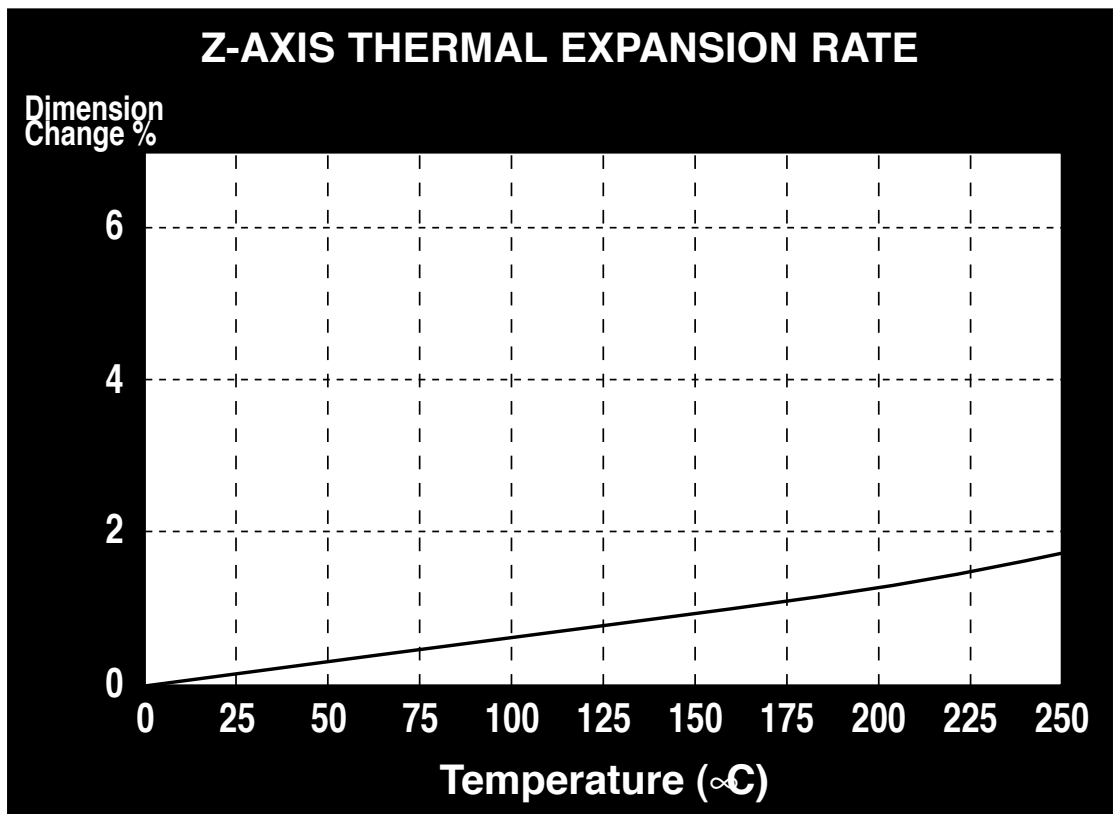

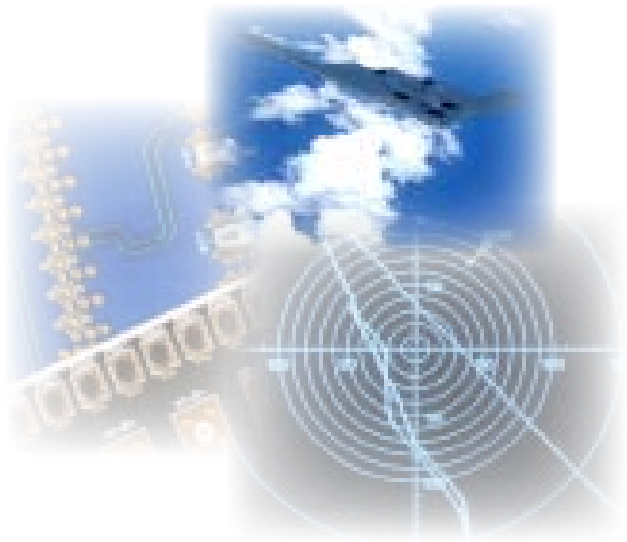

**North & South America**
Taconic - Headquarters
Petersburgh, NY 12138
Tel: 518-658-3202 / 1-800-833-1805
addinfo@4taconic.com

**Europe/Middle East/Australia**
Taconic International Ltd.
Republic of Ireland
Tel: +353-44-9395600
add@4taconic.com

**Asia**
Korea Taconic Company
Republic of Korea
Tel: +82-31-704-1858
sales@taconic.co.kr

**China**
Taconic Advanced Material (Suzhou) Co., Ltd.
Suzhou City, China
Tel: +86-512-8718-9678
tssales@taconic.co.kr

An ISO 9001:2000 Registered Company
www.taconic-add.com

| TLC-32 Typical Values | | | | | |
|---|---|---|---|---|---|
| Property | Test Method | Unit | Value | Unit | Value |
| Dk @ 10 GHz | IPC-650 2.5.5.5 | | 3.20 | | 3.20 |
| Df @ 10 GHz | IPC-650 2.5.5.5 | | 0.0030 | | 0.0030 |
| Moisture Absorption | IPC-650 2.6.2.1 | % | <0.02 | % | <0.02 |
| Dielectric Breakdown | IPC-650 2.5.6 | Kv | >60 | Kv | >60 |
| Volume Resistivity | IPC-650 2.5.17.1 | Mohms/cm | $10^7$ | Mohms/cm | $10^7$ |
| Surface Resistivity | IPC-650 2.5.17.1 | Mohms | $10^7$ | Mohms | $10^7$ |
| Arc Resistance | IPC-650 2.5.1 | seconds | >180 | seconds | >180 |
| Flexural Strength (MD) | IPC-650 2.4.4 | lbs./inch | >40,000 | N/mm$^2$ | >276 |
| Flexural Strength (CD) | IPC-650 2.4.4 | lbs./inch | >35,000 | N/mm$^2$ | >241 |
| Peel Strength (1 oz. copper) | IPC-650 2.4.8 | lbs./linear inch | 12.0 | N/mm | 2.1 |
| Thermal Conductivity | ASTM F 433 | W/m/K | 0.24 | W/m/K | 0.24 |
| CTE (x-y axis) | ASTM D 3386/TMA | ppm/°C | 9 - 12 | ppm/°C | 9 - 12 |
| CTE (z axis) | ASTM D 3386/TMA | ppm/°C | 70 | ppm/°C | 70 |
| UL-94 Flammability Rating | UL-94 | | V-0 | | V-0 |

*All reported values are typical and should not be used for specification purposes. In all instances, the user shall determine suitability in any given application.*

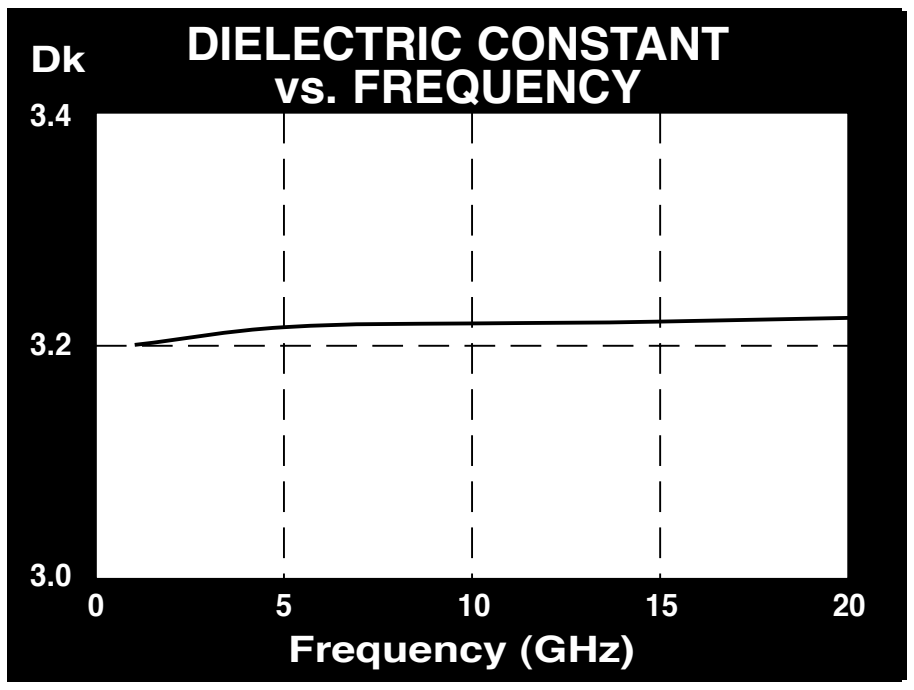

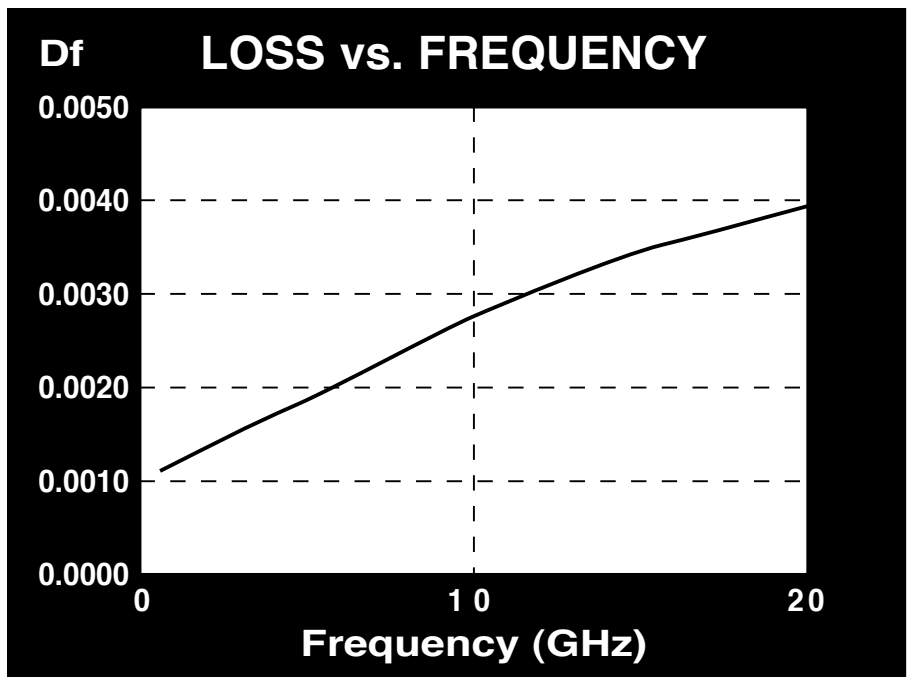

| Designation | Dk | Dielectric Thickness inches | Dielectric Thickness mm |
|---|---|---|---|
| TLC-27 | 2.75 +/-0.05 | 0.0145 | 0.37 |
| TLC-30 | 3.00 +/- 0.05 | 0.0200<br>0.0300 - 0.0620 | 0.50<br>0.78 - 1.5 |
| TLC-32 | 3.20 +/- 0.05 | ≥0.0300 | ≥0.78 |

| Available Copper Cladding | | | | |
|---|---|---|---|---|
| Designation | Weight | Copper Thickness inches | Copper Thickness µm | Description |
| CH | 1/2 oz / ft$^2$ | ~0.0007 | ~18 | Electrodeposited |
| C1 | 1 oz / ft$^2$ | ~0.0014 | ~35 | Electrodeposited |
| C2 | 2 oz / ft$^2$ | ~0.0028 | ~70 | Electrodeposited |

Heavy metal claddings (aluminum, brass & copper) may also be available upon request. Standard sheet size is 36” x 48” (914 mm x 1220 mm). Please contact our Customer Service Department for the availability of other sizes and claddings.

| Available Sheet Sizes Inches | mm |
|---|---|
| 12 x 18 | 304 x 457 |
| 16 x 18 | 406 x 457 |
| 18 x 24 | 457 x 610 |
| 16 x 36 | 406 x 914 |
| 24 x 36 | 610 x 914 |
| 18 x 48 | 457 x 1220 |

RoHS WEEE Compliant

An example of our part number is: **TLC-32-0620-CH/CH - 18” x 24” (457 mm x 610 mm)**

rev. 1/09